\documentclass[prd,twocolumn,showpacs,floatfix,amsmath,amssymb,floatfix]{revtex4}
\usepackage{graphicx,dcolumn,booktabs,bm}
\usepackage{longtable,lscape}
\usepackage{txfonts}
\usepackage{overpic}
\usepackage{multirow}
\usepackage{amssymb}
\usepackage{indentfirst}
\usepackage{feynmf}   
\usepackage{slashed}  
\usepackage{cases}
\usepackage{color,ulem}
\usepackage{graphicx}

\graphicspath{{Figures/}} 

\begin{document}


\title{Strong decays of the $XYZ$ states}
\author{Li Ma$^{1}$}\email{lima@pku.edu.cn}
\author{Xiao-Hai Liu$^{1}$}\email{liuxiaohai@pku.edu.cn}
\author{Xiang Liu$^{2,3}$}\email{xiangliu@lzu.edu.cn}
\author{Shi-Lin Zhu$^{1,4}$}\email{zhusl@pku.edu.cn}

\affiliation{
$^1$Department of Physics and State Key Laboratory of Nuclear Physics and Technology and Center of High Energy Physics, Peking University, Beijing 100871, China\\
$^2$Research Center for Hadron and CSR Physics, Lanzhou University and Institute of Modern Physics of CAS, Lanzhou 730000, China\\
$^3$School of Physical Science and Technology, Lanzhou University, Lanzhou 730000, China\\
$^4$Collaborative Innovation Center of Quantum Matter, Beijing
100871, China}

\date{\today}

\begin{abstract}

Through the spin rearrangement scheme in the heavy quark limit, we
have performed a comprehensive investigation of the decay pattern
and production mechanism of the hidden beauty di-meson states, which
are either composed of a P-wave bottom meson and an S-wave bottom
meson or two S-wave bottom mesons. We further extend the
corresponding formula to discuss the decay behavior of some
charmonium-like states by combining the experimental information
with our numerical results. The typical ratios presented in this
work can be measured by future experiments like BESIII, Belle, LHCb
and the forthcoming BelleII, which shall provide important clues to
the inner structures of the exotic states.

\end{abstract}

 \maketitle

\section{Introduction}\label{sec1}

Many so called $XYZ$ states have been discovered in the past decade
by the Belle, BaBar, CLEO-c, CDF, D0, CMS, LHCb and BESIII
collaborations \cite{Beringer:1900zz,zhu-review,Liu:2013waa}. Among
these observed $XYZ$ states, some of them are good candidates of the
exotic states. Especially, the charged charmonium-like state
$Z_1(4475)$ was firstly observed by Belle \cite{belle1,belle2} and
recently confirmed as a genuine resonance by LHCb \cite{lhcb}. Due
to the peculiarity of $Z_1(4475)$, the number of the quark component
of $Z_1(4475)$ is at least four. Up to now, $Z_1(4475)$ seems to be
the best candidate of the four-quark states.

These $XYZ$ states have stimulated theorist's extensive interest in
revealing their underlying structures. Various theoretical schemes
were proposed, which include the "exotic" explanations like the
tetraquark state, molecular state and charmonium hybrid and
non-resonant interpretations like the cusp effect and initial single
pion emission mechanism
\cite{Swanson:2004pp,Maiani:2004vq,Bugg:2004rk,Rosner:2006vc,Li:2004sta,Hogaasen:2005jv,Ebert:2005nc,Barnea:2006sd,Cui:2006mp,Zhu:2005hp,Kou:2005gt,Close:2005iz,Chen:2011xk,Chen:2013wca}.

Let's take the two charmonium-like states $X(3872)$ and $Y(4260)$ as
an example. Since the mass of $X(3872)$ \cite{Abe:2005ix} is
slightly below the $D\bar{D}^\ast$ threshold, the $D\bar{D}^*$
molecular picture was proposed in Refs.
\cite{Close:2003sg,Voloshin:2003nt,Wong:2003xk,Swanson:2003tb,Tornqvist:2004qy,Suzuki:2005ha,Liu:2008fh,Thomas:2008ja,Lee:2009hy,Li:2012cs}.
The charmonium-like state $Y(4260)$ was reported by BaBar in the
$e^+e^-\to \pi^+\pi^- J/\psi$ process \cite{Aubert:2005rm}. Later
many theoretical explanations were proposed, which include the
traditional charmonium assignment
\cite{LlanesEstrada:2005hz,Eichten:2005ga,Segovia:2008zz,Li:2009zu},
charmonium hybrid \cite{Zhu:2005hp,Kou:2005gt,Close:2005iz},
diquark-antidiaquark state \cite{Maiani:2005pe,Ebert:2008wm},
$D_1\bar{D}^\ast$ molecule or other molecular state assignments
\cite{Liu:2005ay,Yuan:2005dr,Qiao:2005av,Ding:2008gr,MartinezTorres:2009xb,Close:2010wq},
and non-resonant explanation \cite{Chen:2010nv}.

In order to answer whether these $XYZ$ states can be explained as
the candidate of the exotic states, one need carry out the dynamical
calculation by adopting the specific dynamical model. For example,
the one boson exchange model is often applied to study the loosely
molecular state.

On the other hand, the symmetry analysis, which does not depend on
the dynamical model, can be an effective approach to explore the
molecular state. The spin rearrangement scheme based on the heavy
quark symmetry provides another approach to shed light on the inner
structures of the $XYZ$ states through investigating their decay and
production behaviors. There were some discussions on $Z_c(3900)$,
$Z_c(4025)$ and $Z_b(10610)/Z_b(10650)$ using the spin rearrangement
scheme in the heavy quark limit
\cite{Braaten:2013boa,Ohkoda:2012rj,He:2013nwa}. The selection rules
in the meson-antimeson states under the heavy quark symmetry were
discussed in Ref. \cite{Liu:2013rxa}. The relations between the
rates of the radiative transitions from $\Upsilon(5S)$ to the
hypothetical isovector molecular bottomonium resonances with
negative $G$-parity via the spin rearrangement scheme were presented
in Ref. \cite{Voloshin:2011qa}.

Very recently, Ma {\it et al.} discussed the radiative decays of the
$XYZ$ states \cite{Ma:2014ofa,Ma:2014zua}, where the spin
rearrangement scheme in the heavy quark limit was adopted. Besides
their radiative decays, the strong decay patterns and production
behaviors of the XYZ states are crucial to probe their inner
structures. The experimental information of the strong decay modes
of $XYZ$ states is more abundant than that of the radiative decays
of the $XYZ$ states.

In this work we will adopt the spin rearrangement scheme and extend
the formalism in Refs. \cite{Ma:2014ofa,Ma:2014zua} to study the
strong decays of the $XYZ$ states. We will consider the following
three classes of strong decays
\begin{eqnarray*}
B_{(1,2)}\bar{B}^{(\ast)}\rightarrow(b\bar{b})+light\,\,meson,\\
(b\bar{b})\rightarrow B_{(1,2)}\bar{B}^{(\ast)}+light\,\,meson,\\
B_{(1,2)}\bar{B}^{(\ast)}\rightarrow
B_{(1,2)}\bar{B}^{(\ast)}+light\,\,meson.
\end{eqnarray*}
corresponding to the strong decays from one molecular (resonant)
state into a bottomonia, a bottomonia decaying into a
molecular/resonant state, and strong decays from one
molecular/resonant state into another molecular/resonant state,
respectively, where we use the notations $B_{(1,2)}\bar{B}^{(\ast)}$
and $(b\bar{b})$ to denote the molecular/resonant states and
bottomonium, respectively.

This paper is organized as follows. After the introduction, we give
the calculation details of the above three classes of strong decays
in Sec. \ref{sec2}. We present the numerical results in Sec.
\ref{sec3}. With the same method, we discuss the possible
hidden-charm molecules/resonances in Sec. \ref{sec4}. The last
section is the summary.

\section{The formalism of the strong decays in the heavy quark limit}
\label{sec2}

Heavy quark symmetry is a good tool in the study of the structures
of hadrons containing heavy quarks. In the heavy quark limit, heavy
quarks only have spin-independent chromoelectric interactions with
gluons, while the spin-dependent chromomagnetic interaction is
proportional to $1/m_{Q}$ and suppressed. Thus, the conserved
angular momentum operators of a hadron containing the heavy quarks
are the total angular momentum $J$, spin of heavy quarks $S_H$,
which are also named as the "heavy spin" and the spin of light
degrees of freedom $S_l$ with the definition $\vec{S}_l\equiv
\vec{J}-\vec{S}_H$. The spin of the light degrees of freedom
includes all the orbital angular momenta and spin of light quarks
within a hadron, where we simply denote it as the "light spin" in
the following.

In this work, we investigate the strong decays of the hidden beauty
molecular/resonant states composed of two bottom meson. We discuss
two dimeson systems. The first ne is a molecular/resonant state
which is composed of one P-wave bottom meson like
$B_0$, $B_1'$, $B_1$, $B_2$ and one S-wave bottom meson like
$\bar{B}$, $\bar{B}^\ast$. The other is a molecular/resonant state
composed of two S-wave bottom mesons.

In the heavy quark limit, ($B, B^\ast$), ($B_0,B_1'$) and
($B_1,B_2$) belong to the doublets $H=(0^-,1^-)$, $S=(0^+,1^+)$ and
$T=(1^+,2^+)$, respectively. Adopting the same definition of the
$C$-parity eigenstate of the molecular states in Refs.
\cite{Liu:2013rxa,Ma:2014ofa,Ma:2014zua}, we list all relevant
molecular/resonant states in Table \ref{state}. Since in this work
we introduce no dynamical models, we will call them the "hidden
beauty molecules/resonances" for simplicity. In the spin
rearrangement scheme, the decay patterns of a hadron is determined
by their spin configurations which are only determined by the spin
structures of their constitutes.

\renewcommand{\arraystretch}{1.5}
\begin{table}[htbp]
  \caption{The hidden beauty molecular states with different $J^{PC}$ quantum numbers.\label{state}}
\begin{center}
   \begin{tabular}{ c c c c } \toprule[1pt]
   $J^{PC}$&\multicolumn{3}{c}{States}\\
   \midrule[1pt]

     \multirow{2}{*}{$1^{--}$} & $\frac{1}{\sqrt{2}}(B_0\bar{B}^\ast-B^\ast\bar{B}_0)$ &$\frac{1}{\sqrt{2}}(B_1'\bar{B}-B\bar{B}_1')$ & $\frac{1}{\sqrt{2}}(B_1\bar{B}-B\bar{B}_1)$ \\
        &$\frac{1}{\sqrt{2}}(B_1'\bar{B}^\ast+B^\ast\bar{B}_1')$ & $\frac{1}{\sqrt{2}}(B_1\bar{B}^\ast+B^\ast\bar{B}_1)$  & $\frac{1}{\sqrt{2}}(B_2\bar{B}^\ast-B^\ast\bar{B}_2)$ \\
      \multirow{2}{*}{$1^{-+}$} &  $\frac{1}{\sqrt{2}}(B_0\bar{B}^\ast+B^\ast\bar{B}_0)$ & $\frac{1}{\sqrt{2}}(B_1'\bar{B}+B\bar{B}_1')$ & $\frac{1}{\sqrt{2}}(B_1\bar{B}+B\bar{B}_1)$ \\
      & $\frac{1}{\sqrt{2}}(B_1'\bar{B}^\ast-B^\ast\bar{B}_1')$
      & $\frac{1}{\sqrt{2}}(B_1\bar{B}^\ast-B^\ast\bar{B}_1)$ & $\frac{1}{\sqrt{2}}(B_2\bar{B}^\ast+B^\ast\bar{B}_2)$ \\
      $1^{++}$ &   $\frac{1}{\sqrt{2}}(B\bar{B}^\ast+B^\ast\bar{B})$ &  &  \\
      $1^{+-}$ &  $\frac{1}{\sqrt{2}}(B\bar{B}^\ast-B^\ast\bar{B})$ & $B^\ast\bar{B}^\ast$ &   \\
      {$0^{--}$} & $\frac{1}{\sqrt{2}}(B_0\bar{B}-B\bar{B}_0)$ &$\frac{1}{\sqrt{2}}(B_1'\bar{B}^\ast-B^\ast\bar{B}_1')$ & $\frac{1}{\sqrt{2}}(B_1\bar{B}^\ast-B^\ast\bar{B}_1)$ \\
      $0^{-+}$ &  $\frac{1}{\sqrt{2}}(B_0\bar{B}+B\bar{B}_0)$ &$\frac{1}{\sqrt{2}}(B_1'\bar{B}^\ast+B^\ast\bar{B}_1')$ & $\frac{1}{\sqrt{2}}(B_1\bar{B}^\ast+B^\ast\bar{B}_1)$ \\
      $0^{++}$ &  $B\bar{B}$ & $B^\ast\bar{B}^\ast$ &   \\
      \multirow{2}{*}{$2^{--}$} & $\frac{1}{\sqrt{2}}(B_1'\bar{B}^\ast-B^\ast\bar{B}_1')$ &$\frac{1}{\sqrt{2}}(B_1\bar{B}^\ast-B^\ast\bar{B}_1)$ & $\frac{1}{\sqrt{2}}(B_2\bar{B}-B\bar{B}_2)$ \\
        &$\frac{1}{\sqrt{2}}(B_2\bar{B}^\ast+B^\ast\bar{B}_2)$ &  &  \\
      \multirow{2}{*}{$2^{-+}$} &  $\frac{1}{\sqrt{2}}(B_1'\bar{B}^\ast+B^\ast\bar{B}_1')$ &$\frac{1}{\sqrt{2}}(B_1\bar{B}^\ast+B^\ast\bar{B}_1)$ & $\frac{1}{\sqrt{2}}(B_2\bar{B}+B\bar{B}_2)$ \\
        &$\frac{1}{\sqrt{2}}(B_2\bar{B}^\ast-B^\ast\bar{B}_2)$ &  &  \\
      $2^{++}$ &   $B^\ast\bar{B}^\ast$ &  &  \\\bottomrule[1pt]
       \end{tabular}
 \end{center}
\end{table}

In the strong decays of hadrons containing the heavy quarks, not
only the heavy spin, light spin, total angular momentum, $C$-parity
and parity, but also $G$-parity and isospin are conserved. We need
to distinguish the different isospin of a system. The
$B_{(1,2)}\bar{B}^{(\ast)}$ system with one P-wave bottom meson and
one S-wave bottom meson can be categorized as the isovector and
isoscalar states with the corresponding spin wave functions
\begin{eqnarray*}
  |B_{(1,2)}\bar{B}^{(\ast)}\rangle_1^{+} &=& \frac{1}{\sqrt{2}}[|B_{(1,2)}^+\bar{B}^{(\ast)0}\rangle+\tilde{c}|B^{(\ast)+}\bar{B}_{(1,2)}^0\rangle], \\
  |B_{(1,2)}\bar{B}^{(\ast)}\rangle_1^{-} &=&
  \frac{1}{\sqrt{2}}[|B_{(1,2)}^0B^{(\ast)-}\rangle+\tilde{c}|B^{(\ast)0}B_{(1,2)}^-\rangle], \\
  |B_{(1,2)}\bar{B}^{(\ast)}\rangle_1^{0} &=& \frac{1}{2}\bigg\{[|B_{(1,2)}^0\bar{B}^{(\ast)0}\rangle-|B_{(1,2)}^+B^{(\ast)-}\rangle]\\
  &&+\tilde{c}[|B^{(\ast)0}\bar{B}_{(1,2)}^0\rangle-|B^{(\ast)+}B_{(1,2)}^{-}\rangle]\bigg\},
\\
  |B_{(1,2)}\bar{B}^{(\ast)}\rangle_0^{0} &=& \frac{1}{2}\bigg\{[|B_{(1,2)}^0\bar{B}^{(\ast)0}\rangle+|B_{(1,2)}^+B^{(\ast)-}\rangle]\\
  &&+\tilde{c}[|B^{(\ast)0}\bar{B}_{(1,2)}^0\rangle+|B^{(\ast)+}B_{(1,2)}^{-}\rangle]\bigg\},
\end{eqnarray*}
where $\tilde{c}=c(-1)^{L+K-J}$ and $c=\pm 1$ corresponds to
$C$-parity $C=\mp$. The factor $(-1)^{L+K-J}$ is due to the exchange
of the spin vectors of the two bottoms in the systems. The
spin-flavor wave functions of the $B\bar{B}$ systems can be
constructed as
\begin{eqnarray*}
  |B\bar{B}\rangle_1^{+} &=& |B^{+}\bar{B}^{0}\rangle, \\
  |B\bar{B}\rangle_1^{-} &=& |B^{-}\bar{B}^{0}\rangle, \\
  |B\bar{B}\rangle_1^{0} &=& \frac{1}{\sqrt{2}}[|B^{+}\bar{B}^{-}\rangle-|B^{0}\bar{B}^{0}\rangle],
\\
 |B\bar{B}\rangle_0^{0} &=& \frac{1}{\sqrt{2}}[|B^{+}\bar{B}^{-}\rangle+|B^{0}\bar{B}^{0}\rangle].
\end{eqnarray*}
The spin-flavor wave functions of the $B^\ast\bar{B}^\ast$ system
can be categorized as
\begin{eqnarray*}
  |B^\ast\bar{B}^\ast\rangle_1^{+} &=& |B^{\ast+}\bar{B}^{\ast 0}\rangle, \\
  |B^\ast\bar{B}^\ast\rangle_1^{-} &=& |B^{\ast-}\bar{B}^{\ast 0}\rangle, \\
  |B^\ast\bar{B}^\ast\rangle_1^{0} &=& \frac{1}{\sqrt{2}}[|B^{\ast+}\bar{B}^{\ast-}\rangle-|B^{\ast 0}\bar{B}^{\ast 0}\rangle],
\\
  |B^\ast\bar{B}^\ast\rangle_0^{0} &=& \frac{1}{\sqrt{2}}[|B^{\ast+}\bar{B}^{\ast-}\rangle+|B^{\ast 0}\bar{B}^{\ast 0}\rangle],
\end{eqnarray*}

We decompose the total angular momentum of the above systems into
their heavy spin and light spin. We adopt the spin re-coupling
formula in analyzing the general spin structure as in Refs.
\cite{Ohkoda:2012rj,Ma:2014ofa,Ma:2014zua}. For instance, we can
decompose each part in the isoscalar states of the
$B_{(1,2)}\bar{B}^{(\ast)}$ system, i.e.,
\begin{eqnarray}
 && |B_{(1,2)}^0\bar{B}^{(\ast)0}\rangle\nonumber\\ &&= \left[[\bar{b}\otimes(d\otimes 1)_s]_K\otimes[b\otimes \bar{d}]_L\right]_J |(\bar{b}d)(b\bar{d})\rangle  \nonumber\\
 &&=\sum_{g=0}^1 \sum_{m=0}^1 \sum_{h=|s-\frac{1}{2}|}^{s+\frac{1}{2}} \mathcal{A}^{s,L,K,J}_{g,m,h}\Bigg|\left[\left[\bar{b}b\right]_g\otimes\left(\left[d\bar{d}\right]_m\otimes1\right)_h\right]_J \Bigg\rangle |(\bar{b}d)(b\bar{d})\rangle,\nonumber\\
 && |B^{(\ast)0}\bar{B}_{(1,2)}^0\rangle\nonumber\\ &&= \left[\left[\bar{b}\otimes q\right]_L\otimes\left[b\otimes(\bar{d}\otimes 1)_s\right]_K\right]_J |(b\bar{d})(\bar{b}d)\rangle\nonumber\\
 &&=\sum_{g=0}^1 \sum_{m=0}^1 \sum_{h=|s-\frac{1}{2}|}^{s+\frac{1}{2}} \mathcal{B}^{s,L,K,J}_{g,m,h}\Bigg|\left[\left[\bar{b}b\right]_g\otimes\left(\left[d\bar{d}\right]_m\otimes1\right)_h\right]_J \Bigg\rangle |({b}\bar{d})(\bar{b}{d})\rangle,\nonumber\\
 && |B_{(1,2)}^+B^{(\ast)-}\rangle\nonumber\\ &&= \left[[\bar{b}\otimes(u\otimes 1)_s]_K\otimes[b\otimes \bar{u}]_L\right]_J |(\bar{b}u)(b\bar{u})\rangle  \nonumber\\
 &&=\sum_{g=0}^1 \sum_{m=0}^1 \sum_{h=|s-\frac{1}{2}|}^{s+\frac{1}{2}} \mathcal{A}^{s,L,K,J}_{g,m,h}\Bigg|\left[\left[\bar{b}b\right]_g\otimes\left(\left[u\bar{u}\right]_m\otimes1\right)_h\right]_J \Bigg\rangle |(\bar{b}u)(b\bar{u})\rangle.\nonumber\\
 && |B^{(\ast)+}B_{(1,2)}^-\rangle\nonumber\\ &&= \left[\left[\bar{b}\otimes u\right]_L\otimes\left[b\otimes(\bar{u}\otimes 1)_s\right]_K\right]_J |(b\bar{u})(\bar{b}u)\rangle\nonumber\\
 &&=\sum_{g=0}^1 \sum_{m=0}^1 \sum_{h=|s-\frac{1}{2}|}^{s+\frac{1}{2}} \mathcal{B}^{s,L,K,J}_{g,m,h}\Bigg|\left[\left[\bar{b}b\right]_g\otimes\left(\left[u\bar{u}\right]_m\otimes1\right)_h\right]_J \Bigg\rangle |({b}\bar{u})(\bar{b}{u})\rangle,\nonumber\label{h1}
\end{eqnarray}
In the above equations, the indices $b$, $\bar{b}$, $u$, $d$,
$\bar{u}$ and $\bar{d}$ in the square brackets represent the
corresponding quark spin wave functions. The notation
$\left[[\bar{b}\otimes(q_i\otimes 1)_s]_K\otimes[b\otimes
\bar{q_j}]_L\right]_J$ denotes the spin structures of
$B_{(1,2)}\bar{B}^{(\ast)}$. In the heavy quark limit, the spin of
the light quark $q_i(u,d)$ in $B_{(1,2)}$ couples with the P-wave
orbital angular momentum to form the light spin $s$, which further
couples with $\bar{b}$ to form the total angular momentum $K$.
Similarly, the spin of the light quark $\bar{q}_j(u,d)$ in
$\bar{B}^{(\ast)}$ couples with $b$, which corresponds to the total
angular momentum $L$. Then, the coupling between $K$ and $L$ leads
to the total angular momentum $J$ of the systems. The notation
$[[\bar{b}b]_g\otimes([u\bar{u}]_m\otimes1)_h]_J$ means that the two
heavy quark spin $\bar{b}$ and $b$ couple into the heavy spin $g$
and the two light quark spin $u$ and $\bar{u}$ couple into the total
light quark spin $m$. And then, the coupling of $m$ with the orbital
angular momentum from the P-wave bottom meson forms the light spin
$h$. The spin re-coupling coefficients
$\mathcal{A}^{s,L,K,J}_{g,m,h}$ and $\mathcal{B}^{s,L,K,J}_{g,m,h}$
are the same as that in Ref. \cite{Ma:2014ofa}.

We need to emphasize that we explicitly include the flavor wave
function $|(\bar{b}q_i)(b\bar{q_j})\rangle$ in Eq. (\ref{h1}). The
symbol $|(\bar{b}q_i)\rangle$ represents
$|(\bar{b}q_i)\equiv\frac{1}{\sqrt{2}}(|\bar{b}q_i\rangle+|q_i\bar{b}\rangle)$.
Here, the position ordering of the $\bar{b}$ and $b$ cannot be
interchanged in order to guarantee the orthogonalization of the
heavy meson and anti-meson wave functions at the quark level. This
treatment ensures the normalization of the spin configurations after
performing the spin rearrangement, which was discussed in details in
Ref. \cite{Ma:2014ofa}.

The spin structure of the isoscalar states of the
$B_{(1,2)}\bar{B}^{(\ast)}$ systems can be expressed as
\begin{eqnarray*}
  &&|B_{(1,2)}\bar{B}^{(\ast)}\rangle_0^{0} \\
  &&=\frac{1}{\sqrt{2}}\sum_{g,m,h} \Bigg\{\mathcal{A}^{s,L,K,J}_{g,m,h}\Bigg|\left[\left[\bar{b}b\right]_g\otimes\left(\left[\frac{d\bar{d}+u\bar{u}}{\sqrt{2}}\right]_m\otimes1\right)_h\right]_J \Bigg\rangle\\
  &&\times\bigg(\frac{|(\bar{b}d)(b\bar{d})\rangle+|(\bar{b}u)(b\bar{u})\rangle}{\sqrt{2}}\bigg)\nonumber\\
  &&+\tilde{c}\mathcal{B}^{s,L,K,J}_{g,m,h}\Bigg|\left[\left[\bar{b}b\right]_g\otimes\left(\left[\frac{d\bar{d}+u\bar{u}}{\sqrt{2}}\right]_m\otimes1\right)_h\right]_J \Bigg\rangle\\
  &&\times\bigg(\frac{|({b}\bar{d})(\bar{b}{d})\rangle+|({b}\bar{u})(\bar{b}{u})\rangle}{\sqrt{2}}\bigg)\Bigg\}.\nonumber\\\label{h2}
\end{eqnarray*}
Similarly, we obtain the re-coupled spin structures of the isovector
states of the $B_{(1,2)}\bar{B}^{(\ast)}$ systems as
\begin{eqnarray*}
  &&|B_{(1,2)}\bar{B}^{(\ast)}\rangle_1^{0} \\
  &&=\frac{1}{\sqrt{2}}\sum_{g,m,h} \Bigg\{\mathcal{A}^{s,L,K,J}_{g,m,h}\Bigg|\left[\left[\bar{b}b\right]_g\otimes\left(\left[\frac{d\bar{d}-u\bar{u}}{\sqrt{2}}\right]_m\otimes1\right)_h\right]_J \Bigg\rangle\\
  &&\quad\times\bigg(\frac{|(\bar{b}d)(b\bar{d})\rangle-|(\bar{b}u)(b\bar{u})\rangle}{\sqrt{2}}\bigg)\nonumber\\
  &&\quad+\tilde{c}\mathcal{B}^{s,L,K,J}_{g,m,h}\Bigg|\left[\left[\bar{b}b\right]_g\otimes\left(\left[\frac{d\bar{d}-u\bar{u}}{\sqrt{2}}\right]_m\otimes1\right)_h\right]_J \Bigg\rangle\\
  &&\quad\times\bigg(\frac{|({b}\bar{d})(\bar{b}{d})\rangle-|({b}\bar{u})(\bar{b}{u})\rangle}{\sqrt{2}}\bigg)\Bigg\},\nonumber\\\label{h2}
\end{eqnarray*}
\begin{eqnarray*}
  &&|B_{(1,2)}\bar{B}^{(\ast)}\rangle_1^{+} \\
  &&=\frac{1}{\sqrt{2}}\sum_{g,m,h} \Bigg\{\mathcal{A}^{s,L,K,J}_{g,m,h}\Bigg|\left[\left[\bar{b}b\right]_g\otimes\left(\left[u\bar{d}\right]_m\otimes1\right)_h\right]_J \Bigg\rangle\\
  &&\quad+\tilde{c}\mathcal{B}^{s,L,K,J}_{g,m,h}\Bigg|\left[\left[\bar{b}b\right]_g\otimes\left(\left[u\bar{d}\right]_m\otimes1\right)_h\right]_J \Bigg\rangle\Bigg\}|(\bar{b}u)(b\bar{d})\rangle,\nonumber\\\label{h2}
\end{eqnarray*}
and
\begin{eqnarray*}
  &&|B_{(1,2)}\bar{B}^{(\ast)}\rangle_1^{-} \\
  &&=\frac{1}{\sqrt{2}}\sum_{g,m,h} \Bigg\{\mathcal{A}^{s,L,K,J}_{g,m,h}\Bigg|\left[\left[\bar{b}b\right]_g\otimes\left(\left[-d\bar{u}\right]_m\otimes1\right)_h\right]_J \Bigg\rangle\\
  &&\quad+\tilde{c}\mathcal{B}^{s,L,K,J}_{g,m,h}\Bigg|\left[\left[\bar{b}b\right]_g\otimes\left(\left[u\bar{d}\right]_m\otimes1\right)_h\right]_J \Bigg\rangle\Bigg\}|-(\bar{b}d)(b\bar{u})\rangle.\nonumber\\\label{h2}
\end{eqnarray*}

In heavy quark limit, the bottomonia can also be decomposed into the
heavy spin and light spin
\begin{eqnarray}
   |\eta_b(1^1S_0) \rangle &=& |(0_H^-\otimes0_l^+)_0^{-+}\rangle|(b\bar{b})\rangle,\label{k1}\\
  | \Upsilon(1^3S_1)\rangle &=& |(1_H^-\otimes0_l^+)_1^{--}\rangle| |(b\bar{b})\rangle,\\
  | h_b(1^1P_1)\rangle &=&| (0_H^-\otimes1_l^-)_1^{+-}\rangle|(b\bar{b})\rangle,\\
  | \chi_{b0}(1^3P_0) \rangle&=&| (1_H^-\otimes1_l^-)_0^{++}\rangle|(b\bar{b})\rangle,\\
   |\chi_{b1}(1^3P_1) \rangle&=&| (1_H^-\otimes1_l^-)_1^{++}\rangle|(b\bar{b})\rangle,\\
  | \chi_{b2}(1^3P_2)\rangle &=&| (1_H^-\otimes1_l^-)_2^{++}\rangle|(b\bar{b})\rangle,\\
   |\eta_{b2}(1^1D_2)\rangle &=&| (0_H^-\otimes2_l^+)_2^{-+}\rangle|(b\bar{b})\rangle,\\
   |\Upsilon(1^3D_1)\rangle &=&| (1_H^-\otimes2_l^+)_1^{--}\rangle|(b\bar{b})\rangle,\\
  | \Upsilon(1^3D_2)\rangle &=&| (1_H^-\otimes2_l^+)_2^{--}\rangle|(b\bar{b})\rangle,\\
  | \Upsilon(1^3D_3)\rangle &=&| (1_H^-\otimes2_l^+)_3^{--}\rangle|(b\bar{b})\rangle,\label{k2}
  \end{eqnarray}
where the flavor wave function is defined as
$|(b\bar{b})\rangle\equiv\frac{1}{\sqrt{2}}(|\bar{b}b\rangle+|b\bar{b}\rangle)$.
The superscripts $+$ and $-$ inside the parentheses denote the
positive and negative parity of the corresponding parts,
respectively, while the superscripts $-+$ and subscripts $0,1,2,3$
outside the parentheses correspond to the quantum numbers $PC$ and
$J$ of $J^{PC}$ of the bottomonium. The subscripts $H$ and $l$ are
used to distinguish the heavy and light spins of a bottomonia. Here,
the spin wave functions reflect the $C$ parity of the bottomonia,
i.e., $C=(-1)^{S_H+S_l}$.

We also need the spin structures of the light mesons, i.e.,
 \begin{eqnarray}
   |\pi^+ \rangle &=&| (0_H^+\otimes 0_l^-)_0^{-+}\rangle|(u\bar{d})  \rangle, \\
   |\pi^0 \rangle &=&| (0_H^+\otimes 0_l^-)_0^{-+}\rangle|\frac{1}{\sqrt{2}}(d\bar{d}-u\bar{u})  \rangle, \\
   |\pi^- \rangle &=&| (0_H^+\otimes 0_l^-)_0^{-+}\rangle|-(d\bar{u})  \rangle, \label{n1}\\
   |\rho^+\rangle &=& |(0_H^+\otimes 1_l^-)_1^{--}\rangle|(u\bar{d})  \rangle,\\
   |\rho^0\rangle &=& |(0_H^+\otimes 1_l^-)_1^{--}\rangle|\frac{1}{\sqrt{2}}(d\bar{d}-u\bar{u})  \rangle,\\
   |\rho^-\rangle &=& |(0_H^+\otimes 1_l^-)_1^{--}\rangle|-(d\bar{u})  \rangle,\label{n2}\\
   |\eta \rangle &=&| (0_H^+\otimes 0_l^-)_0^{-+}\rangle|\frac{1}{\sqrt{2}}(d\bar{d}+u\bar{u})  \rangle, \label{n3}\\
   |\omega\rangle &=& |(0_H^+\otimes 1_l^-)_1^{--}\rangle|\frac{1}{\sqrt{2}}(d\bar{d}+u\bar{u})  \rangle,\label{n4}\\
   |\sigma \rangle &=&| (0_H^+\otimes 0_l^+)_0^{++}\rangle|\frac{1}{\sqrt{2}}(d\bar{d}+u\bar{u})  \rangle, \label{n5}
 \end{eqnarray}

The orthogonalization of the spin wave functions are defined as
 \begin{eqnarray}
   \langle(a_H\otimes b_L)_J^{pc}|(c_H\otimes d_L)_{J'}^{p'c'}\rangle =\delta_{ac}\delta_{bd}\delta_{JJ'}\delta_{pp'}\delta_{cc'},
  \end{eqnarray}
where the superscripts $p^{(\prime)}$ and $c^{(\prime)}$ represent
the parity and $C$ parity, respectively. This formula reflects the
conservation of the parity, $C$ parity, the total angular momentum,
heavy spin, and light spin.

In addition, the orthogonalization of the flavor wave functions
leads to
\begin{eqnarray*}
    \langle (\bar{b}q_i)(b\bar{q_m})|(\bar{b}q_j)(b\bar{q_n})\rangle &=& \delta_{ij}\delta_{mn}, \\
    \langle (b\bar{q_i})(\bar{b}q_m)|(\bar{b}q_j)(b\bar{q_n})\rangle &=& 0, \\
    \langle (b\bar{q_i})(\bar{b}q_m)|(b\bar{q_j})(\bar{b}q_n)\rangle &=& \delta_{ij}\delta_{mn}, \\
    \langle( \bar{b}q_i)(b\bar{q_m})|(b\bar{q_j})(\bar{b}q_n)\rangle &=& 0,
\end{eqnarray*}
where $q_i$, $q_j$, $q_m$ and $q_n$ can be $u$ or $d$ quark. We need
to specify that the position ordering of the $\bar{b}$ and $b$,
$q_i$ and $\bar{q_m}$ cannot be interchanged. This definition
guarantees the orthogonalization of their total wave functions.
Moreover, the above definition guarantees that
$|B_{(1,2)}^0\bar{B}^{(\ast)0}\rangle$ and
$|B^{(\ast)0}\bar{B}_{(1,2)}^0\rangle$ are different physical
states.

The effective strong decay Hamiltonian $H_{eff}$ conserves the heavy
spin, light spin, isospin, parity, C parity and $G$-parity
separately, which can be decomposed into the spatial and flavor
parts,
\begin{equation}
  H_{eff}=H_{eff}^{spatial}\otimes H_{eff}^{flavor},
\end{equation}
For the decays
$B_{(1,2)}\bar{B}^{(\ast)}\rightarrow(b\bar{b})+light\,\,meson$, the
transition matrix elements related to the flavor wave functions can
be written as
\begin{eqnarray*}
    \langle q_i\bar{q_m}|\langle\bar{b}b|H_{eff}^{flavor}|(\bar{b}q_j)(b\bar{q_n})\rangle &=& \delta_{ij}\delta_{mn}, \\
    \langle q_i\bar{q_m}|\langle b\bar{b}|H_{eff}^{flavor}|(\bar{b}q_j)(b\bar{q_n})\rangle &=& 0,\\
    \langle \bar{q_i}q_m|\langle b\bar{b}|H_{eff}^{flavor}|(b\bar{q_j})(\bar{b}q_n)\rangle &=& \delta_{ij}\delta_{mn}, \\
    \langle \bar{q_i}q_m|\langle \bar{b}b|H_{eff}^{flavor}|(b\bar{q_j})(\bar{b}q_n)\rangle &=& 0.
\end{eqnarray*}

To calculate the strong decays, we also introduce the rearranged
spin structure of the final state. Its general expression is
\begin{eqnarray}
 && |Bottomionia\rangle\otimes|\textsf{light}\,\, \textsf{meson}\rangle\nonumber\\
 & &= \left[[(\bar{b}b)_g\otimes L ]_K\otimes Q\right]_J |(\bar{b}b)\rangle|(q_i\bar{q_j})\rangle  \nonumber\nonumber\\
  &&=\sum_{h=|L-Q|}^{L+Q}  \mathcal{D}^{g,L,K,J}_{g,h}\Bigg|\left[\left(\bar{b}b\right)_g\otimes\left[L\otimes Q\right]_h\right]_J \Bigg\rangle |(\bar{b}b)\rangle|(q_i\bar{q_j})\rangle,\label{11}
   \end{eqnarray}
where the indices $b$, $\bar{b}$ and $q_i$, $\bar{q_j}$ in the
square brackets denote the corresponding spin wave functions. And
$g$ and $L$ denote the heavy and light spin of the bottomonium,
respectively. We collect the coefficients
$\mathcal{D}^{L,K,J}_{g,h}$ in Table \ref{final state}.

For the decays $(b\bar{b})\rightarrow
B_{(1,2)}\bar{B}^{(\ast)}+light\,\,meson$, the transition matrix
elements relevant to the flavor wave functions read as
\begin{eqnarray*}
  \langle q_i\bar{q_j}|\langle(\bar{b}q_m)(b\bar{q_n})|H_{eff}^{flavor}|b\bar{b}\rangle &=& 0 \\
  \langle q_i\bar{q_j}|\langle(\bar{b}q_m)(b\bar{q_n})|H_{eff}^{flavor}|\bar{b}b\rangle &=& \delta_{in}\delta_{jm}+\delta_{ij}\delta_{mn} \\
  \langle q_i\bar{q_j}|\langle(b\bar{q_m})(\bar{b}q_n)|H_{eff}^{flavor}|b\bar{b}\rangle &=& \delta_{im}\delta_{jn}+\delta_{ij}\delta_{mn} \\
  \langle q_i\bar{q_j}|\langle(b\bar{q_m})(\bar{b}q_n)|H_{eff}^{flavor}|\bar{b}b\rangle &=& 0 \\
  \langle \bar{q_i}q_j|\langle(\bar{b}q_m)(b\bar{q_n})|H_{eff}^{flavor}|b\bar{b}\rangle &=& 0 \\
  \langle \bar{q_i}q_j|\langle(\bar{b}q_m)(b\bar{q_n})|H_{eff}^{flavor}|\bar{b}b\rangle &=& \delta_{im}\delta_{jn}+\delta_{ij}\delta_{mn} \\
  \langle \bar{q_i}q_j|\langle(b\bar{q_m})(\bar{b}q_n)|H_{eff}^{flavor}|b\bar{b}\rangle &=& \delta_{in}\delta_{jm}+\delta_{ij}\delta_{mn} \\
  \langle \bar{q_i}q_j|\langle(b\bar{q_m})(\bar{b}q_n)|H_{eff}^{flavor}|\bar{b}b\rangle &=& 0.
\end{eqnarray*}

In the decays $(b\bar{b})\rightarrow
B_{(1,2)}\bar{B}^{(\ast)}+light\,\,meson$, the final states need to
be decomposed in the similar way,
\begin{eqnarray*}
 && |B_{1,2}\bar{B}^{(\ast)}\rangle\otimes|\textsf{light}\,\, \textsf{meson}\rangle\nonumber\\ &&= [[\bar{b}\otimes(q\otimes 1)_s]_K\otimes[b\otimes \bar{q}]_L]_J\otimes(0_H^+\otimes Q_L^\pm)\\
  &&= \sum_{g=0}^1 \sum_{m=0}^1 \sum_{h=|s-\frac{1}{2}|}^{s+\frac{1}{2}} \sum_{h_0=|h-Q|}^{h+Q}\mathcal{E}^{s,L,K,J,Q,J_0}_{g,m,h,h_0}\\
   &&\quad\times\{[\bar{b}b]_g\otimes[([q\bar{q}]_m\otimes1)_h\otimes Q]_{h_0}\}_{J_0},
\end{eqnarray*}
and
\begin{eqnarray*}
 && |B^{(\ast)}\bar{B_{1,2}}\rangle\otimes|\textsf{light}\,\, \textsf{meson}\rangle\nonumber\\ &&= [[b\otimes \bar{q}]_L\otimes[\bar{b}\otimes(q\otimes 1)_s]_K]_J\otimes(0_H^+\otimes Q_L^\pm)\\
  &&= \sum_{g=0}^1 \sum_{m=0}^1 \sum_{h=|s-\frac{1}{2}|}^{s+\frac{1}{2}} \sum_{h_0=|h-Q|}^{h+Q} \mathcal{F}^{s,L,K,J,Q,J_0}_{g,m,h,h_0} \\
  &&\quad\times\{[\bar{b}b]_g\otimes[([q\bar{q}]_m\otimes1)_h\otimes Q]_{h_0}\}_{J_0},
\end{eqnarray*}
which will be applied in the following calculation.

For the decays $B_{(1,2)}\bar{B}^{(\ast)}\rightarrow
B_{(1,2)}\bar{B}^{(\ast)}+light\,\,meson$, the transition matrix
elements are
\begin{eqnarray*}
  &&\langle q_i\bar{q_j}|\langle(\bar{b}q_m)(b\bar{q_n})|H_{eff}^{flavor}|(b\bar{q_k})(\bar{b}q_l)\rangle = 0, \\
  &&\langle q_i\bar{q_j}|\langle(\bar{b}q_m)(b\bar{q_n})|H_{eff}^{flavor}|(\bar{b}q_k)(b\bar{q_l})\rangle\nonumber\\&& = \delta_{in}\delta_{jm}\delta_{kl}+\delta_{il}\delta_{jk}\delta_{mn} +\delta_{ij}\delta_{ml}\delta_{nk}+\delta_{ij}\delta_{mn}\delta_{kl}+\delta_{in}\delta_{jk}\delta_{ml}\nonumber\\&&\quad+\delta_{il}\delta_{jm}\delta_{nk},\\
  &&\langle q_i\bar{q_j}|\langle(b\bar{q_m})(\bar{b}q_n)|H_{eff}^{flavor}|(b\bar{q_k})(\bar{b}q_l)\rangle \nonumber\\&&= \delta_{im}\delta_{jn}\delta_{kl}+\delta_{ik}\delta_{jl}\delta_{mn}+\delta_{ij}\delta_{nk}\delta_{ml}+\delta_{ij}\delta_{mn}\delta_{kl}+\delta_{im}\delta_{jl}\delta_{nk}\nonumber\\&&\quad+\delta_{ik}\delta_{jn}\delta_{ml},\\
  &&\langle q_i\bar{q_j}|\langle(b\bar{q_m})(\bar{b}q_n)|H_{eff}^{flavor}|\bar{b}b\rangle = 0, \\
  &&\langle \bar{q_i}q_j|\langle(\bar{b}q_m)(b\bar{q_n})|H_{eff}^{flavor}|(b\bar{q_k})(\bar{b}q_l)\rangle = 0, \\
  &&\langle \bar{q_i}q_j|\langle(\bar{b}q_m)(b\bar{q_n})|H_{eff}^{flavor}|(\bar{b}q_k)(b\bar{q_l})\rangle \nonumber\\&&= \delta_{jn}\delta_{im}\delta_{kl}+\delta_{jl}\delta_{ik}\delta_{mn}+\delta_{ij}\delta_{ml}\delta_{nk}+\delta_{ij}\delta_{mn}\delta_{kl}+\delta_{jn}\delta_{ik}\delta_{ml} \\
  &&\quad+\delta_{jl}\delta_{im}\delta_{nk},
  \end{eqnarray*}
  \begin{eqnarray*}
  &&\langle \bar{q_i}q_j|\langle(b\bar{q_m})(\bar{b}q_n)|H_{eff}^{flavor}|(b\bar{q_k})(\bar{b}q_l)\rangle\nonumber\\&& = \delta_{jm}\delta_{in}\delta_{kl}+\delta_{jk}\delta_{il}\delta_{mn} +\delta_{ij}\delta_{nk}\delta_{ml}+\delta_{ij}\delta_{mn}\delta_{kl}+\delta_{jm}\delta_{il}\delta_{nk} \\
  &&\quad+\delta_{jk}\delta_{in}\delta_{ml},\\
  &&\langle \bar{q_i}q_j|\langle(b\bar{q_m})(\bar{b}q_n)|H_{eff}^{flavor}|(\bar{b}q_k)(b\bar{q_l})\rangle = 0.
\end{eqnarray*}

\renewcommand{\arraystretch}{1.4}
\begin{table*}[htbp]
 \caption{The coefficient $\mathcal{D}^{L,K,J}_{g,h}$ in Eq. (\ref{11}) corresponding to different combinations of $[g,h]$.}\label{final state}
\begin{center}
   \begin{tabular}{c|cc|c c c c|cccc} \toprule[1pt]
     &\multicolumn{2}{c|}{$J=0$}  &&\multicolumn{2}{c}{$J=1$}&  &&\multicolumn{2}{c}{$J=2$}&\\\midrule[1pt]

    & $[0,0]$   &  $[1,1]$
      & $[0,1]$  &  $[1,0]$ &  $[1,1]$ &  $[1,2]$ & $[0,2]$   &  $[1,1]$ &  $[1,2]$  &  $[1,3]$\\\midrule[1pt]

      $|\eta_b(1^1S_0)\pi/\eta/\sigma\rangle$ & 1 & 0
      & -- & -- & -- & -- &--&--&--&--\\

      $|\Upsilon(1^3S_1)\pi/\eta/\sigma\rangle$ & -- & --
       & 0 & 1 & 0 & 0
       & -- & -- & -- & --\\

      $|h_b(1^1P_1)\pi/\eta/\sigma\rangle$ & -- & --
       & 1 & 0 & 0 & 0
       & -- & -- & -- & --\\

      $|\chi_{b0}(1^3P_0)\pi/\eta/\sigma\rangle$ & 0 & 1
       & -- & -- & -- & -- &--&--&--&-- \\

      $|\chi_{b1}(1^3P_1)\pi/\eta/\sigma\rangle$ & -- & --
       & 0 & 0 & 1 & 0
       & -- & -- & -- & --\\

      $|\chi_{b2}(1^3P_2)\pi/\eta/\sigma\rangle$ &--&--
       & -- & -- & -- & --
       & 0 & 1 & 0 & 0\\

      $|\eta_{b2}(1^1D_2)\pi/\eta/\sigma\rangle$ &--&--
       & -- & -- & -- & --
       & 1 & 0 & 0 & 0\\

      $|\Upsilon(1^3D_1)\pi/\eta/\sigma\rangle$ & -- &--
       & 0 & 0 & 0 & 1
       & -- & -- & -- & --\\

      $|\Upsilon(1^3D_2)\pi/\eta/\sigma\rangle$ &--&--
       & -- & -- & -- & --
       & 0 & 0 & 1 & 0 \\\hline

      $|\eta_b(1^1S_0)\rho/\omega\rangle$ &--&--
      & 1 & 0 & 0 & 0 &--&--&--&--\\

      $|\Upsilon(1^3S_1)\rho/\omega\rangle$& 0 & 1
       & 0 & 0 & 1 & 0
       & 0 & 1 & 0 & 0\\

      $|h_b(1^1P_1)\rho/\omega\rangle$& 1 & 0
       & 1 & 0 & 0 & 0
       & 1 & 0 & 0 & 0\\
      $|\chi_{b0}(1^3P_0)\rho/\omega\rangle$&--&--
       & 0 & $\frac{1}{3}$ & $-\frac{\sqrt{3}}{3}$ & $\frac{\sqrt{5}}{3}$&--&--&--&-- \\

      $|\chi_{b1}(1^3P_1)\rho/\omega\rangle$& 0 & 1
       & 0 & $-\frac{\sqrt{3}}{3}$ & $\frac{1}{2}$ & $\frac{\sqrt{15}}{6}$
       & 0 & $-\frac{1}{2}$ & $\frac{\sqrt{3}}{2}$ & 0\\

      $|\chi_{b2}(1^3P_2)\rho/\omega\rangle$&--&--
       & 0 & $\frac{\sqrt{5}}{3}$ & $\frac{\sqrt{15}}{6}$ & $\frac{1}{6}$
       & 0 & $\frac{\sqrt{3}}{2}$ & $\frac{1}{2}$ & 0\\

      $|\eta_{b2}(1^1D_2)\rho/\omega\rangle$&--&--
       & 1 & 0 & 0 & 0
       & 1 & 0 & 0 & 0\\

      $|\Upsilon(1^3D_1)\rho/\omega\rangle$& 0 &1
       & 0 & 0 & $-\frac{1}{2}$ & $\frac{\sqrt{3}}{2}$
       & 0 & $\frac{1}{10}$ & $-\frac{\sqrt{15}}{10}$ & $\frac{\sqrt{21}}{5}$\\

      $|\Upsilon(1^3D_2)\rho/\omega\rangle$&--&--
       & 0 & 0 & $\frac{\sqrt{3}}{2}$ & $\frac{1}{2}$
       & 0 & $-\frac{\sqrt{15}}{10}$ & $\frac{5}{6}$ & $\frac{\sqrt{35}}{15}$\\

      \bottomrule[1pt]
      \end{tabular}
  \end{center}
\end{table*}

\section{Numerical results}\label{sec3}

With the help of heavy quark symmetry, the conservations of parity,
$C$-parity and $G$-parity, we are ready to discuss the strong decays
of the $B_{(1,2)}\bar{B}^{(\ast)}
(B\bar{B}\,\,\mathrm{or}\,\,B^\ast\bar{B}^\ast)$ systems, which can
be categorized into three groups:
\begin{eqnarray*}
&&B_{(1,2)}\bar{B}^{(\ast)}(B\bar{B}\,\,\mathrm{or}\,\,B^\ast\bar{B}^\ast)\rightarrow(b\bar{b})+light\,\,meson,\\
&&(b\bar{b})\rightarrow B_{(1,2)}\bar{B}^{(\ast)} (B\bar{B}\,\,\mathrm{or}\,\,B^\ast\bar{B}^\ast)+light\,\,meson,\\
&&B_{(1,2)}\bar{B}^{(\ast)}
(B\bar{B}\,\,\mathrm{or}\,\,B^\ast\bar{B}^\ast)+light\,\,meson\\&&\rightarrow
B_{(1,2)}\bar{B}^{(\ast)}
(B\bar{B}\,\,\mathrm{or}\,\,B^\ast\bar{B}^\ast)+light\,\,meson.
\end{eqnarray*}

We collect the typical decay ratios in Tables
\ref{tab:3}-\ref{tab:8}, where the ratios in the brackets are the
results considering the contribution from the phase space factors.
We need to specify that in this work we do not introduce any
dynamics model for the strong decays. Generally the decay width of a
specific decay channel is proportional to the spatial matrix
elements which are related to its spatial wave functions. Only if
the initial systems and final states belong to the same heavy spin
multiplet, the spatial matrix elements of these strong decays are
the same, which leads to quite simple ratios between their decay
widths, as we have discussed in our former work
\cite{Ma:2014ofa,Ma:2014zua}. Since the masses of $B_0$ meson and
D-wave bottomonia are still absent experimentally, we ignore the
contribution of the phase space factors when calculating the
corresponding ratios. In the following, we present the numerical
results.

\subsection{$B_{(1,2)}\bar{B}^{(\ast)} (B\bar{B}\,\,\mathrm{or}\,\,B^\ast\bar{B}^\ast)\rightarrow(b\bar{b})+light\,\,meson$}

\renewcommand{\arraystretch}{1.0}
\begin{table*}[htbp]
\begin{center}
\caption{\label{tab:1}  The typical relations between the decay
widths
$\Gamma(B_{(1,2)}\bar{B}^{(\ast)}\rightarrow(b\bar{b})+light\,\,meson)$
and the reduced matrix elements $H_{\alpha}^{ij}\propto\langle
Q,i\|H_{eff}(\alpha)\|j\rangle$, where the indices $i$ and $j$
denote the light spin of the final and initial hadron respectively,
and $Q$ is the angular momentum of the final light meson.}
   \begin{tabular}{c cccccccccccc} \toprule[1pt]
    $I^G(J^{pc})$ & {Initial state} & \multicolumn{4}{c}{Final state} \\\midrule[1pt]
      & & $h_b\pi$ & $\chi_{b0}\rho$ & $\chi_{b1}\rho$ & $\chi_{b2}\rho$   \\
      \multirow{6}{*}{$1^+(1^{--})$} & $\frac{1}{\sqrt{2}}(B_0\bar{B}^\ast-B^\ast\bar{B}_0)$     &$\frac{\sqrt{6}}{3}H_{\pi}^{11}$    & $-\frac{\sqrt{2}}{3}H_{\rho}^{11}$      &$-\frac{\sqrt{6}}{6}H_{\rho}^{11}$     &$-\frac{\sqrt{10}}{6}H_{\rho}^{11}$ \\

      &$\frac{1}{\sqrt{2}}(B_1'\bar{B}-B\bar{B}_1')$   &   $-\frac{\sqrt{6}}{3}H_{\pi}^{11}$    & $-\frac{\sqrt{2}}{3}H_{\rho}^{11}$      &$-\frac{\sqrt{6}}{6}H_{\rho}^{11}$     &$-\frac{\sqrt{10}}{6}H_{\rho}^{11}$ \\

      &$\frac{1}{\sqrt{2}}(B_1\bar{B}-B\bar{B}_1)$ &$-\frac{\sqrt{3}}{3}H_{\pi}^{11}$    &$-\frac{1}{3}H_{\rho}^{11}$    & $\frac{\sqrt{3}}{6}H_{\rho}^{11}$     & $\frac{\sqrt{5}}{6}H_{\rho}^{11}$  \\

      &$\frac{1}{\sqrt{2}}(B_1'\bar{B}^\ast+B^\ast\bar{B}_1')$ & $\frac{2\sqrt{3}}{3}H_{\pi}^{11}$    &$0$    & $0$      &$0$  \\

      &$\frac{1}{\sqrt{2}}(B_1\bar{B}^\ast+B^\ast\bar{B}_1)$ & $-\frac{\sqrt{6}}{6}H_{\pi}^{11}$    &$-\frac{\sqrt{2}}{2}H_{\rho}^{11}$    & $\frac{\sqrt{6}}{4}H_{\rho}^{11}$      & $\frac{\sqrt{10}}{4}H_{\rho}^{11}$  \\

      &$\frac{1}{\sqrt{2}}(B_2\bar{B}^\ast-B^\ast\bar{B}_2)$ & $-\frac{\sqrt{30}}{6}H_{\pi}^{11}$    &$\frac{\sqrt{10}}{6}H_{\rho}^{11}$    & $-\frac{\sqrt{30}}{12}H_{\rho}^{11}$      & $-\frac{5\sqrt{2}}{12}H_{\rho}^{11}$  \\

      \midrule[1pt]

      $I^G(J^{pc})$ & {Initial state} & \multicolumn{4}{c}{Final state} \\\midrule[1pt]
      & & $\Upsilon(1^3S_1)\pi$ & $\Upsilon(1^3D_1)\pi$ & $\eta_b\rho$ &  $\eta_{b2}\rho$ & \\
      \multirow{2}{*}{$1^+(1^{+-})$} & $\frac{1}{\sqrt{2}}(B\bar{B}^\ast-B^\ast\bar{B})$     &$-H_{\pi}^{00}$    &$0$      &$H_{\rho}^{01}$     &$H_{\rho}^{21}$ \\

      &$B^\ast\bar{B}^\ast$ &$H_{\pi}^{00}$    &$0$      &$H_{\rho}^{01}$     &$H_{\rho}^{21}$ \\

      \midrule[1pt]

     $I^G(J^{pc})$ & {Initial state} & \multicolumn{3}{c}{Final state} \\\midrule[1pt]

     & & $\Upsilon(1^3S_1)\rho$ & $\Upsilon(1^3D_1)\rho$  & $\Upsilon(1^3D_2)\rho$ \\

     $1^-(1^{++})$     &$\frac{1}{\sqrt{2}}(B\bar{B}^\ast+B^\ast\bar{B})$ & $\sqrt{2}H_{\rho}^{01}$ & $-\frac{\sqrt{2}}{2}H_{\rho}^{21}$ & $\frac{\sqrt{6}}{2}H_{\rho}^{21}$ \\\midrule[1pt]

     $I^G(J^{pc})$ & {Initial state} & \multicolumn{2}{c}{Final state}  \\\midrule[1pt]
      &  & $\chi_{b1}\pi$ & $h_b\rho$   \\
      \multirow{6}{*}{$1^-(1^{-+})$} & $\frac{1}{\sqrt{2}}(B_0\bar{B}^\ast+B^\ast\bar{B}_0)$     &$\frac{2\sqrt{3}}{3}H_{\pi}^{11}$    &$\frac{\sqrt{3}}{3}H_{\rho}^{11}$     \\

      &$\frac{1}{\sqrt{2}}(B_1'\bar{B}+B\bar{B}_1')$ &$\frac{2\sqrt{3}}{3}H_{\pi}^{11}$    &$-\frac{\sqrt{3}}{3}H_{\rho}^{11}$      \\

      &$\frac{1}{\sqrt{2}}(B_1\bar{B}+B\bar{B}_1)$ &$-\frac{\sqrt{6}}{6}H_{\pi}^{11}$    &$\frac{\sqrt{6}}{3}H_{\rho}^{11}$      \\

      &$\frac{1}{\sqrt{2}}(B_1'\bar{B}^\ast-B^\ast\bar{B}_1')$ & $0$    &$\frac{\sqrt{6}}{3}H_{\rho}^{11}$      \\

      &$\frac{1}{\sqrt{2}}(B_1\bar{B}^\ast-B^\ast\bar{B}_1)$ & $-\frac{\sqrt{3}}{2}H_{\pi}^{11}$    &$\frac{\sqrt{3}}{3}H_{\rho}^{11}$      \\

      &$\frac{1}{\sqrt{2}}(B_2\bar{B}^\ast+B^\ast\bar{B}_2)$ & $\frac{\sqrt{15}}{6}H_{\pi}^{11}$    &$\frac{\sqrt{15}}{3}H_{\rho}^{11}$      \\

      \midrule[1pt]

    $I^G(J^{pc})$ & {Initial state} & \multicolumn{6}{c}{Final state} \\\midrule[1pt]
      & & $\Upsilon(1^3S_1)\sigma$ & $\Upsilon(1^3D_1)\sigma$ & $h_b\eta$ & $\chi_{b0}\omega$ & $\chi_{b1}\omega$ & $\chi_{b2}\omega$   \\
      \multirow{6}{*}{$0^-(1^{--})$} & $\frac{1}{\sqrt{2}}(B_0\bar{B}^\ast-B^\ast\bar{B}_0)$     &$0$    &$0$   &$\frac{\sqrt{6}}{3}H_{\eta}^{11}$   & $-\frac{\sqrt{2}}{3}H_{\omega}^{11}$      &$-\frac{\sqrt{6}}{6}H_{\omega}^{11}$     &$-\frac{\sqrt{10}}{6}H_{\omega}^{11}$ \\

      &$\frac{1}{\sqrt{2}}(B_1'\bar{B}-B\bar{B}_1')$ &$0$    &$0$  &   $-\frac{\sqrt{6}}{3}H_{\eta}^{11}$  & $-\frac{\sqrt{2}}{3}H_{\omega}^{11}$      &$-\frac{\sqrt{6}}{6}H_{\omega}^{11}$     &$-\frac{\sqrt{10}}{6}H_{\omega}^{11}$  \\

      &$\frac{1}{\sqrt{2}}(B_1\bar{B}-B\bar{B}_1)$ & $0$    &$0$  &$-\frac{\sqrt{3}}{3}H_{\eta}^{11}$   &$-\frac{1}{3}H_{\omega}^{11}$    & $\frac{\sqrt{3}}{6}H_{\omega}^{11}$     & $\frac{\sqrt{5}}{6}H_{\omega}^{11}$  \\

      &$\frac{1}{\sqrt{2}}(B_1'\bar{B}^\ast+B^\ast\bar{B}_1')$ & $0$    &$0$  & $\frac{2\sqrt{3}}{3}H_{\eta}^{11}$  & $0$      &$0$   &  $0$  \\

      &$\frac{1}{\sqrt{2}}(B_1\bar{B}^\ast+B^\ast\bar{B}_1)$ & $0$    &$0$  & $-\frac{\sqrt{6}}{6}H_{\eta}^{11}$  &$-\frac{\sqrt{2}}{2}H_{\omega}^{11}$    & $\frac{\sqrt{6}}{4}H_{\omega}^{11}$      & $\frac{\sqrt{10}}{4}H_{\omega}^{11}$  \\

      &$\frac{1}{\sqrt{2}}(B_2\bar{B}^\ast-B^\ast\bar{B}_2)$ & $0$    &$0$  & $-\frac{\sqrt{30}}{6}H_{\eta}^{11}$  &$\frac{\sqrt{10}}{6}H_{\omega}^{11}$    & $-\frac{\sqrt{30}}{12}H_{\omega}^{11}$      & $-\frac{5\sqrt{2}}{12}H_{\omega}^{11}$  \\

      \midrule[1pt]

      $I^G(J^{pc})$ & {Initial state} & \multicolumn{5}{c}{Final state} & \\\midrule[1pt]
      & &  $h_b\sigma$ & $\Upsilon(1^3S_1)\eta$ & $\Upsilon(1^3D_1)\eta$ & $\eta_b\omega$ &  $\eta_{b2}\omega$ & \\
      \multirow{2}{*}{$0^-(1^{+-})$} & $\frac{1}{\sqrt{2}}(B\bar{B}^\ast-B^\ast\bar{B})$     &$H_{\sigma}^{11}$    &$-H_{\pi}^{00}$    &$0$ &$H_{\omega}^{01}$     &$H_{\omega}^{21}$    & \\

      &$B^\ast\bar{B}^\ast$ &$H_{\sigma}^{11}$  &$H_{\pi}^{00}$    &$0$   &$H_{\omega}^{01}$     &$H_{\omega}^{21}$   &  \\

      \midrule[1pt]

       $I^G(J^{pc})$ & {Initial state} & \multicolumn{4}{c}{Final state} \\\midrule[1pt]

       & & $\chi_{b1}\sigma$ & $\Upsilon(1^3S_1)\omega$ & $\Upsilon(1^3D_1)\omega$  & $\Upsilon(1^3D_2)\omega$ \\

       $0^+(1^{++})$     &$\frac{1}{\sqrt{2}}(B\bar{B}^\ast+B^\ast\bar{B})$ & $\sqrt{2}H_{\sigma}^{01}$ & $\sqrt{2}H_{\omega}^{01}$  & $-\frac{\sqrt{2}}{2}H_{\omega}^{21}$ & $\frac{\sqrt{6}}{2}H_{\omega}^{21}$ \\\midrule[1pt]

     $I^G(J^{pc})$ & {Initial state} & \multicolumn{2}{c}{Final state}  \\\midrule[1pt]
      & & $\chi_{b1}\eta$  & $h_b\omega$    \\
      \multirow{6}{*}{$0^+(1^{-+})$} & $\frac{1}{\sqrt{2}}(B_0\bar{B}^\ast+B^\ast\bar{B}_0)$     &$\frac{2\sqrt{3}}{3}H_{\eta}^{11}$  &$\frac{\sqrt{3}}{3}H_{\omega}^{11}$        \\

      &$\frac{1}{\sqrt{2}}(B_1'\bar{B}+B\bar{B}_1')$ &$\frac{2\sqrt{3}}{3}H_{\eta}^{11}$ &$-\frac{\sqrt{3}}{3}H_{\omega}^{11}$     \\

      &$\frac{1}{\sqrt{2}}(B_1\bar{B}+B\bar{B}_1)$  &$-\frac{\sqrt{6}}{6}H_{\eta}^{11}$ &$\frac{\sqrt{6}}{3}H_{\omega}^{11}$      \\

      &$\frac{1}{\sqrt{2}}(B_1'\bar{B}^\ast-B^\ast\bar{B}_1')$ & $0$ &$\frac{\sqrt{6}}{3}H_{\omega}^{11}$     \\

      &$\frac{1}{\sqrt{2}}(B_1\bar{B}^\ast-B^\ast\bar{B}_1)$  & $-\frac{\sqrt{3}}{2}H_{\eta}^{11}$ & $\frac{\sqrt{3}}{3}H_{\omega}^{11}$       \\

      &$\frac{1}{\sqrt{2}}(B_2\bar{B}^\ast+B^\ast\bar{B}_2)$  & $\frac{\sqrt{15}}{6}H_{\eta}^{11}$ & $\frac{\sqrt{15}}{3}H_{\omega}^{11}$      \\

      \midrule[1pt]

      \end{tabular}
\end{center}
\end{table*}

\begin{table*}[htbp]
\begin{center}
\caption{\label{tab:2}  The typical relations between the decay
widths
$\Gamma(B_{(1,2)}\bar{B}^{(\ast)}\rightarrow(b\bar{b})+light\,\,meson)$
and the reduced matrix elements $H_{\alpha}^{ij}\propto\langle
Q,i\|H_{eff}(\alpha)\|j\rangle$, where the indices $i$ and $j$
denote the light spin of the final and initial hadron respectively,
and $Q$ means the angular momentum of the final light meson.}
   \begin{tabular}{c ccccccccccccc} \toprule[1pt]
    $I^G(J^{pc})$ & {Initial state} & \multicolumn{5}{c}{Final state} \\\midrule[1pt]
      & & $\eta_{b2}\pi$ & $\Upsilon(1^3S_1)\rho$ & $\Upsilon(1^3D_1)\rho$ & $\Upsilon(1^3D_2)\rho$ & $\Upsilon(1^3D_3)\rho$  \\
      $1^-(2^{++})$ & $B^\ast\bar{B}^\ast$     &$0$    &$\sqrt{2}H_{\rho}^{01}$      &$\frac{\sqrt{3}}{10}H_{\rho}^{21}$     &$-\frac{\sqrt{30}}{10}H_{\rho}^{21}$ &$\frac{\sqrt{42}}{5}H_{\rho}^{21}$ \\

      \midrule[1pt]

      $I^G(J^{pc})$ & {Initial state} & \multicolumn{3}{c}{Final state} & \\\midrule[1pt]
      &  & $\chi_{b1}\rho$ &  $\chi_{b2}\rho$ & \\
      \multirow{4}{*}{$1^+(2^{--})$} & $\frac{1}{\sqrt{2}}(B_1'\bar{B}^\ast-B^\ast\bar{B}_1')$      &$-\frac{\sqrt{3}}{3}H_{\rho}^{11}$    &$H_{\rho}^{11}$           & \\

      &$\frac{1}{\sqrt{2}}(B_1\bar{B}^\ast-B^\ast\bar{B}_1)$ &$\frac{\sqrt{6}}{12}H_{\rho}^{11}$    &$-\frac{\sqrt{2}}{4}H_{\rho}^{11}$          &  \\

      &$\frac{1}{\sqrt{2}}(B_2\bar{B}-B\bar{B}_2)$ &$-\frac{1}{2}H_{\rho}^{11}$    &$\frac{\sqrt{3}}{2}H_{\rho}^{11}$         &  \\

      &$\frac{1}{\sqrt{2}}(B_2\bar{B}^\ast+B^\ast\bar{B}_2)$ &$-\frac{\sqrt{6}}{4}H_{\rho}^{11}$    &$\frac{3\sqrt{2}}{4}H_{\rho}^{11}$        &  \\

      \midrule[1pt]

      $I^G(J^{pc})$ & {Initial state} & \multicolumn{3}{c}{Final state} & \\\midrule[1pt]
      &  & $\chi_{b2}\pi$ &  $h_b\rho$ & \\
      \multirow{4}{*}{$1^-(2^{-+})$} & $\frac{1}{\sqrt{2}}(B_1'\bar{B}^\ast+B^\ast\bar{B}_1')$      &$-\frac{2\sqrt{6}}{3}H_{\pi}^{11}$    &$0$           & \\

      &$\frac{1}{\sqrt{2}}(B_1\bar{B}^\ast+B^\ast\bar{B}_1)$ &$-\frac{\sqrt{3}}{6}H_{\pi}^{11}$    &$0$          &  \\

      &$\frac{1}{\sqrt{2}}(B_2\bar{B}+B\bar{B}_2)$ &$-\frac{\sqrt{2}}{2}H_{\pi}^{11}$    &$0$         &  \\

      &$\frac{1}{\sqrt{2}}(B_2\bar{B}^\ast-B^\ast\bar{B}_2)$ &$-\frac{\sqrt{3}}{2}H_{\pi}^{11}$    &$0$         &  \\

      \midrule[1pt]

       $I^G(J^{pc})$ & {Initial state} & \multicolumn{3}{c}{Final state} \\\midrule[1pt]
      & & $\eta_b\pi$ & $\Upsilon(1^3S_1)\rho$ & $\Upsilon(1^3D_1)\rho$ \\
      \multirow{2}{*}{$1^-(0^{++})$} & $B\bar{B}$     &$\frac{\sqrt{2}}{2}H_{\pi}^{00}$    &$\frac{\sqrt{6}}{2}H_{\rho}^{01}$      & $\frac{\sqrt{6}}{20}H_{\rho}^{21}$      \\

      &$B^\ast\bar{B}^\ast$  & $\frac{\sqrt{3}}{2}H_{\pi}^{00}$    &$-\frac{\sqrt{2}}{2}H_{\rho}^{01}$         &$-\frac{\sqrt{2}}{20}H_{\rho}^{21}$  \\

      \midrule[1pt]

      $I^G(J^{pc})$ & {Initial state} & \multicolumn{2}{c}{Final state} &  $I^G(J^{pc})$ & {Initial state} & \multicolumn{1}{c}{Final state} \\\midrule[1pt]
      &  & $\chi_{b1}\pi$ & $h_b\rho$  & & & $\chi_{b1}\rho$  \\
      \multirow{3}{*}{$1^-(0^{-+})$} & $\frac{1}{\sqrt{2}}(B_0\bar{B}+B\bar{B}_0)$     &$\sqrt{2}H_{\pi}^{11}$    &$0$      &\multirow{3}{*}{$1^+(0^{--})$ }    &$\frac{1}{\sqrt{2}}(B_0\bar{B}-B\bar{B}_0)$ & $H_{\rho}^{11}$ \\

      &$\frac{1}{\sqrt{2}}(B_1'\bar{B}^\ast+B^\ast\bar{B}_1')$ &$-\frac{\sqrt{6}}{3}H_{\pi}^{11}$   &$0$   & & $\frac{1}{\sqrt{2}}(B_1'\bar{B}^\ast-B^\ast\bar{B}_1')$ & $-\frac{\sqrt{3}}{3}H_{\rho}^{11}$ \\

      &$\frac{1}{\sqrt{2}}(B_1\bar{B}^\ast+B^\ast\bar{B}_1)$ &$\frac{2\sqrt{3}}{3}H_{\pi}^{11}$    &$0$  &  & $\frac{1}{\sqrt{2}}(B_1\bar{B}^\ast-B^\ast\bar{B}_1)$ & $-\frac{2\sqrt{6}}{3}H_{\rho}^{11}$  \\

      \midrule[1pt]

    $I^G(J^{pc})$ & {Initial state} & \multicolumn{4}{c}{Final state} \\\midrule[1pt]
      & & $\chi_{b2}\sigma$ & $\eta_{b2}\eta$ & $\Upsilon(1^3S_1)\omega$ & $\Upsilon(1^3D_1)\omega$ & $\Upsilon(1^3D_2)\omega$ & $\Upsilon(1^3D_3)\omega$  \\
      $0^+(2^{++})$ & $B^\ast\bar{B}^\ast$     &$\sqrt{2}H_{\sigma}^{01}$   & $0$  & $\sqrt{2}H_{\omega}^{01}$ &$\frac{\sqrt{3}}{10}H_{\omega}^{21}$     &$-\frac{\sqrt{30}}{10}H_{\omega}^{21}$ &$\frac{\sqrt{42}}{5}H_{\omega}^{21}$ \\

      \midrule[1pt]

      $I^G(J^{pc})$ & {Initial state} & \multicolumn{3}{c}{Final state} & \\\midrule[1pt]
      &  &  $\Upsilon(1^3D_2)\sigma$ & $\chi_{b1}\omega$ &  $\chi_{b2}\omega$ & \\
      \multirow{4}{*}{$0^-(2^{--})$} & $\frac{1}{\sqrt{2}}(B_1'\bar{B}^\ast-B^\ast\bar{B}_1')$      &$0$    &$-\frac{\sqrt{3}}{3}H_{\omega}^{11}$    &$H_{\omega}^{11}$ \\

      &$\frac{1}{\sqrt{2}}(B_1\bar{B}^\ast-B^\ast\bar{B}_1)$ &$0$    &$\frac{\sqrt{6}}{12}H_{\omega}^{11}$    &$-\frac{\sqrt{2}}{4}H_{\omega}^{11}$    \\

      &$\frac{1}{\sqrt{2}}(B_2\bar{B}-B\bar{B}_2)$ &$0$    &$-\frac{1}{2}H_{\omega}^{11}$    &$\frac{\sqrt{3}}{2}H_{\omega}^{11}$  \\

      &$\frac{1}{\sqrt{2}}(B_2\bar{B}^\ast+B^\ast\bar{B}_2)$ &$0$    &$-\frac{\sqrt{6}}{4}H_{\omega}^{11}$    &$\frac{3\sqrt{2}}{4}H_{\omega}^{11}$   \\

      \midrule[1pt]

      $I^G(J^{pc})$ & {Initial state} & \multicolumn{3}{c}{Final state} & \\\midrule[1pt]
      &  & $\eta_{b2}\sigma$   & $\chi_{b2}\eta$  &  $h_b\omega$ & \\
      \multirow{4}{*}{$0^+(2^{-+})$} & $\frac{1}{\sqrt{2}}(B_1'\bar{B}^\ast+B^\ast\bar{B}_1')$      &$0$  &$-\frac{2\sqrt{6}}{3}H_{\eta}^{11}$  &$0$           & \\

      &$\frac{1}{\sqrt{2}}(B_1\bar{B}^\ast+B^\ast\bar{B}_1)$ &$0$  &$-\frac{\sqrt{3}}{6}H_{\eta}^{11}$  &$0$          &  \\

      &$\frac{1}{\sqrt{2}}(B_2\bar{B}+B\bar{B}_2)$ &$0$  &$-\frac{\sqrt{2}}{2}H_{\eta}^{11}$  &$0$         &  \\

      &$\frac{1}{\sqrt{2}}(B_2\bar{B}^\ast-B^\ast\bar{B}_2)$ &$0$  &$-\frac{\sqrt{3}}{2}H_{\eta}^{11}$  &$0$         &  \\

      \midrule[1pt]

       $I^G(J^{pc})$ & {Initial state} & \multicolumn{4}{c}{Final state} \\\midrule[1pt]
      & & $\chi_{b0}\sigma$ & $\eta_b\eta$ & $\Upsilon(1^3S_1)\omega$ & $\Upsilon(1^3D_1)\omega$ \\
      \multirow{2}{*}{$0^+(0^{++})$} & $B\bar{B}$     &$\frac{\sqrt{6}}{2}H_{\sigma}^{11}$ &$\frac{\sqrt{2}}{2}H_{\eta}^{00}$   &$\frac{\sqrt{6}}{2}H_{\omega}^{01}$      & $\frac{\sqrt{6}}{20}H_{\omega}^{21}$      \\

      &$B^\ast\bar{B}^\ast$ &$-\frac{\sqrt{2}}{2}H_{\sigma}^{11}$  & $\frac{\sqrt{3}}{2}H_{\eta}^{00}$  &$-\frac{\sqrt{2}}{2}H_{\omega}^{01}$         &$-\frac{\sqrt{2}}{20}H_{\omega}^{21}$   \\

      \midrule[1pt]

      $I^G(J^{pc})$ & {Initial state} & \multicolumn{3}{c}{Final state} &  $I^G(J^{pc})$ & {Initial state} & \multicolumn{1}{c}{Final state} \\\midrule[1pt]
      &  & $\eta_b\sigma$ & $\chi_{b1}\eta$ & $h_b\omega$  & & & $\chi_{b1}\omega$  \\
      \multirow{3}{*}{$0^+(0^{-+})$} & $\frac{1}{\sqrt{2}}(B_0\bar{B}+B\bar{B}_0)$     &$0$  &$\sqrt{2}H_{\eta}^{11}$  &$0$      &\multirow{3}{*}{$0^-(0^{--})$ }    &$\frac{1}{\sqrt{2}}(B_0\bar{B}-B\bar{B}_0)$ & $H_{\omega}^{11}$ \\

      &$\frac{1}{\sqrt{2}}(B_1'\bar{B}^\ast+B^\ast\bar{B}_1')$ &$0$  &$-\frac{\sqrt{6}}{3}H_{\eta}^{11}$  &$0$   & & $\frac{1}{\sqrt{2}}(B_1'\bar{B}^\ast-B^\ast\bar{B}_1')$ & $-\frac{\sqrt{3}}{3}H_{\omega}^{11}$ \\

      &$\frac{1}{\sqrt{2}}(B_1\bar{B}^\ast+B^\ast\bar{B}_1)$ &$0$  &$\frac{2\sqrt{3}}{3}H_{\eta}^{11}$  &$0$  &  & $\frac{1}{\sqrt{2}}(B_1\bar{B}^\ast-B^\ast\bar{B}_1)$ & $-\frac{2\sqrt{6}}{3}H_{\omega}^{11}$ \\

      \midrule[1pt]

      \end{tabular}
\end{center}
\end{table*}

\begin{table*}[htbp]
\begin{center}
\caption{\label{tab:3} The typical ratios of the
$B_{(1,2)}\bar{B}^{(\ast)}\rightarrow(b\bar{b})+light\,\,meson$
decay widths.}
   \begin{tabular}{c|ccccccccccc} \toprule[1pt]
      &$I^G(J^{PC})$ &  & Final state\\\cline{2-4}
   \multirow{17}{*}{\rotatebox{90}{Initial \,\, state}}&&& $\Gamma(\chi_{b0}\rho):\Gamma(\chi_{b1}\rho):\Gamma(\chi_{b2}\rho)$  \\
      &\multirow{6}{*}{$1^+(1^{--})$} & $\frac{1}{\sqrt{2}}(B_0\bar{B}^\ast-B^\ast\bar{B}_0)$     & $4:3:5$   \\
      &&$\frac{1}{\sqrt{2}}(B_1'\bar{B}-B\bar{B}_1')$     & $4:3:5$ $(1.60:1:1.48)$   \\
      &&$\frac{1}{\sqrt{2}}(B_1\bar{B}-B\bar{B}_1)$     & $4:3:5$ $(1.57:1:1.50)$   \\
      &&$\frac{1}{\sqrt{2}}(B_1'\bar{B}^\ast+B^\ast\bar{B}_1')$     & $0:0:0$   \\
      &&$\frac{1}{\sqrt{2}}(B_1\bar{B}^\ast+B^\ast\bar{B}_1)$     & $4:3:5$ $(1.54:1:1.52)$    \\
      &&$\frac{1}{\sqrt{2}}(B_2\bar{B}^\ast-B^\ast\bar{B}_2)$     & $4:3:5$ $(1.53:1:1.53)$  \\\cline{2-4}
     &&& $\Gamma(\Upsilon(1^3D_1)\rho):\Gamma(\Upsilon(1^3D_2)\rho)$  \\

     &\multirow{1}{*}{$1^-(1^{++})$} & $\frac{1}{\sqrt{2}}(B\bar{B}^\ast+B^\ast\bar{B})$     & $1:3$   \\\cline{2-4}
      && & $\Gamma(\chi_{b1}\rho):\Gamma(\chi_{b2}\rho)$  \\

    &\multirow{4}{*}{$1^+(2^{--})$} & $\frac{1}{\sqrt{2}}(B_1\bar{B}^\ast-B^\ast\bar{B}_1)$     & $1:3$ $(1:2.73)$   \\
      &&$\frac{1}{\sqrt{2}}(B_1'\bar{B}^\ast-B^\ast\bar{B}_1')$     & $1:3$ $(1:2.70)$   \\
      &&$\frac{1}{\sqrt{2}}(B_2\bar{B}-B\bar{B}_2)$     & $1:3$ $(1:2.72)$  \\
      &&$\frac{1}{\sqrt{2}}(B_2\bar{B}^\ast+B^\ast\bar{B}_2)$     & $1:3$ $(1:2.75)$ \\\cline{2-4}
      && & $\Gamma(\Upsilon(1^3D_1)\rho):\Gamma(\Upsilon(1^3D_2)\rho):\Gamma(\Upsilon(1^3D_3)\rho)$   \\
    &$1^-(2^{++})$&  $B^\ast\bar{B}^\ast$     & $1:15:84$    \\
      \cline{2-4}

   \multirow{17}{*}{\rotatebox{90}{Initial \,\, state}}&&& $\Gamma(\chi_{b0}\omega):\Gamma(\chi_{b1}\omega):\Gamma(\chi_{b2}\omega)$  \\
      &\multirow{6}{*}{$0^-(1^{--})$} & $\frac{1}{\sqrt{2}}(B_0\bar{B}^\ast-B^\ast\bar{B}_0)$     & $4:3:5$   \\
      &&$\frac{1}{\sqrt{2}}(B_1'\bar{B}-B\bar{B}_1')$     & $4:3:5$ $(1.61:1:1.46)$   \\
      &&$\frac{1}{\sqrt{2}}(B_1\bar{B}-B\bar{B}_1)$     & $4:3:5$ $(1.57:1:1.50)$   \\
      &&$\frac{1}{\sqrt{2}}(B_1'\bar{B}^\ast+B^\ast\bar{B}_1')$     & $0:0:0$   \\
      &&$\frac{1}{\sqrt{2}}(B_1\bar{B}^\ast+B^\ast\bar{B}_1)$     & $4:3:5$ $(1.55:1:1.51)$    \\
      &&$\frac{1}{\sqrt{2}}(B_2\bar{B}^\ast-B^\ast\bar{B}_2)$     & $4:3:5$ $(1.53:1:1.52)$  \\\cline{2-4}
     &&& $\Gamma(\Upsilon(1^3D_1)\omega):\Gamma(\Upsilon(1^3D_2)\omega)$  \\

     &\multirow{1}{*}{$0^+(1^{++})$} & $\frac{1}{\sqrt{2}}(B\bar{B}^\ast+B^\ast\bar{B})$     & $1:3$   \\\cline{2-4}
      && & $\Gamma(\chi_{b1}\omega):\Gamma(\chi_{b2}\omega)$  \\

    &\multirow{4}{*}{$0^-(2^{--})$} & $\frac{1}{\sqrt{2}}(B_1\bar{B}^\ast-B^\ast\bar{B}_1)$     & $1:3$ $(1:2.73)$   \\
      &&$\frac{1}{\sqrt{2}}(B_1'\bar{B}^\ast-B^\ast\bar{B}_1')$     & $1:3$ $(1:2.69)$   \\
      &&$\frac{1}{\sqrt{2}}(B_2\bar{B}-B\bar{B}_2)$     & $1:3$ $(1:2.71)$  \\
      &&$\frac{1}{\sqrt{2}}(B_2\bar{B}^\ast+B^\ast\bar{B}_2)$     & $1:3$ $(1:2.74)$ \\\cline{2-4}
      && & $\Gamma(\Upsilon(1^3D_1)\omega):\Gamma(\Upsilon(1^3D_2)\omega):\Gamma(\Upsilon(1^3D_3)\omega)$   \\
    &$0^+(2^{++})$&  $B^\ast\bar{B}^\ast$     & $1:15:84$    \\

     \bottomrule[1pt]
       \end{tabular}
\end{center}
\end{table*}

\begin{table*}[htbp]
\begin{center}
   \caption{\label{tab:4}
The typical ratios
$\frac{\Gamma(B_{(1,2)}\bar{B}^{(\ast)}\rightarrow(b\bar{b})+light\,\,meson)}{
\Gamma(B_{(1,2)}\bar{B}^{(\ast)}\rightarrow(b\bar{b})+light\,\,meson)}$,
where the initial molecular states are different while the final
states are the same.}
   \begin{tabular}{c|ccccccccccc} \toprule[1pt]
     \multirow{26}{*}{\rotatebox{90}{Initial \,\, state}}& $I^G(J^{pc})$ & \multicolumn{6}{c}{Final state} \\\cline{2-9}

      & & & $h_b\pi$ & $\chi_{b0}\rho$ & $\chi_{b1}\rho$  & $\chi_{b2}\rho$ & \\

      &\multirow{6}{*}{$1^+(1^{--})$} & $\frac{\frac{1}{\sqrt{2}}(B_0\bar{B}^\ast-B^\ast\bar{B}_0)}{\frac{1}{\sqrt{2}}(B_1'\bar{B}^\ast+B^\ast\bar{B}_1')}$     &$1:2$    &$4:0$   & $4:0$   & $4:0$  &   \\

      &&$\frac{\frac{1}{\sqrt{2}}(B_1\bar{B}^\ast+B^\ast\bar{B}_1)}{\frac{1}{\sqrt{2}}(B_2\bar{B}^\ast-B^\ast\bar{B}_2)}$ &$1:5$ $(1:5.09)$    &$9:5$ $(1.62:1)$  & $9:5$ $(1.61:1)$  & $9:5$ $(1.60:1)$  &  \\

      &&$\frac{\frac{1}{\sqrt{2}}(B_1'\bar{B}-B\bar{B}_1')}{\frac{1}{\sqrt{2}}(B_1'\bar{B}^\ast+B^\ast\bar{B}_1')}$      &$1:2$ $(1:2.07)$   &$4:0$   & $4:0$  &  $4:0$  &  \\

      &&$\frac{\frac{1}{\sqrt{2}}(B_1\bar{B}-B\bar{B}_1)}{\frac{1}{\sqrt{2}}(B_1\bar{B}^\ast+B^\ast\bar{B}_1)}$           &$2:1$ $(1.94:1)$    &$2:9$ $(1:5.45)$  & $2:9$ $(1:5.54)$ & $2:9$ $(1:5.61)$   & \\

      &&$\frac{\frac{1}{\sqrt{2}}(B_0\bar{B}^\ast-B^\ast\bar{B}_0)}{\frac{1}{\sqrt{2}}(B_1'\bar{B}-B\bar{B}_1')}$ &$1:1$ &$1:1$ &$1:1$ &$1:1$  &  \\

      &&$\frac{\frac{1}{\sqrt{2}}(B_1\bar{B}-B\bar{B}_1)}{\frac{1}{\sqrt{2}}(B_2\bar{B}^\ast-B^\ast\bar{B}_2)}$ & $2:5$ $(1:2.62)$ &$2:5$ $(1:3.36)$ &$2:5$ $(1:3.44)$ &$2:5$ $(1:3.50)$  & \\\cline{2-9}


      & & & $\Upsilon(1^3S_1)\sigma$ & $\Upsilon(1^3D_1)\sigma$ & $h_b\eta$ & $\chi_{b0}\omega$ & $\chi_{b1}\omega$  & $\chi_{b2}\omega$  \\

      &\multirow{6}{*}{$0^-(1^{--})$} & $\frac{\frac{1}{\sqrt{2}}(B_0\bar{B}^\ast-B^\ast\bar{B}_0)}{\frac{1}{\sqrt{2}}(B_1'\bar{B}^\ast+B^\ast\bar{B}_1')}$     &$0:0$ &$0:0$ &$1:2$  &$4:0$   & $4:0$   & $4:0$  \\

      &&$\frac{\frac{1}{\sqrt{2}}(B_1\bar{B}^\ast+B^\ast\bar{B}_1)}{\frac{1}{\sqrt{2}}(B_2\bar{B}^\ast-B^\ast\bar{B}_2)}$ &$0:0$  &$0:0$  &$1:5$ $(1:5.12)$  &$9:5$ $(1.62:1)$  & $9:5$ $(1.61:1)$  & $9:5$ $(1.60:1)$   \\

      &&$\frac{\frac{1}{\sqrt{2}}(B_1'\bar{B}-B\bar{B}_1')}{\frac{1}{\sqrt{2}}(B_1'\bar{B}^\ast+B^\ast\bar{B}_1')}$      &$0:0$ &$0:0$ &$1:2$ $(1:2.10)$  &$4:0$   & $4:0$  &  $4:0$   \\

      &&$\frac{\frac{1}{\sqrt{2}}(B_1\bar{B}-B\bar{B}_1)}{\frac{1}{\sqrt{2}}(B_1\bar{B}^\ast+B^\ast\bar{B}_1)}$           &$0:0$ &$0:0$  &$2:1$ $(1.92:1)$  &$2:9$ $(1:5.48)$  & $2:9$ $(1:5.58)$ & $2:9$ $(1:5.65)$   \\

      &&$\frac{\frac{1}{\sqrt{2}}(B_0\bar{B}^\ast-B^\ast\bar{B}_0)}{\frac{1}{\sqrt{2}}(B_1'\bar{B}-B\bar{B}_1')}$ &$0:0$ &$0:0$ &$1:1$ &$1:1$ &$1:1$ &$1:1$    \\

      &&$\frac{\frac{1}{\sqrt{2}}(B_1\bar{B}-B\bar{B}_1)}{\frac{1}{\sqrt{2}}(B_2\bar{B}^\ast-B^\ast\bar{B}_2)}$        &$0:0$  &$0:0$ & $2:5$ $(1:2.67)$ &$2:5$ $(1:3.39)$ &$2:5$ $(1:3.47)$ &$2:5$ $(1:3.54)$  \\\cline{2-9}

      & & & $\chi_{b1}\pi$ & $h_b\rho$ & & & $\chi_{b1}\eta$ & $h_b\omega$  \\

      &\multirow{6}{*}{$1^-(1^{-+})$} &$\frac{\frac{1}{\sqrt{2}}(B_0\bar{B}^\ast+B^\ast\bar{B}_0)}{\frac{1}{\sqrt{2}}(B_1'\bar{B}^\ast-B^\ast\bar{B}_1')}$     &$16:0$    &$1:2$  &  \multirow{6}{*}{$0^+(1^{-+})$} & $\frac{\frac{1}{\sqrt{2}}(B_0\bar{B}^\ast+B^\ast\bar{B}_0)}{\frac{1}{\sqrt{2}}(B_1'\bar{B}^\ast-B^\ast\bar{B}_1')}$ &$16:0$ & $1:2$  \\

      &&$\frac{\frac{1}{\sqrt{2}}(B_1\bar{B}^\ast-B^\ast\bar{B}_1)}{\frac{1}{\sqrt{2}}(B_2\bar{B}^\ast+B^\ast\bar{B}_2)}$ &$9:5$ $(1.77:1)$   &$1:5$ $(1:5.59)$  & &  $\frac{\frac{1}{\sqrt{2}}(B_1\bar{B}^\ast-B^\ast\bar{B}_1)}{\frac{1}{\sqrt{2}}(B_2\bar{B}^\ast+B^\ast\bar{B}_2)}$  &$9:5$ $(1.76:1)$ &$1:5$ $(1:5.61)$ \\

      &&$\frac{\frac{1}{\sqrt{2}}(B_1'\bar{B}+B\bar{B}_1')}{\frac{1}{\sqrt{2}}(B_1'\bar{B}^\ast-B^\ast\bar{B}_1')}$      &$16:0$    &$1:2$ $(1:2.56)$ & &  $\frac{\frac{1}{\sqrt{2}}(B_1'\bar{B}+B\bar{B}_1')}{\frac{1}{\sqrt{2}}(B_1'\bar{B}^\ast-B^\ast\bar{B}_1')}$  &$16:0$ &$1:2$ $(1:2.58)$\\

      &&$\frac{\frac{1}{\sqrt{2}}(B_1\bar{B}+B\bar{B}_1)}{\frac{1}{\sqrt{2}}(B_1\bar{B}^\ast-B^\ast\bar{B}_1)}$           &$2:9$ $(1:4.64)$    &$2:1$ $(1.62:1)$  & & $\frac{\frac{1}{\sqrt{2}}(B_1\bar{B}+B\bar{B}_1)}{\frac{1}{\sqrt{2}}(B_1\bar{B}^\ast-B^\ast\bar{B}_1)}$  &$2:9$ $(1:4.69)$ &$2:1$ $(1.61:1)$\\

      &&$\frac{\frac{1}{\sqrt{2}}(B_0\bar{B}^\ast+B^\ast\bar{B}_0)}{\frac{1}{\sqrt{2}}(B_1'\bar{B}+B\bar{B}_1')}$ &$1:1$ &$1:1$  & &  $\frac{\frac{1}{\sqrt{2}}(B_0\bar{B}^\ast+B^\ast\bar{B}_0)}{\frac{1}{\sqrt{2}}(B_1'\bar{B}+B\bar{B}_1')}$ &$1:1$ &$1:1$ \\

      &&$\frac{\frac{1}{\sqrt{2}}(B_1\bar{B}+B\bar{B}_1)}{\frac{1}{\sqrt{2}}(B_2\bar{B}^\ast+B^\ast\bar{B}_2)}$ &$2:5$ $(1:2.62)$ &$2:5$ $(1:3.46)$ & &  $\frac{\frac{1}{\sqrt{2}}(B_1\bar{B}+B\bar{B}_1)}{\frac{1}{\sqrt{2}}(B_2\bar{B}^\ast+B^\ast\bar{B}_2)}$  &$2:5$ $(1:2.67)$ &$2:5$ $(1:3.49)$\\\cline{2-9}

      & & &  $\Upsilon(1^3S_1)\pi$  & $\Upsilon(^3D_1)\pi$ & $\eta_b\rho$ & $\eta_{b2}(1^1D_2)\rho$\\

      & { $1^+(1^{+-})$} & $\frac{\frac{1}{\sqrt{2}}(B\bar{B}^\ast-B^\ast\bar{B})}{B^\ast\bar{B}^\ast}$     &$1:1$ $(1:1.03)$   &$0:0$  &$1:1$ $(1:1.18)$  &$1:1$  \\\cline{2-9}

      & & &  $h_b\sigma$  &  $\Upsilon(1^3S_1)\eta$  & $\Upsilon(^3D_1)\eta$  & $\eta_b\omega$ & $\eta_{b2}(1^1D_2)\omega$\\

      & { $0^-(1^{+-})$} & $\frac{\frac{1}{\sqrt{2}}(B\bar{B}^\ast-B^\ast\bar{B})}{B^\ast\bar{B}^\ast}$     &$1:1$ $(1:1.20)$ &$1:1$ $(1:1.04)$    &$0:0$   &$1:1$ $(1:1.19)$  &$1:1$  \\

      \bottomrule[1pt]

      \end{tabular}
\end{center}
\end{table*}

There are six $B_{(1,2)}\bar{B}^{(\ast)}
(B\bar{B}\,\,or\,\,B^\ast\bar{B}^\ast)$ systems with
$J^{PC}=1^{--}$. Except
$\frac{1}{\sqrt{2}}(B_1'\bar{B}^\ast+B^\ast\bar{B}_1')$ with isospin
$I=1$, all the other systems with $I=1$ can decay into
$\chi_{bJ}\,\,(J=0,1,2)$ via the $\rho$ emission. These decays are
governed by the spin configuration $(1_H^-\otimes
1_l^+)|_{J=1}^{--}$. Since
$\frac{1}{\sqrt{2}}(B_1'\bar{B}^\ast+B^\ast\bar{B}_1')$ contains the
configuration $(0_H^-\otimes 1_l^+)|_{J=1}^{--}$ only, the isovector
decay mode $\chi_{bJ}\rho$ is not allowed in the heavy quark limit.

All the isovector states of the $B_{(1,2)}\bar{B}^{(\ast)}
(B\bar{B}\,\,\mathrm{or}\,\,B^\ast\bar{B}^\ast)$ systems can decay
into $h_b\pi$. This decay mode depends on the spin configuration
$(0_H^-\otimes 1_l^+)|_{J=1}^{--}$. And their decay widths are
proportional to the parameter $H_{\pi}^{11}$ as listed in Table
\ref{tab:1}, which is defined as $H_{\pi}^{ij}\propto\langle
0,i\|H_{eff}(\pi)\|j\rangle$. And the indices $i$ and $j$ represent
the light spin of the final and initial hadron, respectively, where
the $H_{eff}(\pi)$ denotes the effective Hamiltonian for the
one-pion decay.

We also calculate the strong decay ratios of the isovector
$B_{(1,2)}\bar{B}^{(\ast)}(B\bar{B}\,\,\mathrm{or}\,\,B^\ast\bar{B}^\ast)$
systems with $J^{PC}=1^{--}$, which are listed in Table \ref{tab:3}.
Except the isovector
$\frac{1}{\sqrt{2}}(B_1'\bar{B}^\ast+B^\ast\bar{B}_1')$ system, the
remaining five systems have the same ratio
$\Gamma(\chi_{b0}\rho):\Gamma(\chi_{b1}\rho):\Gamma(\chi_{b2}\rho)=4:3:5$
without considering the contribution of the phase space factor.

The isovector states of
$\frac{1}{\sqrt{2}}(B\bar{B}^\ast-B^\ast\bar{B})$ and
$B^\ast\bar{B}^\ast$ with $J=1^{+-}$ were considered as the most
possible candidates of $Z_b(10610)$ and $Z_b(10650)$, respectively
\cite{Sun:2011uh}. Both of them can decay into $\eta_b\rho$ and
$\eta_{b2}\rho$ via the spin configuration $(0_H^-\otimes
1_l^-)|_{J=1}^{+-}$ as shown in Table \ref{tab:1}. They can also
decay into $\Upsilon(1^3S_1)$ via the one-pion
emission through the spin configuration $(1_H^-\otimes
1_l^-)|_{J=1}^{+-}$. Their one-pion decay mode $\Upsilon(1^3D_1)\pi$
depends on the spin configuration $(1_H^-\otimes
2_l^-)|_{J=1}^{+-}$. According to the spin structure of
$\frac{1}{\sqrt{2}}(B\bar{B}^\ast-B^\ast\bar{B})$ and
$B^\ast\bar{B}^\ast$ with $J=1^{+-}$, we conclude that the decay
mode $\Upsilon(1^3D_1)\pi$ of $Z_b(10610)$ and $Z_b(10650)$ are
strongly suppressed due to the heavy quark symmetry as shown in
Table \ref{tab:1}.

For the $\frac{1}{\sqrt{2}}(B\bar{B}^\ast+B^\ast\bar{B})$ system
with $J=1^{++}$, its isovector states have decay modes
$\Upsilon(1^3S_1)\rho$ , $\Upsilon(1^3D_1)\rho$  and
$\Upsilon(1^3D_2)\rho$, where the spin configuration $(1_H^-\otimes
1_l^-)|_{J=1}^{++}$ is dominant. The branching ratio of the
isovector states relevant to the
$\frac{1}{\sqrt{2}}(B\bar{B}^\ast+B^\ast\bar{B})$ system with
$J=1^{++}$ is
$\Gamma(\Upsilon(1^3D_1)\rho):\Gamma(\Upsilon(1^3D_2)\rho)=1:3$ as
listed in Table \ref{tab:3}.

From Table \ref{tab:1}, we notice that except
$\frac{1}{\sqrt{2}}(B_1'\bar{B}^\ast-B^\ast\bar{B}_1')$ the
remaining the isovector states of the
$B_{(1,2)}\bar{B}^{(\ast)}(B\bar{B}\,\,\mathrm{or}\,\,B^\ast\bar{B}^\ast)$
systems with $J^{PC}=1^{-+}$ can decay into $\chi_{b1}\pi$ and
$h_b\rho$, which depend on the spin configuration $(1_H^-\otimes
1_l^+)|_{J=1}^{-+}$ and $(0_H^-\otimes 1_l^+)|_{J=1}^{-+}$
respectively. However,
$\frac{1}{\sqrt{2}}(B_1'\bar{B}^\ast-B^\ast\bar{B}_1')$ only has the
spin configurations $(0_H^-\otimes 1_l^+)|_{J=1}^{-+}$ and
$(1_H^-\otimes 0_l^+)|_{J=1}^{-+}$. Thus, the decays of these
isovector states into $\chi_{b1}\pi$ are suppressed due to the
conservations of heavy spin, light spin and $C$-parity.

Except $\frac{1}{\sqrt{2}}(B_1'\bar{B}^\ast+B^\ast\bar{B}_1')$, the
isoscalar states relevant to the
$B_{(1,2)}\bar{B}^{(\ast)}(B\bar{B}\,\,\mathrm{or}\,\,B^\ast\bar{B}^\ast)$
systems with $J^{PC}=1^{--}$ have the allowed decay modes $h_b\eta$
, $\chi_{b0}\omega$ , $\chi_{b1}\omega$ and $\chi_{b2}\omega$, which
is similar to these decays of its isovector partners into $h_b\pi$
and $\chi_{bJ}\rho$, where $h_b\pi$ is related to the spin
configuration $(0_H^-\otimes 1_l^+)|_{J=1}^{--}$ and $\chi_{bJ}\rho$
is due to the contribution of the spin configuration $(1_H^-\otimes
1_l^+)|_{J=1}^{--}$. The decay mode $h_b\eta$ of the isoscalar
states relevant to the
$\frac{1}{\sqrt{2}}(B_1'\bar{B}^\ast+B^\ast\bar{B}_1')$ system is
also allowed through the configuration $(0_H^-\otimes
1_l^+)|_{J=1}^{--}$, but its decay into $\chi_{bJ}\omega$ is
suppressed in the heavy quark limit. As shown in Table \ref{tab:1},
these isoscalar partners relevant to the
$B_{(1,2)}\bar{B}^{(\ast)}(B\bar{B}\,\,\mathrm{or}\,\,B^\ast\bar{B}^\ast)$
systems with $J^{PC}=1^{--}$ cannot decay into $\Upsilon\sigma$ and
$\Upsilon(1^3D_1)\sigma$. This phenomena can be understood well
since the $\Upsilon(1^3S_1)\sigma$ and
$\Upsilon(1^3D_1)\sigma$ decay modes are governed by the spin
configuration $(1_H^-\otimes 0_l^+)|_{J=1}^{--}$ and $(1_H^-\otimes
2_l^+)|_{J=1}^{--}$, respectively. These two spin configurations do
not appear in the spin structures of the
$B_{(1,2)}\bar{B}^{(\ast)}(B\bar{B}\,\,\mathrm{or}\,\,B^\ast\bar{B}^\ast)$
systems. Therefore their decays into
$\Upsilon(1^3S_1)\sigma$ and $\Upsilon(1^3D_1)\sigma$
are strongly suppressed due to the heavy quark symmetry. The
branching ratios of them into $\chi_{bJ}\omega$ is
$\Gamma(\chi_{b0}\rho):\Gamma(\chi_{b1}\rho):\Gamma(\chi_{b2}\rho)=4:3:5$
if ignoring the phase space difference. The other suppressed
channels in Table \ref{tab:1} are similar to the corresponding decay
channels of their isovector partners.

As listed in Table \ref{tab:2}, the isovector states of the $B^\ast
\bar{B}^\ast$ system with $J^{PC}=2^{++}$ can decay into
$\Upsilon(1^3S_1)\rho$ and $\Upsilon(1^3D_J)\rho$
through the spin configuration $(1_H^-\otimes 1_l^-)|_{J=1}^{++}$.
But its decay mode $\eta_{b2}\pi$ is suppressed which depends on the
spin configuration $(0_H^-\otimes 2_l^-)|_{J=1}^{++}$. However,
$B^\ast \bar{B}^\ast$ doesn't contain this spin configuration. The
typical ratio of $B^\ast \bar{B}^\ast$ decays into
$\Upsilon(1^3D_J)\rho$ is
$\Gamma(\Upsilon(1^3D_1)\rho):\Gamma(\Upsilon(1^3D_2)\rho):\Gamma(\Upsilon(1^3D_3)\rho)=1:15:84$.

There are four $B_{(1,2)}\bar{B}^{(\ast)}$ systems with
$J^{PC}=2^{--}$. Their isovector partners can decay into $\chi_{b1}$
and $\chi_{b2}$ through the emission of the $\rho$ meson, where the
spin configuration $(1_H^-\otimes 1_l^+)|_{J=2}^{--}$ is dominant.
Our result indicates that
$\Gamma(\chi_{b1}\rho):\Gamma(\chi_{b2}\rho)=1:3$ for all four
states. When the initial states are the $B_{(1,2)}\bar{B}^{(\ast)}$
systems with $J^{PC}=2^{-+}$, their decays into $\chi_{b2}\pi$
depend on the spin configuration $(1_H^-\otimes 1_l^+)|_{J=2}^{-+}$.
The decay mode $h_b\rho$ is suppressed as shown in Table \ref{tab:2}
since these decays are only governed by the spin configuration
$(0_H^-\otimes 2_l^+)|_{J=1}^{-+}$. Similar conclusions hold for the
three $B_{(1,2)}\bar{B}^{(\ast)}$ systems with $J^{PC}=0^{-+}$.

The isoscalar partner of $B^\ast \bar{B}^\ast$ system only contain
the spin configuration $(1_H^-\otimes 1_l^-)|_{J=2}^{++}$ with
$J^{PC}=2^{++}$. Thus its decay into $\eta_{b2}\eta$ is suppressed.
The $\chi_{b1}\omega$ and $\chi_{b2}\omega$ are the allowed decay
modes of the isoscalar partners relevant to the
$B_{(1,2)}\bar{B}^{(\ast)}$ systems with $J^{PC}=2^{--}$, where the
$(1_H^-\otimes 1_l^+)|_{J=2}^{--}$ component is dominant. We have
the typical ratios
$\Gamma(\chi_{b1}\omega):\Gamma(\chi_{b2}\omega)=1:3$, which is the
same as that of their isovector partners. However, the decay mode
$\Upsilon(1^3D_2)\sigma$ of all the isoscalar partners of the
$B_{(1,2)}\bar{B}^{(\ast)}$ systems with $J^{PC}=2^{--}$ is
suppressed due to the absence of the spin configuration
$(1_H^-\otimes 2_l^+)|_{J=2}^{--}$.

From Table \ref{tab:2}, we can see that the decay modes
$\eta_{b2}\sigma$ and $h_b\omega$ of the four isoscalar partners
relevant to the $B_{(1,2)}\bar{B}^{(\ast)}$ systems with
$J^{PC}=2^{-+}$ are suppressed. These decays are governed by the
$(0_H^-\otimes 2_l^+)|_{J=2}^{-+}$ configuration, while the four
isoscalar states only contain spin configuration with heavy spin
equal to $1$. Similar situations occur in the suppressed decay
channels $\eta_b\sigma$ and $h_b\omega$ of the isoscalar states
relevant to the $B_{(1,2)}\bar{B}^{(\ast)}$ systems with
$J^{PC}=0^{-+}$ shown in Table \ref{tab:2}.

We need to specify that the ratios shown in Tables \ref{tab:3} and
\ref{tab:4} are also suitable for the strong decays involved with
the higher radially excited bottomonia as long as these decays are
kinematically allowed. Since the dominant decay modes of the
$\sigma/\rho$ and $\omega$ mesons are $2\pi$ and $3\pi$, the above
numerical results can be easily extended to discuss the di-pion and
tri-pion strong decays.

\subsection{$(b\bar{b})\rightarrow B_{(1,2)}\bar{B}^{(\ast)} (B\bar{B}\,\,\mathrm{or}\,\,B^\ast\bar{B}^\ast)+light\,\,meson$}

\begin{table*}[htbp]
\begin{center}
\caption{\label{tab:5}  The typical relations between the decay
widths $\Gamma((b\bar{b})\rightarrow
B_{(1,2)}\bar{B}^{(\ast)}+light\,\,meson)$ and the reduced matrix
elements $H_{\alpha}^{ij}\propto\langle
Q,i\|H_{eff}(\alpha)\|j\rangle$, where the indices $i$ and $j$
denote the light spin of the final and initial hadron respectively,
and $Q$ denotes the angular momentum of the final light meson.}
   \begin{tabular}{c cccccccccccc} \toprule[1pt]
    {Initial state} & \multicolumn{6}{c}{Final state\,\,$I^G(J^{pc})=1^+(1^{--})$} \\\midrule[1pt]

       & $\frac{1}{\sqrt{2}}(B_0\bar{B}^\ast-B^\ast\bar{B}_0)\pi$  & $\frac{1}{\sqrt{2}}(B_1'\bar{B}-B\bar{B}_1')\pi$ & $\frac{1}{\sqrt{2}}(B_1\bar{B}-B\bar{B}_1)\pi$  & $\frac{1}{\sqrt{2}}(B_1'\bar{B}^\ast+B^\ast\bar{B}_1')\pi$ & $\frac{1}{\sqrt{2}}(B_1\bar{B}^\ast+B^\ast\bar{B}_1)\pi$  & $\frac{1}{\sqrt{2}}(B_2\bar{B}^\ast-B^\ast\bar{B}_2)\pi$ \\

      $h_b(n^1P_1)$ & $\frac{\sqrt{6}}{3}H_{\pi}^{11}$ & $-\frac{\sqrt{6}}{3}H_{\pi}^{11}$ & $-\frac{\sqrt{3}}{3}H_{\pi}^{11}$ & $\frac{2\sqrt{3}}{3}H_{\pi}^{11}$ &  $-\frac{\sqrt{6}}{6}H_{\pi}^{11}$ &  $-\frac{\sqrt{30}}{6}H_{\pi}^{11}$\\

      --&--&--&--&--&--&--\\

      & $\frac{1}{\sqrt{2}}(B_0\bar{B}^\ast-B^\ast\bar{B}_0)\rho$  & $\frac{1}{\sqrt{2}}(B_1'\bar{B}-B\bar{B}_1')\rho$ & $\frac{1}{\sqrt{2}}(B_1\bar{B}-B\bar{B}_1)\rho$  & $\frac{1}{\sqrt{2}}(B_1'\bar{B}^\ast+B^\ast\bar{B}_1')\rho$ & $\frac{1}{\sqrt{2}}(B_1\bar{B}^\ast+B^\ast\bar{B}_1)\rho$  & $\frac{1}{\sqrt{2}}(B_2\bar{B}^\ast-B^\ast\bar{B}_2)\rho$ \\

      $\chi_{b0}(n^3P_0)$ & $-\frac{\sqrt{2}}{3}H_{\rho}^{11}$ & $-\frac{\sqrt{2}}{3}H_{\rho}^{11}$ & $-\frac{1}{3}H_{\rho}^{11}$ & $0$ & $-\frac{\sqrt{2}}{2}H_{\rho}^{11}$ & $\frac{\sqrt{10}}{6}H_{\rho}^{11}$ \\

      $\chi_{b1}(n^3P_1)$ & $-\frac{\sqrt{6}}{6}H_{\rho}^{11}$ & $-\frac{\sqrt{6}}{6}H_{\rho}^{11}$ & $\frac{\sqrt{3}}{6}H_{\rho}^{11}$ & $0$ & $\frac{\sqrt{6}}{4}H_{\rho}^{11}$ & $-\frac{\sqrt{30}}{12}H_{\rho}^{11}$ \\

      $\chi_{b2}(n^3P_2)$ & $-\frac{\sqrt{10}}{6}H_{\rho}^{11}$ & $-\frac{\sqrt{10}}{6}H_{\rho}^{11}$ & $\frac{\sqrt{5}}{6}H_{\rho}^{11}$ & $0$ & $\frac{\sqrt{10}}{4}H_{\rho}^{11}$ & $-\frac{5\sqrt{2}}{12}H_{\rho}^{11}$ \\

      \midrule[1pt]

      {Initial state} & \multicolumn{6}{c}{Final state\,\,$I^G(J^{pc})=1^-(1^{-+})$} \\\midrule[1pt]

      & $\frac{1}{\sqrt{2}}(B_0\bar{B}^\ast+B^\ast\bar{B}_0)\pi$ & $\frac{1}{\sqrt{2}}(B_1'\bar{B}+B\bar{B}_1')\pi$ & $\frac{1}{\sqrt{2}}(B_1\bar{B}+B\bar{B}_1)\pi$ & $\frac{1}{\sqrt{2}}(B_1'\bar{B}^\ast-B^\ast\bar{B}_1')\pi$
      & $\frac{1}{\sqrt{2}}(B_1\bar{B}^\ast-B^\ast\bar{B}_1)\pi$ & $\frac{1}{\sqrt{2}}(B_2\bar{B}^\ast+B^\ast\bar{B}_2)\pi$ \\

      $\chi_{b1}(n^3P_1)$ & $\frac{2\sqrt{3}}{3}H_{\pi}^{11}$ & $\frac{2\sqrt{3}}{3}H_{\pi}^{11}$ & $-\frac{\sqrt{6}}{6}H_{\pi}^{11}$ & $0$ & $-\frac{\sqrt{3}}{2}H_{\pi}^{11}$ & $\frac{\sqrt{15}}{6}H_{\pi}^{11}$  \\

      --&--&--&--&--&--&--\\

      & $\frac{1}{\sqrt{2}}(B_0\bar{B}^\ast+B^\ast\bar{B}_0)\rho$ & $\frac{1}{\sqrt{2}}(B_1'\bar{B}+B\bar{B}_1')\rho$ & $\frac{1}{\sqrt{2}}(B_1\bar{B}+B\bar{B}_1)\rho$ & $\frac{1}{\sqrt{2}}(B_1'\bar{B}^\ast-B^\ast\bar{B}_1')\rho$
      & $\frac{1}{\sqrt{2}}(B_1\bar{B}^\ast-B^\ast\bar{B}_1)\rho$ & $\frac{1}{\sqrt{2}}(B_2\bar{B}^\ast+B^\ast\bar{B}_2)\rho$ \\

      $h_b(n^1P_1)$ & $\frac{\sqrt{3}}{3}H_{\rho}^{11}$ & $-\frac{\sqrt{3}}{3}H_{\rho}^{11}$ & $\frac{\sqrt{6}}{3}H_{\rho}^{11}$ & $\frac{\sqrt{6}}{3}H_{\rho}^{11}$ & $\frac{\sqrt{3}}{3}H_{\rho}^{11}$ & $\frac{\sqrt{15}}{3}H_{\rho}^{11}$ \\\midrule[1pt]

      {Initial state} & \multicolumn{2}{c}{Final state\,\,$I^G(J^{pc})=1^+(1^{+-})$} & & {Initial state} & {Final state\,\,$I^G(J^{pc})=1^-(1^{++})$} \\\midrule[1pt]

      &  $\frac{1}{\sqrt{2}}(B\bar{B}^\ast-B^\ast\bar{B})\pi$ & $B^\ast\bar{B}^\ast\pi$ & & & $\frac{1}{\sqrt{2}}(B\bar{B}^\ast+B^\ast\bar{B})\rho$  \\

      $\Upsilon(n^3S_1)$ & $-H_{\pi}^{00}$ & $H_{\pi}^{00}$ & &  $\Upsilon(n^3S_1)$ & $-\frac{\sqrt{6}}{3}H_{\rho}^{10}$\\

      $\Upsilon(n^3D_1)$ & $0$ & $0$ & & $\Upsilon(n^3D_1)$ & $\frac{\sqrt{30}}{6}H_{\rho}^{12}$ \\

      & & & & $\Upsilon(n^3D_2)$ & $-\frac{\sqrt{10}}{2}H_{\rho}^{12}$ \\

      --&--&--&--&--&--&--\\

      &  $\frac{1}{\sqrt{2}}(B\bar{B}^\ast-B^\ast\bar{B})\rho$ & $B^\ast\bar{B}^\ast\rho$ & & & \\

      $\eta_b(n^1S_0)$ & $H_{\rho}^{10}$ & $H_{\rho}^{10}$ \\

      $\eta_{b2}(n^1D_2)$ & $H_{\rho}^{12}$ & $H_{\rho}^{12}$ \\\midrule[1pt]

    {Initial state} & \multicolumn{6}{c}{Final state\,\,$I^G(J^{pc})=0^-(1^{--})$} \\\midrule[1pt]

       & $\frac{1}{\sqrt{2}}(B_0\bar{B}^\ast-B^\ast\bar{B}_0)\sigma$  & $\frac{1}{\sqrt{2}}(B_1'\bar{B}-B\bar{B}_1')\sigma$ & $\frac{1}{\sqrt{2}}(B_1\bar{B}-B\bar{B}_1)\sigma$  & $\frac{1}{\sqrt{2}}(B_1'\bar{B}^\ast+B^\ast\bar{B}_1')\sigma$ & $\frac{1}{\sqrt{2}}(B_1\bar{B}^\ast+B^\ast\bar{B}_1)\sigma$  & $\frac{1}{\sqrt{2}}(B_2\bar{B}^\ast-B^\ast\bar{B}_2)\sigma$ \\

      $\Upsilon(n^3S_1)$ & $0$ & $0$ & $0$ & $0$ & $0$ & $0$ \\

      $\Upsilon(n^3D_1)$ & $0$ & $0$ & $0$ & $0$ & $0$ & $0$ \\

      --&--&--&--&--&--&--\\

      & $\frac{1}{\sqrt{2}}(B_0\bar{B}^\ast-B^\ast\bar{B}_0)\eta$  & $\frac{1}{\sqrt{2}}(B_1'\bar{B}-B\bar{B}_1')\eta$ & $\frac{1}{\sqrt{2}}(B_1\bar{B}-B\bar{B}_1)\eta$  & $\frac{1}{\sqrt{2}}(B_1'\bar{B}^\ast+B^\ast\bar{B}_1')\eta$ & $\frac{1}{\sqrt{2}}(B_1\bar{B}^\ast+B^\ast\bar{B}_1)\eta$  & $\frac{1}{\sqrt{2}}(B_2\bar{B}^\ast-B^\ast\bar{B}_2)\eta$ \\

      $h_b(n^1P_1)$ & $\frac{\sqrt{6}}{3}H_{\eta}^{11}$ & $-\frac{\sqrt{6}}{3}H_{\eta}^{11}$ & $-\frac{\sqrt{3}}{3}H_{\eta}^{11}$ & $\frac{2\sqrt{3}}{3}H_{\eta}^{11}$ &  $-\frac{\sqrt{6}}{6}H_{\eta}^{11}$ &  $-\frac{\sqrt{30}}{6}H_{\eta}^{11}$\\

      --&--&--&--&--&--&--\\

      & $\frac{1}{\sqrt{2}}(B_0\bar{B}^\ast-B^\ast\bar{B}_0)\omega$  & $\frac{1}{\sqrt{2}}(B_1'\bar{B}-B\bar{B}_1')\rho$ & $\frac{1}{\sqrt{2}}(B_1\bar{B}-B\bar{B}_1)\omega$  & $\frac{1}{\sqrt{2}}(B_1'\bar{B}^\ast+B^\ast\bar{B}_1')\omega$ & $\frac{1}{\sqrt{2}}(B_1\bar{B}^\ast+B^\ast\bar{B}_1)\omega$  & $\frac{1}{\sqrt{2}}(B_2\bar{B}^\ast-B^\ast\bar{B}_2)\omega$ \\

      $\chi_{b0}(n^3P_0)$ & $-\frac{\sqrt{2}}{3}H_{\omega}^{11}$ & $-\frac{\sqrt{2}}{3}H_{\omega}^{11}$ & $-\frac{1}{3}H_{\omega}^{11}$ & $0$ & $-\frac{\sqrt{2}}{2}H_{\omega}^{11}$ & $\frac{\sqrt{10}}{6}H_{\omega}^{11}$ \\

      $\chi_{b1}(n^3P_1)$ & $-\frac{\sqrt{6}}{6}H_{\omega}^{11}$ & $-\frac{\sqrt{6}}{6}H_{\omega}^{11}$ & $\frac{\sqrt{3}}{6}H_{\omega}^{11}$ & $0$ & $\frac{\sqrt{6}}{4}H_{\omega}^{11}$ & $-\frac{\sqrt{30}}{12}H_{\omega}^{11}$ \\

      $\chi_{b2}(n^3P_2)$ & $-\frac{\sqrt{10}}{6}H_{\omega}^{11}$ & $-\frac{\sqrt{10}}{6}H_{\omega}^{11}$ & $\frac{\sqrt{5}}{6}H_{\omega}^{11}$ & $0$ & $\frac{\sqrt{10}}{4}H_{\omega}^{11}$ & $-\frac{5\sqrt{2}}{12}H_{\omega}^{11}$ \\

      \midrule[1pt]

      {Initial state} & \multicolumn{6}{c}{Final state\,\,$I^G(J^{pc})=0^+(1^{-+})$} \\\midrule[1pt]

      & $\frac{1}{\sqrt{2}}(B_0\bar{B}^\ast+B^\ast\bar{B}_0)\eta$ & $\frac{1}{\sqrt{2}}(B_1'\bar{B}+B\bar{B}_1')\eta$ & $\frac{1}{\sqrt{2}}(B_1\bar{B}+B\bar{B}_1)\eta$ & $\frac{1}{\sqrt{2}}(B_1'\bar{B}^\ast-B^\ast\bar{B}_1')\eta$
      & $\frac{1}{\sqrt{2}}(B_1\bar{B}^\ast-B^\ast\bar{B}_1)\eta$ & $\frac{1}{\sqrt{2}}(B_2\bar{B}^\ast+B^\ast\bar{B}_2)\eta$ \\

      $\chi_{b1}(n^3P_1)$ & $\frac{2\sqrt{3}}{3}H_{\eta}^{11}$ & $\frac{2\sqrt{3}}{3}H_{\eta}^{11}$ & $-\frac{\sqrt{6}}{6}H_{\eta}^{11}$ & $0$ & $-\frac{\sqrt{3}}{2}H_{\eta}^{11}$ & $\frac{\sqrt{15}}{6}H_{\eta}^{11}$  \\

      --&--&--&--&--&--&--\\

      & $\frac{1}{\sqrt{2}}(B_0\bar{B}^\ast+B^\ast\bar{B}_0)\omega$ & $\frac{1}{\sqrt{2}}(B_1'\bar{B}+B\bar{B}_1')\omega$ & $\frac{1}{\sqrt{2}}(B_1\bar{B}+B\bar{B}_1)\omega$ & $\frac{1}{\sqrt{2}}(B_1'\bar{B}^\ast-B^\ast\bar{B}_1')\omega$
      & $\frac{1}{\sqrt{2}}(B_1\bar{B}^\ast-B^\ast\bar{B}_1)\omega$ & $\frac{1}{\sqrt{2}}(B_2\bar{B}^\ast+B^\ast\bar{B}_2)\omega$ \\

       $h_b(n^1P_1)$ & $\frac{\sqrt{3}}{3}H_{\omega}^{11}$ & $-\frac{\sqrt{3}}{3}H_{\omega}^{11}$ & $\frac{\sqrt{6}}{3}H_{\omega}^{11}$ & $\frac{\sqrt{6}}{3}H_{\omega}^{11}$ & $\frac{\sqrt{3}}{3}H_{\omega}^{11}$ & $\frac{\sqrt{15}}{3}H_{\omega}^{11}$ \\\midrule[1pt]

      {Initial state} & \multicolumn{2}{c}{Final state\,\,$I^G(J^{pc})=1^+(1^{+-})$} & & {Initial state} & {Final state\,\,$I^G(J^{pc})=1^-(1^{++})$} \\\midrule[1pt]

      &  $\frac{1}{\sqrt{2}}(B\bar{B}^\ast-B^\ast\bar{B})\sigma$ & $B^\ast\bar{B}^\ast\sigma$ & & & $\frac{1}{\sqrt{2}}(B\bar{B}^\ast+B^\ast\bar{B})\sigma$  \\

      $h_b(n^1P_1)$ & $H_{\sigma}^{11}$ & $H_{\sigma}^{11}$ & & $\chi_{b1}(n^3P_1)$ &  $-\frac{\sqrt{6}}{3}H_{\sigma}^{10}$ \\

      --&--&--&--&--&--&--\\

      &  $\frac{1}{\sqrt{2}}(B\bar{B}^\ast-B^\ast\bar{B})\eta$ & $B^\ast\bar{B}^\ast\eta$ & & & \\

      $\Upsilon(n^3S_1)$ & $-H_{\eta}^{00}$ & $H_{\eta}^{00}$ & &   & \\

      $\Upsilon(n^3D_1)$ & $0$ & $0$ & &  &  \\

       --&--&--&--&--&--&--\\

      &  $\frac{1}{\sqrt{2}}(B\bar{B}^\ast-B^\ast\bar{B})\omega$ & $B^\ast\bar{B}^\ast\omega$ & & & $\frac{1}{\sqrt{2}}(B\bar{B}^\ast+B^\ast\bar{B})\omega$  \\

      $\eta_b(n^1S_0)$ & $H_{\rho}^{10}$ &  $H_{\rho}^{10}$ & & $\Upsilon(n^3S_1)$ & $-\frac{\sqrt{6}}{3}H_{\omega}^{10}$\\

      $\eta_{b2}(n^1D_2)$ & $H_{\rho}^{12}$ & $H_{\rho}^{12}$ & & $\Upsilon(n^3D_1)$ & $\frac{\sqrt{30}}{6}H_{\omega}^{12}$ \\

      & & & & $\Upsilon(n^3D_2)$ & $-\frac{\sqrt{10}}{2}H_{\omega}^{12}$ \\

      \midrule[1pt]

      \end{tabular}
\end{center}
\end{table*}

\begin{table*}[htbp]
\scriptsize
\begin{center}
\caption{\label{tab:6}  The typical relations between the decay
widths $\Gamma((b\bar{b})\rightarrow
B_{(1,2)}\bar{B}^{(\ast)}+light\,\,meson)$ and the reduced matrix
elements $H_{\alpha}^{ij}\propto\langle
Q,i\|H_{eff}(\alpha)\|j\rangle$, where the indices $i$ and $j$
denote the light spin of the final and initial hadron respectively,
and $Q$ denotes the angular momentum of the final light meson.}
   \begin{tabular}{c cccccccccccc} \toprule[1pt]
    {Initial state} & \multicolumn{4}{c}{Final state\,\,$I^G(J^{pc})=1^+(2^{--})$} \\\midrule[1pt]

    & $\frac{1}{\sqrt{2}}(B_1'\bar{B}^\ast-B^\ast\bar{B}_1')\rho$ &$\frac{1}{\sqrt{2}}(B_1\bar{B}^\ast-B^\ast\bar{B}_1)\rho$ & $\frac{1}{\sqrt{2}}(B_2\bar{B}-B\bar{B}_2)\rho$ & $\frac{1}{\sqrt{2}}(B_2\bar{B}^\ast+B^\ast\bar{B}_2)\rho$ \\

    $\chi_{b1}(n^3P_1)$ & $-\frac{\sqrt{3}}{3}H_{\rho}^{11}$ & $\frac{\sqrt{6}}{12}H_{\rho}^{11}$ & $-\frac{1}{2}H_{\rho}^{11}$ & $-\frac{\sqrt{6}}{4}H_{\rho}^{11}$ \\

    $\chi_{b2}(n^3P_2)$ & $H_{\rho}^{11}$ & $-\frac{\sqrt{2}}{4}H_{\rho}^{11}$ & $\frac{\sqrt{3}}{2}H_{\rho}^{11}$ & $\frac{3\sqrt{2}}{4}H_{\rho}^{11}$ \\\midrule[1pt]

    {Initial state} & \multicolumn{4}{c}{Final state\,\,$I^G(J^{pc})=1^-(2^{-+})$} \\\midrule[1pt]

    &  $\frac{1}{\sqrt{2}}(B_1'\bar{B}^\ast+B^\ast\bar{B}_1')\pi$ &$\frac{1}{\sqrt{2}}(B_1\bar{B}^\ast+B^\ast\bar{B}_1)\pi$ & $\frac{1}{\sqrt{2}}(B_2\bar{B}+B\bar{B}_2)\pi$ & $\frac{1}{\sqrt{2}}(B_2\bar{B}^\ast-B^\ast\bar{B}_2)\pi$ \\

    $\chi_{b2}(n^3P_2)$ & $-\frac{2\sqrt{6}}{3}H_{\pi}^{11}$ & $-\frac{\sqrt{3}}{6}H_{\pi}^{11}$ & $-\frac{\sqrt{2}}{2}H_{\pi}^{11}$ & $-\frac{\sqrt{3}}{2}H_{\pi}^{11}$ \\

    --&--&--&--&--\\

    &  $\frac{1}{\sqrt{2}}(B_1'\bar{B}^\ast+B^\ast\bar{B}_1')\rho$ &$\frac{1}{\sqrt{2}}(B_1\bar{B}^\ast+B^\ast\bar{B}_1)\rho$ & $\frac{1}{\sqrt{2}}(B_2\bar{B}+B\bar{B}_2)\rho$ & $\frac{1}{\sqrt{2}}(B_2\bar{B}^\ast-B^\ast\bar{B}_2)\rho$ \\

    $h_b(n^1P_1)$ & $0$ & $0$ & $0$ & $0$ \\\midrule[1pt]

    {Initial state} & \multicolumn{3}{c}{Final state\,\,$I^G(J^{pc})=1^-(0^{-+})$} \\\midrule[1pt]

    &  $\frac{1}{\sqrt{2}}(B_0\bar{B}+B\bar{B}_0)\pi$ &$\frac{1}{\sqrt{2}}(B_1'\bar{B}^\ast+B^\ast\bar{B}_1')\pi$ & $\frac{1}{\sqrt{2}}(B_1\bar{B}^\ast+B^\ast\bar{B}_1)\pi$ \\

    $\chi_{b1}(n^3P_1)$ & $\sqrt{2}H_{\pi}^{11}$ & $-\frac{\sqrt{6}}{3}H_{\pi}^{11}$ & $\frac{2\sqrt{3}}{3}H_{\pi}^{11}$  \\

     --&--&--&--\\

    &  $\frac{1}{\sqrt{2}}(B_0\bar{B}+B\bar{B}_0)\rho$ &$\frac{1}{\sqrt{2}}(B_1'\bar{B}^\ast+B^\ast\bar{B}_1')\rho$ & $\frac{1}{\sqrt{2}}(B_1\bar{B}^\ast+B^\ast\bar{B}_1)\rho$ \\

    $h_b(n^1P_1)$ & $0$ & $0$ & $0$ \\\midrule[1pt]

    {Initial state} & \multicolumn{3}{c}{Final state\,\,$I^G(J^{pc})=1^+(0^{--})$} \\\midrule[1pt]

    & $\frac{1}{\sqrt{2}}(B_0\bar{B}-B\bar{B}_0)\rho$ &$\frac{1}{\sqrt{2}}(B_1'\bar{B}^\ast-B^\ast\bar{B}_1')\rho$ & $\frac{1}{\sqrt{2}}(B_1\bar{B}^\ast-B^\ast\bar{B}_1)\rho$ \\

    $\chi_{b1}(n^3P_1)$  & $H_{\rho}^{11}$ & $-\frac{\sqrt{3}}{3}H_{\rho}^{11}$ & $-\frac{2\sqrt{6}}{3}H_{\rho}^{11}$ \\\midrule[1pt]

    {Initial state} & {Final state\,\,$I^G(J^{pc})=1^-(2^{++})$} & & {Initial state} & \multicolumn{2}{c}{Final state\,\,$I^G(J^{pc})=1^-(0^{++})$} \\\midrule[1pt]

    &   $B^\ast\bar{B}^\ast\pi$ & & & $B\bar{B}\pi$ & $B^\ast\bar{B}^\ast\pi$    \\

    $\eta_{b2}(n^1D_2)$ & $0$ & & $\eta_b(n^1S_0)$ &  $\frac{\sqrt{2}}{2}H_{\pi}^{00}$ & $\frac{\sqrt{3}}{2}H_{\pi}^{00}$ \\

     --&--&--&--&--&--\\

    &   $B^\ast\bar{B}^\ast\rho$ & & & $B\bar{B}\rho$ & $B^\ast\bar{B}^\ast\rho$    \\

    $\Upsilon(n^3S_1)$ & $-\frac{\sqrt{3}}{6}H_{\rho}^{10}$ &  & $\Upsilon(n^3S_1)$  & $-\frac{\sqrt{2}}{2}H_{\rho}^{10}$ & $\frac{\sqrt{6}}{6}H_{\rho}^{10}$ \\

    $\Upsilon(n^3D_1)$ & $-\frac{\sqrt{5}}{10}H_{\rho}^{12}$ & & $\Upsilon(n^3D_1)$  & $-\frac{\sqrt{10}}{20}H_{\rho}^{12}$ & $\frac{\sqrt{30}}{60}H_{\rho}^{12}$  \\

    $\Upsilon(n^3D_2)$ & $\frac{\sqrt{50}}{10}H_{\rho}^{12}$ & &   & &   \\

    $\Upsilon(n^3D_3)$ & $-\frac{\sqrt{70}}{5}H_{\rho}^{12}$ & &   & &   \\\midrule[1pt]

    {Initial state} & \multicolumn{4}{c}{Final state\,\,$I^G(J^{pc})=0^-(2^{--})$} \\\midrule[1pt]

    & $\frac{1}{\sqrt{2}}(B_1'\bar{B}^\ast-B^\ast\bar{B}_1')\sigma$ &$\frac{1}{\sqrt{2}}(B_1\bar{B}^\ast-B^\ast\bar{B}_1)\sigma$ & $\frac{1}{\sqrt{2}}(B_2\bar{B}-B\bar{B}_2)\sigma$ & $\frac{1}{\sqrt{2}}(B_2\bar{B}^\ast+B^\ast\bar{B}_2)\sigma$ \\

    $\Upsilon(n^3D_2)$ & $0$ & $0$ & $0$ & $0$\\

    --&--&--&--&--\\

    & $\frac{1}{\sqrt{2}}(B_1'\bar{B}^\ast-B^\ast\bar{B}_1')\omega$ &$\frac{1}{\sqrt{2}}(B_1\bar{B}^\ast-B^\ast\bar{B}_1)\omega$ & $\frac{1}{\sqrt{2}}(B_2\bar{B}-B\bar{B}_2)\omega$ & $\frac{1}{\sqrt{2}}(B_2\bar{B}^\ast+B^\ast\bar{B}_2)\omega$ \\

    $\chi_{b1}(n^3P_1)$ & $-\frac{\sqrt{3}}{3}H_{\omega}^{11}$ & $\frac{\sqrt{6}}{12}H_{\omega}^{11}$ & $-\frac{1}{2}H_{\omega}^{11}$ & $-\frac{\sqrt{6}}{4}H_{\omega}^{11}$ \\

    $\chi_{b2}(n^3P_2)$ & $H_{\omega}^{11}$ & $-\frac{\sqrt{2}}{4}H_{\omega}^{11}$ & $\frac{\sqrt{3}}{2}H_{\omega}^{11}$ & $\frac{3\sqrt{2}}{4}H_{\omega}^{11}$ \\\midrule[1pt]

    {Initial state} & \multicolumn{4}{c}{Final state\,\,$I^G(J^{pc})=0^+(2^{-+})$} \\\midrule[1pt]

    &  $\frac{1}{\sqrt{2}}(B_1'\bar{B}^\ast+B^\ast\bar{B}_1')\sigma$ &$\frac{1}{\sqrt{2}}(B_1\bar{B}^\ast+B^\ast\bar{B}_1)\sigma$ & $\frac{1}{\sqrt{2}}(B_2\bar{B}+B\bar{B}_2)\sigma$ & $\frac{1}{\sqrt{2}}(B_2\bar{B}^\ast-B^\ast\bar{B}_2)\sigma$ \\

    $\eta_{b2}(n^1D_2)$ & $0$ & $0$ & $0$ & $0$\\

    --&--&--&--&--\\

    &  $\frac{1}{\sqrt{2}}(B_1'\bar{B}^\ast+B^\ast\bar{B}_1')\eta$ &$\frac{1}{\sqrt{2}}(B_1\bar{B}^\ast+B^\ast\bar{B}_1)\eta$ & $\frac{1}{\sqrt{2}}(B_2\bar{B}+B\bar{B}_2)\eta$ & $\frac{1}{\sqrt{2}}(B_2\bar{B}^\ast-B^\ast\bar{B}_2)\eta$ \\

    $\chi_{b2}(n^3P_2)$ & $-\frac{2\sqrt{6}}{3}H_{\eta}^{11}$ & $-\frac{\sqrt{3}}{6}H_{\eta}^{11}$ & $-\frac{\sqrt{2}}{2}H_{\eta}^{11}$ & $-\frac{\sqrt{3}}{2}H_{\eta}^{11}$ \\

    --&--&--&--&--\\

    &  $\frac{1}{\sqrt{2}}(B_1'\bar{B}^\ast+B^\ast\bar{B}_1')\omega$ &$\frac{1}{\sqrt{2}}(B_1\bar{B}^\ast+B^\ast\bar{B}_1)\omega$ & $\frac{1}{\sqrt{2}}(B_2\bar{B}+B\bar{B}_2)\omega$ & $\frac{1}{\sqrt{2}}(B_2\bar{B}^\ast-B^\ast\bar{B}_2)\omega$ \\

    $h_b(n^1P_1)$ & $0$ & $0$ & $0$ & $0$ \\\midrule[1pt]

    {Initial state} & \multicolumn{3}{c}{Final state\,\,$I^G(J^{pc})=0^+(0^{-+})$} \\\midrule[1pt]

    &  $\frac{1}{\sqrt{2}}(B_0\bar{B}+B\bar{B}_0)\sigma$ &$\frac{1}{\sqrt{2}}(B_1'\bar{B}^\ast+B^\ast\bar{B}_1')\sigma$ & $\frac{1}{\sqrt{2}}(B_1\bar{B}^\ast+B^\ast\bar{B}_1)\sigma$ \\

    $\eta_b(n^1S_0)$  & $0$ & $0$ & $0$\\

     --&--&--&--\\

     &  $\frac{1}{\sqrt{2}}(B_0\bar{B}+B\bar{B}_0)\eta$ &$\frac{1}{\sqrt{2}}(B_1'\bar{B}^\ast+B^\ast\bar{B}_1')\eta$ & $\frac{1}{\sqrt{2}}(B_1\bar{B}^\ast+B^\ast\bar{B}_1)\eta$ \\

    $\chi_{b1}(n^3P_1)$ & $\sqrt{2}H_{\eta}^{11}$ & $-\frac{\sqrt{6}}{3}H_{\eta}^{11}$ & $\frac{2\sqrt{3}}{3}H_{\eta}^{11}$  \\

     --&--&--&--\\

    &  $\frac{1}{\sqrt{2}}(B_0\bar{B}+B\bar{B}_0)\omega$ &$\frac{1}{\sqrt{2}}(B_1'\bar{B}^\ast+B^\ast\bar{B}_1')\omega$ & $\frac{1}{\sqrt{2}}(B_1\bar{B}^\ast+B^\ast\bar{B}_1)\omega$ \\

    $h_b(n^1P_1)$  & $0$ & $0$ & $0$ \\\midrule[1pt]

    {Initial state} & \multicolumn{3}{c}{Final state\,\,$I^G(J^{pc})=0^-(0^{--})$} \\\midrule[1pt]

    & $\frac{1}{\sqrt{2}}(B_0\bar{B}-B\bar{B}_0)\omega$ &$\frac{1}{\sqrt{2}}(B_1'\bar{B}^\ast-B^\ast\bar{B}_1')\omega$ & $\frac{1}{\sqrt{2}}(B_1\bar{B}^\ast-B^\ast\bar{B}_1)\omega$ \\

    $\chi_{b1}(n^3P_1)$  & $H_{\omega}^{11}$ & $-\frac{\sqrt{3}}{3}H_{\omega}^{11}$ & $-\frac{2\sqrt{6}}{3}H_{\omega}^{11}$ \\\midrule[1pt]

    {Initial state} & {Final state\,\,$I^G(J^{pc})=0^+(2^{++})$} & & {Initial state} & \multicolumn{2}{c}{Final state\,\,$I^G(J^{pc})=0^+(0^{++})$} \\\midrule[1pt]

    &   $B^\ast\bar{B}^\ast\sigma$ & & & $B\bar{B}\sigma$ & $B^\ast\bar{B}^\ast\sigma$    \\

    $\chi_{b2}(n^3P_2)$ & $-\frac{\sqrt{3}}{6}H_{\sigma}^{10}$ & & $\chi_{b0}(n^3P_0)$ & $\frac{\sqrt{6}}{2}H_{\sigma}^{11}$ & $-\frac{\sqrt{2}}{2}H_{\sigma}^{11}$ \\

     --&--&--&--&--&--\\

     &   $B^\ast\bar{B}^\ast\eta$ & & & $B\bar{B}\eta$ & $B^\ast\bar{B}^\ast\eta$    \\

    $\eta_{b2}(n^1D_2)$ & $0$ & & $\eta_b(n^1S_0)$ &  $\frac{\sqrt{2}}{2}H_{\eta}^{00}$ & $\frac{\sqrt{3}}{2}H_{\eta}^{00}$ \\

     --&--&--&--&--&--\\

    &   $B^\ast\bar{B}^\ast\omega$ & & & $B\bar{B}\omega$ & $B^\ast\bar{B}^\ast\omega$    \\

     $\Upsilon(n^3S_1)$ & $-\frac{\sqrt{3}}{6}H_{\omega}^{10}$ &  & $\Upsilon(n^3S_1)$  & $-\frac{\sqrt{2}}{2}H_{\omega}^{10}$ & $\frac{\sqrt{6}}{6}H_{\omega}^{10}$ \\

    $\Upsilon(n^3D_1)$ & $-\frac{\sqrt{5}}{10}H_{\omega}^{12}$ & & $\Upsilon(n^3D_1)$  & $-\frac{\sqrt{10}}{20}H_{\omega}^{12}$ & $\frac{\sqrt{30}}{60}H_{\omega}^{12}$  \\

    $\Upsilon(n^3D_2)$ & $\frac{\sqrt{50}}{10}H_{\omega}^{12}$ & &   & &   \\

    $\Upsilon(n^3D_3)$ & $-\frac{\sqrt{70}}{5}H_{\omega}^{12}$ & &   & &   \\\midrule[1pt]

   \end{tabular}
\end{center}
\end{table*}

\begin{table*}[htbp]
\begin{center}
\caption{\label{tab:7} The typical ratios of the
$(b\bar{b})\rightarrow B_{(1,2)}\bar{B}^{(\ast)}+light\,\,meson$
decay widths.}
   \begin{tabular}{c|ccccccccccc} \toprule[1pt]
      &$I^G(J^{PC})$ &  & Initial state\\\cline{2-4}
   \multirow{17}{*}{\rotatebox{90}{Final \,\, state}}&&& $\Gamma(\chi_{b0}(n^3P_0)):\Gamma(\chi_{b1}(n^3P_1)):\Gamma(\chi_{b2}(n^3P_2))$  \\
      &\multirow{6}{*}{$1^+(1^{--})$} & $\frac{1}{\sqrt{2}}(B_0\bar{B}^\ast-B^\ast\bar{B}_0)\rho$     & $4:3:5$   \\
      &&$\frac{1}{\sqrt{2}}(B_1'\bar{B}-B\bar{B}_1')\rho$     & $4:3:5$    \\
      &&$\frac{1}{\sqrt{2}}(B_1\bar{B}-B\bar{B}_1)\rho$     & $4:3:5$    \\
      &&$\frac{1}{\sqrt{2}}(B_1'\bar{B}^\ast+B^\ast\bar{B}_1')\rho$     & $0:0:0$   \\
      &&$\frac{1}{\sqrt{2}}(B_1\bar{B}^\ast+B^\ast\bar{B}_1)\rho$     & $4:3:5$     \\
      &&$\frac{1}{\sqrt{2}}(B_2\bar{B}^\ast-B^\ast\bar{B}_2)\rho$     & $4:3:5$   \\\cline{2-4}

     &&& $\Gamma(\Upsilon(n^3D_1)):\Gamma(\Upsilon(n^3D_2))$  \\

     &\multirow{1}{*}{$1^-(1^{++})$} & $\frac{1}{\sqrt{2}}(B\bar{B}^\ast+B^\ast\bar{B})\rho$     & $1:3$   \\\cline{2-4}

      && & $\Gamma(\chi_{b1}(n^3P_1)):\Gamma(\chi_{b2}(n^3P_2))$  \\

    &\multirow{4}{*}{$1^+(2^{--})$} & $\frac{1}{\sqrt{2}}(B_1\bar{B}^\ast-B^\ast\bar{B}_1)\rho$     & $1:3$   \\
      &&$\frac{1}{\sqrt{2}}(B_1'\bar{B}^\ast-B^\ast\bar{B}_1')\rho$     & $1:3$    \\
      &&$\frac{1}{\sqrt{2}}(B_2\bar{B}-B\bar{B}_2)\rho$     & $1:3$   \\
      &&$\frac{1}{\sqrt{2}}(B_2\bar{B}^\ast+B^\ast\bar{B}_2)\rho$     & $1:3$  \\\cline{2-4}

      && & $\Gamma(\Upsilon(n^3D_1)):\Gamma(\Upsilon(n^3D_2)):\Gamma(\Upsilon(n^3D_3))$   \\
    &$1^-(2^{++})$&  $B^\ast\bar{B}^\ast\rho$     & $1:15:84$    \\
      \cline{2-4}

   \multirow{17}{*}{\rotatebox{90}{Initial \,\, state}}&&& $\Gamma(\chi_{b0}(n^3P_0)):\Gamma(\chi_{b1}(n^3P_1)):\Gamma(\chi_{b2}(n^3P_2))$  \\
      &\multirow{6}{*}{$0^-(1^{--})$} & $\frac{1}{\sqrt{2}}(B_0\bar{B}^\ast-B^\ast\bar{B}_0)\omega$     & $4:3:5$   \\
      &&$\frac{1}{\sqrt{2}}(B_1'\bar{B}-B\bar{B}_1')\omega$     & $4:3:5$    \\
      &&$\frac{1}{\sqrt{2}}(B_1\bar{B}-B\bar{B}_1)\omega$     & $4:3:5$    \\
      &&$\frac{1}{\sqrt{2}}(B_1'\bar{B}^\ast+B^\ast\bar{B}_1')\omega$     & $0:0:0$   \\
      &&$\frac{1}{\sqrt{2}}(B_1\bar{B}^\ast+B^\ast\bar{B}_1)\omega$     & $4:3:5$     \\
      &&$\frac{1}{\sqrt{2}}(B_2\bar{B}^\ast-B^\ast\bar{B}_2)\omega$     & $4:3:5$  \\\cline{2-4}

     &&& $\Gamma(\Upsilon(n^3D_1)):\Gamma(\Upsilon(n^3D_2))$  \\

     &\multirow{1}{*}{$0^+(1^{++})$} & $\frac{1}{\sqrt{2}}(B\bar{B}^\ast+B^\ast\bar{B})\omega$     & $1:3$   \\\cline{2-4}

      && & $\Gamma(\chi_{b1}(n^3P_1)):\Gamma(\chi_{b2}(n^3P_2))$  \\

    &\multirow{4}{*}{$0^-(2^{--})$} & $\frac{1}{\sqrt{2}}(B_1\bar{B}^\ast-B^\ast\bar{B}_1)\omega$     & $1:3$    \\
      &&$\frac{1}{\sqrt{2}}(B_1'\bar{B}^\ast-B^\ast\bar{B}_1')\omega$     & $1:3$    \\
      &&$\frac{1}{\sqrt{2}}(B_2\bar{B}-B\bar{B}_2)\omega$     & $1:3$  \\
      &&$\frac{1}{\sqrt{2}}(B_2\bar{B}^\ast+B^\ast\bar{B}_2)\omega$     & $1:3$ \\\cline{2-4}

      && & $\Gamma(\Upsilon(n^3D_1)):\Gamma(\Upsilon(n^3D_2)):\Gamma(\Upsilon(n^3D_3))$   \\

    &$0^+(2^{++})$&  $B^\ast\bar{B}^\ast\omega$     & $1:15:84$    \\

     \bottomrule[1pt]
       \end{tabular}
\end{center}
\end{table*}

\begin{table*}[htbp]
\begin{center}
   \caption{\label{tab:8}
The typical ratios $\frac{\Gamma((b\bar{b})\rightarrow
B_{(1,2)}\bar{B}^{(\ast)}+light\,\,meson)}{
\Gamma((b\bar{b})\rightarrow
B_{(1,2)}\bar{B}^{(\ast)}+light\,\,meson)}$, where the initial
molecular states are different while the final states are the same.}
   \begin{tabular}{c|ccccccccccc} \toprule[1pt]
     \multirow{26}{*}{\rotatebox{90}{Final \,\, state}}& $I^G(J^{pc})$ & \multicolumn{7}{c}{Initial state} \\\cline{2-10}

      & & & $h_b(n^1P_1)$ & & & $\chi_{b0}(n^3P_0)$ & $\chi_{b1}(n^3P_1)$  & $\chi_{b2}(n^3P_2)$ & \\

      &\multirow{6}{*}{$1^+(1^{--})$} & $\frac{\frac{1}{\sqrt{2}}(B_0\bar{B}^\ast-B^\ast\bar{B}_0)\pi}{\frac{1}{\sqrt{2}}(B_1'\bar{B}^\ast+B^\ast\bar{B}_1')\pi}$     &$1:2$  &\multirow{6}{*}{$1^+(1^{--})$} & $\frac{\frac{1}{\sqrt{2}}(B_0\bar{B}^\ast-B^\ast\bar{B}_0)\rho}{\frac{1}{\sqrt{2}}(B_1'\bar{B}^\ast+B^\ast\bar{B}_1')\rho}$  &$4:0$   & $4:0$   & $4:0$  &   \\

      &&$\frac{\frac{1}{\sqrt{2}}(B_1\bar{B}^\ast+B^\ast\bar{B}_1)\pi}{\frac{1}{\sqrt{2}}(B_2\bar{B}^\ast-B^\ast\bar{B}_2)\pi}$ &$1:5$ & &$\frac{\frac{1}{\sqrt{2}}(B_1\bar{B}^\ast+B^\ast\bar{B}_1)\rho}{\frac{1}{\sqrt{2}}(B_2\bar{B}^\ast-B^\ast\bar{B}_2)\rho}$    &$9:5$   & $9:5$   & $9:5$   &  \\

      &&$\frac{\frac{1}{\sqrt{2}}(B_1'\bar{B}-B\bar{B}_1')\pi}{\frac{1}{\sqrt{2}}(B_1'\bar{B}^\ast+B^\ast\bar{B}_1')\pi}$      &$1:2$ &&$\frac{\frac{1}{\sqrt{2}}(B_1'\bar{B}-B\bar{B}_1')\rho}{\frac{1}{\sqrt{2}}(B_1'\bar{B}^\ast+B^\ast\bar{B}_1')\rho}$   &$4:0$   & $4:0$  &  $4:0$  &  \\

      &&$\frac{\frac{1}{\sqrt{2}}(B_1\bar{B}-B\bar{B}_1)\pi}{\frac{1}{\sqrt{2}}(B_1\bar{B}^\ast+B^\ast\bar{B}_1)\pi}$           &$2:1$ &&$\frac{\frac{1}{\sqrt{2}}(B_1\bar{B}-B\bar{B}_1)\rho}{\frac{1}{\sqrt{2}}(B_1\bar{B}^\ast+B^\ast\bar{B}_1)\rho}$    &$2:9$   & $2:9$  & $2:9$    & \\

      &&$\frac{\frac{1}{\sqrt{2}}(B_0\bar{B}^\ast-B^\ast\bar{B}_0)\pi}{\frac{1}{\sqrt{2}}(B_1'\bar{B}-B\bar{B}_1')\pi}$ &$1:1$ &&$\frac{\frac{1}{\sqrt{2}}(B_0\bar{B}^\ast-B^\ast\bar{B}_0)\rho}{\frac{1}{\sqrt{2}}(B_1'\bar{B}-B\bar{B}_1')\rho}$ &$1:1$ &$1:1$ &$1:1$  &  \\

      &&$\frac{\frac{1}{\sqrt{2}}(B_1\bar{B}-B\bar{B}_1)\pi}{\frac{1}{\sqrt{2}}(B_2\bar{B}^\ast-B^\ast\bar{B}_2)\pi}$ & $2:5$ &&$\frac{\frac{1}{\sqrt{2}}(B_1\bar{B}-B\bar{B}_1)\rho}{\frac{1}{\sqrt{2}}(B_2\bar{B}^\ast-B^\ast\bar{B}_2)\rho}$ &$2:5$  &$2:5$  &$2:5$   & \\\cline{2-10}


      & & & $\Upsilon(n^3S_1)$ & $\Upsilon(n^3D_1)$ & & & $\chi_{b0}(n^3P_0)$ & $\chi_{b1}(n^3P_1)$  & $\chi_{b2}(n^3P_2)$  \\

      &\multirow{6}{*}{$0^-(1^{--})$} & $\frac{\frac{1}{\sqrt{2}}(B_0\bar{B}^\ast-B^\ast\bar{B}_0)\sigma}{\frac{1}{\sqrt{2}}(B_1'\bar{B}^\ast+B^\ast\bar{B}_1')\sigma}$     &$0:0$ &$0:0$ &\multirow{6}{*}{$0^-(1^{--})$} & $\frac{\frac{1}{\sqrt{2}}(B_0\bar{B}^\ast-B^\ast\bar{B}_0)\omega}{\frac{1}{\sqrt{2}}(B_1'\bar{B}^\ast+B^\ast\bar{B}_1')\omega}$  &$4:0$   & $4:0$   & $4:0$  \\

      &&$\frac{\frac{1}{\sqrt{2}}(B_1\bar{B}^\ast+B^\ast\bar{B}_1)\sigma}{\frac{1}{\sqrt{2}}(B_2\bar{B}^\ast-B^\ast\bar{B}_2)\sigma}$ &$0:0$  &$0:0$  &&$\frac{\frac{1}{\sqrt{2}}(B_1\bar{B}^\ast+B^\ast\bar{B}_1)\omega}{\frac{1}{\sqrt{2}}(B_2\bar{B}^\ast-B^\ast\bar{B}_2)\omega}$  &$9:5$   & $9:5$  & $9:5$   \\

      &&$\frac{\frac{1}{\sqrt{2}}(B_1'\bar{B}-B\bar{B}_1')\sigma}{\frac{1}{\sqrt{2}}(B_1'\bar{B}^\ast+B^\ast\bar{B}_1')\sigma}$      &$0:0$ &$0:0$ &&$\frac{\frac{1}{\sqrt{2}}(B_1'\bar{B}-B\bar{B}_1')\omega}{\frac{1}{\sqrt{2}}(B_1'\bar{B}^\ast+B^\ast\bar{B}_1')\omega}$  &$4:0$   & $4:0$  &  $4:0$   \\

      &&$\frac{\frac{1}{\sqrt{2}}(B_1\bar{B}-B\bar{B}_1)\sigma}{\frac{1}{\sqrt{2}}(B_1\bar{B}^\ast+B^\ast\bar{B}_1)\sigma}$           &$0:0$ &$0:0$  &&$\frac{\frac{1}{\sqrt{2}}(B_1\bar{B}-B\bar{B}_1)\omega}{\frac{1}{\sqrt{2}}(B_1\bar{B}^\ast+B^\ast\bar{B}_1)\omega}$  &$2:9$  & $2:9$  & $2:9$    \\

      &&$\frac{\frac{1}{\sqrt{2}}(B_0\bar{B}^\ast-B^\ast\bar{B}_0)\sigma}{\frac{1}{\sqrt{2}}(B_1'\bar{B}-B\bar{B}_1')\sigma}$ &$0:0$ &$0:0$ &&$\frac{\frac{1}{\sqrt{2}}(B_0\bar{B}^\ast-B^\ast\bar{B}_0)\omega}{\frac{1}{\sqrt{2}}(B_1'\bar{B}-B\bar{B}_1')\omega}$ &$1:1$ &$1:1$ &$1:1$    \\

      &&$\frac{\frac{1}{\sqrt{2}}(B_1\bar{B}-B\bar{B}_1)\sigma}{\frac{1}{\sqrt{2}}(B_2\bar{B}^\ast-B^\ast\bar{B}_2)\sigma}$        &$0:0$  &$0:0$ &&$\frac{\frac{1}{\sqrt{2}}(B_1\bar{B}-B\bar{B}_1)\omega}{\frac{1}{\sqrt{2}}(B_2\bar{B}^\ast-B^\ast\bar{B}_2)\omega}$ &$2:5$  &$2:5$  &$2:5$   \\\cline{2-10}

      & & & $h_b(n^1P_1)$ & & & $\chi_{b1}(n^3P_1)$ \\

      &\multirow{6}{*}{$0^-(1^{--})$} & $\frac{\frac{1}{\sqrt{2}}(B_0\bar{B}^\ast-B^\ast\bar{B}_0)\eta}{\frac{1}{\sqrt{2}}(B_1'\bar{B}^\ast+B^\ast\bar{B}_1')\eta}$     &$1:2$  & \multirow{6}{*}{$0^+(1^{-+})$} &$\frac{\frac{1}{\sqrt{2}}(B_0\bar{B}^\ast+B^\ast\bar{B}_0)\eta}{\frac{1}{\sqrt{2}}(B_1'\bar{B}^\ast-B^\ast\bar{B}_1')\eta}$     &$16:0$  \\

      &&$\frac{\frac{1}{\sqrt{2}}(B_1\bar{B}^\ast+B^\ast\bar{B}_1)\eta}{\frac{1}{\sqrt{2}}(B_2\bar{B}^\ast-B^\ast\bar{B}_2)\eta}$ &$1:5$ & &$\frac{\frac{1}{\sqrt{2}}(B_1\bar{B}^\ast-B^\ast\bar{B}_1)\eta}{\frac{1}{\sqrt{2}}(B_2\bar{B}^\ast+B^\ast\bar{B}_2)\eta}$ &$9:5$ \\

      &&$\frac{\frac{1}{\sqrt{2}}(B_1'\bar{B}-B\bar{B}_1')\eta}{\frac{1}{\sqrt{2}}(B_1'\bar{B}^\ast+B^\ast\bar{B}_1')\eta}$      &$1:2$ &&$\frac{\frac{1}{\sqrt{2}}(B_1'\bar{B}+B\bar{B}_1')\eta}{\frac{1}{\sqrt{2}}(B_1'\bar{B}^\ast-B^\ast\bar{B}_1')\eta}$      &$16:0$  \\

      &&$\frac{\frac{1}{\sqrt{2}}(B_1\bar{B}-B\bar{B}_1)\eta}{\frac{1}{\sqrt{2}}(B_1\bar{B}^\ast+B^\ast\bar{B}_1)\eta}$           &$2:1$ & &$\frac{\frac{1}{\sqrt{2}}(B_1\bar{B}+B\bar{B}_1)\eta}{\frac{1}{\sqrt{2}}(B_1\bar{B}^\ast-B^\ast\bar{B}_1)\eta}$           &$2:9$ \\

      &&$\frac{\frac{1}{\sqrt{2}}(B_0\bar{B}^\ast-B^\ast\bar{B}_0)\eta}{\frac{1}{\sqrt{2}}(B_1'\bar{B}-B\bar{B}_1')\eta}$ &$1:1$ &&$\frac{\frac{1}{\sqrt{2}}(B_0\bar{B}^\ast+B^\ast\bar{B}_0)\eta}{\frac{1}{\sqrt{2}}(B_1'\bar{B}+B\bar{B}_1')\eta}$ &$1:1$  \\

      &&$\frac{\frac{1}{\sqrt{2}}(B_1\bar{B}-B\bar{B}_1)\eta}{\frac{1}{\sqrt{2}}(B_2\bar{B}^\ast-B^\ast\bar{B}_2)\eta}$ & $2:5$ & &$\frac{\frac{1}{\sqrt{2}}(B_1\bar{B}+B\bar{B}_1)\eta}{\frac{1}{\sqrt{2}}(B_2\bar{B}^\ast+B^\ast\bar{B}_2)\eta}$ &$2:5$  \\\cline{2-10}

      & & & $\chi_{b1}(n^3P_1)$ & & & $h_b(n^1P_1)$  & & & $h_b(n^1P_1)$   \\

      &\multirow{6}{*}{$1^-(1^{-+})$} &$\frac{\frac{1}{\sqrt{2}}(B_0\bar{B}^\ast+B^\ast\bar{B}_0)\pi}{\frac{1}{\sqrt{2}}(B_1'\bar{B}^\ast-B^\ast\bar{B}_1')\pi}$     &$16:0$  &\multirow{6}{*}{$1^-(1^{-+})$} &$\frac{\frac{1}{\sqrt{2}}(B_0\bar{B}^\ast+B^\ast\bar{B}_0)\rho}{\frac{1}{\sqrt{2}}(B_1'\bar{B}^\ast-B^\ast\bar{B}_1')\rho}$  &$1:2$  &  \multirow{6}{*}{$0^+(1^{-+})$} & $\frac{\frac{1}{\sqrt{2}}(B_0\bar{B}^\ast+B^\ast\bar{B}_0)\omega}{\frac{1}{\sqrt{2}}(B_1'\bar{B}^\ast-B^\ast\bar{B}_1')\omega}$ & $1:2$  \\

      &&$\frac{\frac{1}{\sqrt{2}}(B_1\bar{B}^\ast-B^\ast\bar{B}_1)\pi}{\frac{1}{\sqrt{2}}(B_2\bar{B}^\ast+B^\ast\bar{B}_2)\pi}$ &$9:5$  &&$\frac{\frac{1}{\sqrt{2}}(B_1\bar{B}^\ast-B^\ast\bar{B}_1)\rho}{\frac{1}{\sqrt{2}}(B_2\bar{B}^\ast+B^\ast\bar{B}_2)\rho}$  &$1:5$  & &  $\frac{\frac{1}{\sqrt{2}}(B_1\bar{B}^\ast-B^\ast\bar{B}_1)\omega}{\frac{1}{\sqrt{2}}(B_2\bar{B}^\ast+B^\ast\bar{B}_2)\omega}$  &$1:5$ \\

      &&$\frac{\frac{1}{\sqrt{2}}(B_1'\bar{B}+B\bar{B}_1')\pi}{\frac{1}{\sqrt{2}}(B_1'\bar{B}^\ast-B^\ast\bar{B}_1')\pi}$      &$16:0$  &&$\frac{\frac{1}{\sqrt{2}}(B_1'\bar{B}+B\bar{B}_1')\rho}{\frac{1}{\sqrt{2}}(B_1'\bar{B}^\ast-B^\ast\bar{B}_1')\rho}$  &$1:2$  & &  $\frac{\frac{1}{\sqrt{2}}(B_1'\bar{B}+B\bar{B}_1')\omega}{\frac{1}{\sqrt{2}}(B_1'\bar{B}^\ast-B^\ast\bar{B}_1')\omega}$  &$1:2$ \\

      &&$\frac{\frac{1}{\sqrt{2}}(B_1\bar{B}+B\bar{B}_1)\pi}{\frac{1}{\sqrt{2}}(B_1\bar{B}^\ast-B^\ast\bar{B}_1)\pi}$           &$2:9$  &&$\frac{\frac{1}{\sqrt{2}}(B_1\bar{B}+B\bar{B}_1)\rho}{\frac{1}{\sqrt{2}}(B_1\bar{B}^\ast-B^\ast\bar{B}_1)\rho}$  &$2:1$  & & $\frac{\frac{1}{\sqrt{2}}(B_1\bar{B}+B\bar{B}_1)\omega}{\frac{1}{\sqrt{2}}(B_1\bar{B}^\ast-B^\ast\bar{B}_1)\omega}$   &$2:1$ \\

      &&$\frac{\frac{1}{\sqrt{2}}(B_0\bar{B}^\ast+B^\ast\bar{B}_0)\pi}{\frac{1}{\sqrt{2}}(B_1'\bar{B}+B\bar{B}_1')\pi}$ &$1:1$  &&$\frac{\frac{1}{\sqrt{2}}(B_0\bar{B}^\ast+B^\ast\bar{B}_0)\rho}{\frac{1}{\sqrt{2}}(B_1'\bar{B}+B\bar{B}_1')\rho}$ &$1:1$  & &  $\frac{\frac{1}{\sqrt{2}}(B_0\bar{B}^\ast+B^\ast\bar{B}_0)\omega}{\frac{1}{\sqrt{2}}(B_1'\bar{B}+B\bar{B}_1')\omega}$ &$1:1$ \\

      &&$\frac{\frac{1}{\sqrt{2}}(B_1\bar{B}+B\bar{B}_1)\pi}{\frac{1}{\sqrt{2}}(B_2\bar{B}^\ast+B^\ast\bar{B}_2)\pi}$ &$2:5$ &&$\frac{\frac{1}{\sqrt{2}}(B_1\bar{B}+B\bar{B}_1)\rho}{\frac{1}{\sqrt{2}}(B_2\bar{B}^\ast+B^\ast\bar{B}_2)\rho}$ &$2:5$  & &  $\frac{\frac{1}{\sqrt{2}}(B_1\bar{B}+B\bar{B}_1)\omega}{\frac{1}{\sqrt{2}}(B_2\bar{B}^\ast+B^\ast\bar{B}_2)\omega}$  &$2:5$ \\\cline{2-10}

      & & &  $\Upsilon(n^3S_1)$ & $\Upsilon(n^3D_1)$ & & & & $\eta_b(n^1S_0)$ & $\eta_{b2}(n^1D_2)$\\

      & { $1^+(1^{+-})$} & $\frac{\frac{1}{\sqrt{2}}(B\bar{B}^\ast-B^\ast\bar{B})\pi}{B^\ast\bar{B}^\ast\pi}$     &$1:1$    &$0:0$  & & { $1^+(1^{+-})$} & $\frac{\frac{1}{\sqrt{2}}(B\bar{B}^\ast-B^\ast\bar{B})\rho}{B^\ast\bar{B}^\ast\rho}$ &$1:1$  &$1:1$  \\\cline{2-10}

      & & &  $h_b(n^1P_1)$  & & & &  & $\eta_b(n^1S_0)$ & $\eta_{b2}(n^1D_2)$\\

      & { $0^-(1^{+-})$} & $\frac{\frac{1}{\sqrt{2}}(B\bar{B}^\ast-B^\ast\bar{B})\sigma}{B^\ast\bar{B}^\ast\sigma}$     &$1:1$    & & & { $0^-(1^{+-})$} & $\frac{\frac{1}{\sqrt{2}}(B\bar{B}^\ast-B^\ast\bar{B})\omega}{B^\ast\bar{B}^\ast\omega}$  &$1:1$  &$1:1$  \\\cline{2-10}

      & & &  $\Upsilon(n^3S_1)$ & $\Upsilon(n^3D_1)$ & & \\

      & { $0^-(1^{+-})$} & $\frac{\frac{1}{\sqrt{2}}(B\bar{B}^\ast-B^\ast\bar{B})\eta}{B^\ast\bar{B}^\ast\eta}$     &$1:1$    &$0:0$  & & \\\cline{2-10}

      \bottomrule[1pt]

      \end{tabular}
\end{center}
\end{table*}

In this subsection, we investigate the production of the hidden
beauty molecular/resonant states via the strong decays of the higher
radial excitations of bottomonium. Among all the hidden beauty
systems considered here, the lowest mass state is
$\frac{1}{\sqrt{2}}(B\bar{B}^\ast-B^\ast\bar{B})$ which is about
10610 MeV. The pion mass is 135 MeV. Thus, we are interested in the
decays of the bottomonia with the mass around 10745 MeV or higher
radial excited states in the bottomonium family like
$\Upsilon(11020)$.

Using the same spin rearrangement scheme approach, we list the
typical relations between the strong decay widths
$\Gamma((b\bar{b})\rightarrow
B_{(1,2)}\bar{B}^{(\ast)}(B\bar{B}\,\,\mathrm{or}\,\,B^\ast\bar{B}^\ast)+light\,\,meson)$
and its corresponding reduced matrix elements relevant to the light
spin. The parameter $H_{m}^{ij}$ is the reduced matrix element with
$H_{m}^{ij}\propto\langle Q,i\|H_{eff}(m)\|j\rangle$, where the $i$
and $j$ indices denote the light spin of the final and initial
hadron, respectively. $m$ can be $\pi,\eta,\rho,\omega,\sigma$
meson, while $Q$ is the spin of light meson. These results are
collected in Tables \ref{tab:5}-\ref{tab:6}. We also calculate the
strong decay ratios, which are shown in Tables
\ref{tab:7}-\ref{tab:8}.

\subsection{$B_{(1,2)}\bar{B}^{(\ast)}(B\bar{B}^\ast\,\,\mathrm{or}\,\,B^\ast\bar{B}^\ast)\rightarrow B_{(1,2)}\bar{B}^{(\ast)}(B\bar{B}^\ast\,\,\mathrm{or}\,\,B^\ast\bar{B}^\ast)+light\,\,meson$}

\begin{table*}[htbp]
\begin{center}
\caption{\label{tab:9}  The typical ratios
$\frac{\Gamma(B_{(1,2)}\bar{B}^{(\ast)}\rightarrow
B_{(1,2)}\bar{B}^{(\ast)}+light\,\,meson)}{
\Gamma(B_{(1,2)}\bar{B}^{(\ast)}\rightarrow
B_{(1,2)}\bar{B}^{(\ast)}+light\,\,meson)}$, where the initial
molecular states are different while the final states are the same.
The parameter $A'$ is defined as $A'=\frac{H_{10}(m)}{H_{11}(m)}$.}
   \begin{tabular}{c cccccccccccc} \toprule[1pt]
    {Initial state\,\,$1^+(1^{--})$} & \multicolumn{2}{c}{$0^-(1^{+-})+\pi$} & \multicolumn{2}{c}{$1^+(1^{+-})+\eta$}& {$0^+(1^{++})+\rho$} & {$1^-(1^{++})+\omega$} \\\midrule[1pt]

    &  $\frac{1}{\sqrt{2}}(B\bar{B}^\ast-B^\ast\bar{B})\pi$ & $B^\ast\bar{B}^\ast\pi$ &  $\frac{1}{\sqrt{2}}(B\bar{B}^\ast-B^\ast\bar{B})\eta$ & $B^\ast\bar{B}^\ast\eta$ &    $\frac{1}{\sqrt{2}}(B\bar{B}^\ast+B^\ast\bar{B})\rho$ &  $\frac{1}{\sqrt{2}}(B\bar{B}^\ast+B^\ast\bar{B})\omega$ \\

     $\frac{\frac{1}{\sqrt{2}}(B_0\bar{B}^\ast-B^\ast\bar{B}_0)}{\frac{1}{\sqrt{2}}(B_1'\bar{B}^\ast+B^\ast\bar{B}_1')}$     &$1:2$        &$1:2$  &$1:2$        &$1:2$  &$4:0$        &$4:0$          \\

    $\frac{\frac{1}{\sqrt{2}}(B_1\bar{B}^\ast+B^\ast\bar{B}_1)}{\frac{1}{\sqrt{2}}(B_2\bar{B}^\ast-B^\ast\bar{B}_2)}$ &$1:5$      &$1:5$ &$1:5$      &$1:5$ &$9:5$      &$9:5$ \\

    $\frac{\frac{1}{\sqrt{2}}(B_1'\bar{B}-B\bar{B}_1')}{\frac{1}{\sqrt{2}}(B_1'\bar{B}^\ast+B^\ast\bar{B}_1')}$      &$1:2$    &$1:2$ &$1:2$    &$1:2$ &$4:0$     &$4:0$ \\

    $\frac{\frac{1}{\sqrt{2}}(B_1\bar{B}-B\bar{B}_1)}{\frac{1}{\sqrt{2}}(B_1\bar{B}^\ast+B^\ast\bar{B}_1)}$           &$2:1$     &$2:1$ &$2:1$     &$2:1$  &$2:9$      &$2:9$ \\

    $\frac{\frac{1}{\sqrt{2}}(B_0\bar{B}^\ast-B^\ast\bar{B}_0)}{\frac{1}{\sqrt{2}}(B_1'\bar{B}-B\bar{B}_1')}$     &$1:1$  &$1:1$ &$1:1$  &$1:1$ &$1:1$      &$1:1$\\

    $\frac{\frac{1}{\sqrt{2}}(B_1\bar{B}-B\bar{B}_1)}{\frac{1}{\sqrt{2}}(B_2\bar{B}^\ast-B^\ast\bar{B}_2)}$       &$2:5$  &$2:5$  &$2:5$  &$2:5$  &$2:5$    &$2:5$ \\\midrule[1pt]

    {Initial state\,\,$0^-(1^{--})$} & \multicolumn{2}{c}{$1^+(1^{+-})+\pi$} & \multicolumn{2}{c}{$0^-(1^{+-})+\eta$} & {$1^-(1^{++})+\rho$} & {$0^+(1^{++})+\omega$} \\\midrule[1pt]

    &  $\frac{1}{\sqrt{2}}(B\bar{B}^\ast-B^\ast\bar{B})\pi$ & $B^\ast\bar{B}^\ast\pi$ &  $\frac{1}{\sqrt{2}}(B\bar{B}^\ast-B^\ast\bar{B})\eta$ & $B^\ast\bar{B}^\ast\eta$ &    $\frac{1}{\sqrt{2}}(B\bar{B}^\ast+B^\ast\bar{B})\rho$ &  $\frac{1}{\sqrt{2}}(B\bar{B}^\ast+B^\ast\bar{B})\omega$ \\

     $\frac{\frac{1}{\sqrt{2}}(B_0\bar{B}^\ast-B^\ast\bar{B}_0)}{\frac{1}{\sqrt{2}}(B_1'\bar{B}^\ast+B^\ast\bar{B}_1')}$     &$1:2$        &$1:2$ &$1:2$        &$1:2$   &$4:0$        &$4:0$          \\

    $\frac{\frac{1}{\sqrt{2}}(B_1\bar{B}^\ast+B^\ast\bar{B}_1)}{\frac{1}{\sqrt{2}}(B_2\bar{B}^\ast-B^\ast\bar{B}_2)}$ &$1:5$      &$1:5$ &$1:5$      &$1:5$  &$9:5$      &$9:5$ \\

    $\frac{\frac{1}{\sqrt{2}}(B_1'\bar{B}-B\bar{B}_1')}{\frac{1}{\sqrt{2}}(B_1'\bar{B}^\ast+B^\ast\bar{B}_1')}$      &$1:2$    &$1:2$ &$1:2$    &$1:2$ &$4:0$     &$4:0$ \\

    $\frac{\frac{1}{\sqrt{2}}(B_1\bar{B}-B\bar{B}_1)}{\frac{1}{\sqrt{2}}(B_1\bar{B}^\ast+B^\ast\bar{B}_1)}$           &$2:1$     &$2:1$  &$2:1$     &$2:1$ &$2:9$      &$2:9$ \\

    $\frac{\frac{1}{\sqrt{2}}(B_0\bar{B}^\ast-B^\ast\bar{B}_0)}{\frac{1}{\sqrt{2}}(B_1'\bar{B}-B\bar{B}_1')}$     &$1:1$  &$1:1$ &$1:1$  &$1:1$  &$1:1$      &$1:1$\\

    $\frac{\frac{1}{\sqrt{2}}(B_1\bar{B}-B\bar{B}_1)}{\frac{1}{\sqrt{2}}(B_2\bar{B}^\ast-B^\ast\bar{B}_2)}$       &$2:5$  &$2:5$  &$2:5$  &$2:5$  &$2:5$    &$2:5$ \\\midrule[1pt]

    {Initial state\,\,$1^-(1^{-+})$} & {$0^+(1^{++})+\pi$} & {$1^-(1^{++})+\eta$} & \multicolumn{2}{c}{$0^-(1^{+-})+\rho$}  & \multicolumn{2}{c}{$1^+(1^{+-})+\omega$}   \\\midrule[1pt]

     & $\frac{1}{\sqrt{2}}(B\bar{B}^\ast+B^\ast\bar{B})\pi$  & $\frac{1}{\sqrt{2}}(B\bar{B}^\ast+B^\ast\bar{B})\eta$ &  $\frac{1}{\sqrt{2}}(B\bar{B}^\ast-B^\ast\bar{B})\rho$ & $B^\ast\bar{B}^\ast\rho$  &  $\frac{1}{\sqrt{2}}(B\bar{B}^\ast-B^\ast\bar{B})\omega$ & $B^\ast\bar{B}^\ast\omega$ \\

      $\frac{\frac{1}{\sqrt{2}}(B_0\bar{B}^\ast+B^\ast\bar{B}_0)}{\frac{1}{\sqrt{2}}(B_1'\bar{B}^\ast-B^\ast\bar{B}_1')}$     &$16:0$  &$16:0$ &$\frac{(\sqrt{3}+2A')^2}{6}$    &$\frac{(\sqrt{3}-2A')^2}{6}$  &$\frac{(\sqrt{3}+2A')^2}{6}$    &$\frac{(\sqrt{3}-2A')^2}{6}$    \\

      $\frac{\frac{1}{\sqrt{2}}(B_1\bar{B}^\ast-B^\ast\bar{B}_1)}{\frac{1}{\sqrt{2}}(B_2\bar{B}^\ast+B^\ast\bar{B}_2)}$ &$9:5$  &$9:5$ &$\frac{(2\sqrt{3}-3A')^2}{(2\sqrt{15}+\sqrt{5}A')^2}$     &$\frac{(2\sqrt{3}+3A')^2}{(2\sqrt{15}-\sqrt{5}A')^2}$  &$\frac{(2\sqrt{3}-3A')^2}{(2\sqrt{15}+\sqrt{5}A')^2}$     &$\frac{(2\sqrt{3}+3A')^2}{(2\sqrt{15}-\sqrt{5}A')^2}$    \\

      $\frac{\frac{1}{\sqrt{2}}(B_1'\bar{B}+B\bar{B}_1')}{\frac{1}{\sqrt{2}}(B_1'\bar{B}^\ast-B^\ast\bar{B}_1')}$      &$16:0$ &$16:0$  &$\frac{(\sqrt{3}-2A')^2}{6}$    &$\frac{(\sqrt{3}+2A')^2}{6}$  &$\frac{(\sqrt{3}-2A')^2}{6}$    &$\frac{(\sqrt{3}+2A')^2}{6}$   \\

      $\frac{\frac{1}{\sqrt{2}}(B_1\bar{B}+B\bar{B}_1)}{\frac{1}{\sqrt{2}}(B_1\bar{B}^\ast-B^\ast\bar{B}_1)}$           &$2:9$ &$2:9$ &$\frac{(2\sqrt{6}-\sqrt{2}A')^2}{(2\sqrt{3}+3A')^2}$    &$\frac{(2\sqrt{6}+\sqrt{2}A')^2}{(2\sqrt{3}-3A')^2}$  &$\frac{(2\sqrt{6}-\sqrt{2}A')^2}{(2\sqrt{3}+3A')^2}$    &$\frac{(2\sqrt{6}+\sqrt{2}A')^2}{(2\sqrt{3}-3A')^2}$   \\

      $\frac{\frac{1}{\sqrt{2}}(B_0\bar{B}^\ast+B^\ast\bar{B}_0)}{\frac{1}{\sqrt{2}}(B_1'\bar{B}+B\bar{B}_1')}$   &$1:1$ &$1:1$ &$\frac{(\sqrt{3}+2A')^2}{(\sqrt{3}-2A')^2}$ &$\frac{(\sqrt{3}-2A')^2}{(\sqrt{3}+2A')^2}$ &$\frac{(\sqrt{3}+2A')^2}{(\sqrt{3}-2A')^2}$ &$\frac{(\sqrt{3}-2A')^2}{(\sqrt{3}+2A')^2}$  \\

      $\frac{\frac{1}{\sqrt{2}}(B_1\bar{B}+B\bar{B}_1)}{\frac{1}{\sqrt{2}}(B_2\bar{B}^\ast+B^\ast\bar{B}_2)}$     &$2:5$ &$2:5$ &$\frac{(2\sqrt{6}-\sqrt{2}A')^2}{(2\sqrt{15}+\sqrt{5}A')^2}$ &$\frac{(2\sqrt{6}+\sqrt{2}A')^2}{(2\sqrt{15}-\sqrt{5}A')^2}$ &$\frac{(2\sqrt{6}-\sqrt{2}A')^2}{(2\sqrt{15}+\sqrt{5}A')^2}$ &$\frac{(2\sqrt{6}+\sqrt{2}A')^2}{(2\sqrt{15}-\sqrt{5}A')^2}$  \\\midrule[1pt]

      {Initial state\,\,$0^+(1^{-+})$} & {$1^-(1^{++})+\pi$} & {$0^+(1^{++})+\eta$}  & \multicolumn{2}{c}{$1^+(1^{+-})+\rho$}  & \multicolumn{2}{c}{$0^-(1^{+-})+\omega$}   \\\midrule[1pt]

     & $\frac{1}{\sqrt{2}}(B\bar{B}^\ast+B^\ast\bar{B})\pi$  & $\frac{1}{\sqrt{2}}(B\bar{B}^\ast+B^\ast\bar{B})\eta$  &  $\frac{1}{\sqrt{2}}(B\bar{B}^\ast-B^\ast\bar{B})\rho$ & $B^\ast\bar{B}^\ast\rho$  &  $\frac{1}{\sqrt{2}}(B\bar{B}^\ast-B^\ast\bar{B})\omega$ & $B^\ast\bar{B}^\ast\omega$ \\

      $\frac{\frac{1}{\sqrt{2}}(B_0\bar{B}^\ast+B^\ast\bar{B}_0)}{\frac{1}{\sqrt{2}}(B_1'\bar{B}^\ast-B^\ast\bar{B}_1')}$     &$16:0$ &$16:0$  &$\frac{(\sqrt{3}+2A')^2}{6}$    &$\frac{(\sqrt{3}-2A')^2}{6}$  &$\frac{(\sqrt{3}+2A')^2}{6}$    &$\frac{(\sqrt{3}-2A')^2}{6}$    \\

      $\frac{\frac{1}{\sqrt{2}}(B_1\bar{B}^\ast-B^\ast\bar{B}_1)}{\frac{1}{\sqrt{2}}(B_2\bar{B}^\ast+B^\ast\bar{B}_2)}$ &$9:5$ &$9:5$  &$\frac{(2\sqrt{3}-3A')^2}{(2\sqrt{15}+\sqrt{5}A')^2}$     &$\frac{(2\sqrt{3}+3A')^2}{(2\sqrt{15}-\sqrt{5}A')^2}$  &$\frac{(2\sqrt{3}-3A')^2}{(2\sqrt{15}+\sqrt{5}A')^2}$     &$\frac{(2\sqrt{3}+3A')^2}{(2\sqrt{15}-\sqrt{5}A')^2}$    \\

      $\frac{\frac{1}{\sqrt{2}}(B_1'\bar{B}+B\bar{B}_1')}{\frac{1}{\sqrt{2}}(B_1'\bar{B}^\ast-B^\ast\bar{B}_1')}$      &$16:0$ &$16:0$ &$\frac{(\sqrt{3}-2A')^2}{6}$    &$\frac{(\sqrt{3}+2A')^2}{6}$  &$\frac{(\sqrt{3}-2A')^2}{6}$    &$\frac{(\sqrt{3}+2A')^2}{6}$   \\

      $\frac{\frac{1}{\sqrt{2}}(B_1\bar{B}+B\bar{B}_1)}{\frac{1}{\sqrt{2}}(B_1\bar{B}^\ast-B^\ast\bar{B}_1)}$           &$2:9$ &$2:9$  &$\frac{(2\sqrt{6}-\sqrt{2}A')^2}{(2\sqrt{3}+3A')^2}$    &$\frac{(2\sqrt{6}+\sqrt{2}A')^2}{(2\sqrt{3}-3A')^2}$  &$\frac{(2\sqrt{6}-\sqrt{2}A')^2}{(2\sqrt{3}+3A')^2}$    &$\frac{(2\sqrt{6}+\sqrt{2}A')^2}{(2\sqrt{3}-3A')^2}$   \\

      $\frac{\frac{1}{\sqrt{2}}(B_0\bar{B}^\ast+B^\ast\bar{B}_0)}{\frac{1}{\sqrt{2}}(B_1'\bar{B}+B\bar{B}_1')}$   &$1:1$ &$1:1$ &$\frac{(\sqrt{3}+2A')^2}{(\sqrt{3}-2A')^2}$ &$\frac{(\sqrt{3}-2A')^2}{(\sqrt{3}+2A')^2}$ &$\frac{(\sqrt{3}+2A')^2}{(\sqrt{3}-2A')^2}$ &$\frac{(\sqrt{3}-2A')^2}{(\sqrt{3}+2A')^2}$  \\

      $\frac{\frac{1}{\sqrt{2}}(B_1\bar{B}+B\bar{B}_1)}{\frac{1}{\sqrt{2}}(B_2\bar{B}^\ast+B^\ast\bar{B}_2)}$     &$2:5$ &$2:5$ &$\frac{(2\sqrt{6}-\sqrt{2}A')^2}{(2\sqrt{15}+\sqrt{5}A')^2}$ &$\frac{(2\sqrt{6}+\sqrt{2}A')^2}{(2\sqrt{15}-\sqrt{5}A')^2}$ &$\frac{(2\sqrt{6}-\sqrt{2}A')^2}{(2\sqrt{15}+\sqrt{5}A')^2}$ &$\frac{(2\sqrt{6}+\sqrt{2}A')^2}{(2\sqrt{15}-\sqrt{5}A')^2}$  \\\midrule[1pt]

    \end{tabular}
\end{center}
\end{table*}

\begin{table*}[htbp]
\begin{center}
\caption{\label{tab:10} The typical ratios of the
$B_{(1,2)}\bar{B}^{(\ast)}\rightarrow
B_{(1,2)}\bar{B}^{(\ast)}+light\,\,meson$ decay widths. The
parameter $C'$ is defined as $C'=\frac{H_{01}(m)}{H_{11}(m)}$.}
   \begin{tabular}{c cc} \toprule[1pt]
    {Initial state\,\,$1^-(1^{-+})$}  & {$0^-(1^{+-})+\rho$} & {$1^+(1^{+-})+\omega$} \\\midrule[1pt]

      & $\frac{1}{\sqrt{2}}(B\bar{B}^\ast-B^\ast\bar{B})\rho:B^\ast\bar{B}^\ast\rho$ & $\frac{1}{\sqrt{2}}(B\bar{B}^\ast-B^\ast\bar{B})\omega:B^\ast\bar{B}^\ast\omega$  \\

      $\frac{1}{\sqrt{2}}(B_0\bar{B}^\ast+B^\ast\bar{B}_0)$     & $\frac{(1-2C')^2}{(1+2C')^2}$ & $\frac{(1-2C')^2}{(1+2C')^2}$  \\

      $\frac{1}{\sqrt{2}}(B_1'\bar{B}+B\bar{B}_1')$     & $\frac{(1-2C')^2}{(1+2C')^2}$ & $\frac{(1-2C')^2}{(1+2C')^2}$  \\

      $\frac{1}{\sqrt{2}}(B_1\bar{B}+B\bar{B}_1)$     & $\frac{(2+C')^2}{(2-C')^2}$ & $\frac{(2+C')^2}{(2-C')^2}$  \\

      $\frac{1}{\sqrt{2}}(B_1'\bar{B}^\ast-B^\ast\bar{B}_1')$     & $\frac{1}{1}$  & $\frac{1}{1}$ \\

      $\frac{1}{\sqrt{2}}(B_1\bar{B}^\ast-B^\ast\bar{B}_1)$     & $\frac{(2+3C')^2}{(2-3C')^2}$  & $\frac{(2+3C')^2}{(2-3C')^2}$ \\

      $\frac{1}{\sqrt{2}}(B_2\bar{B}^\ast+B^\ast\bar{B}_2)$     & $\frac{(2-C')^2}{(2+C')^2}$ & $\frac{(2-C')^2}{(2+C')^2}$  \\\midrule[1pt]

      {Initial state\,\,$0^+(1^{-+})$}  & {$1^+(1^{+-})+\rho$} & {$0^-(1^{+-})+\omega$} \\\midrule[1pt]

      & $\frac{1}{\sqrt{2}}(B\bar{B}^\ast-B^\ast\bar{B})\rho:B^\ast\bar{B}^\ast\rho$ & $\frac{1}{\sqrt{2}}(B\bar{B}^\ast-B^\ast\bar{B})\omega:B^\ast\bar{B}^\ast\omega$  \\

      $\frac{1}{\sqrt{2}}(B_0\bar{B}^\ast+B^\ast\bar{B}_0)$     & $\frac{(1-2C')^2}{(1+2C')^2}$ & $\frac{(1-2C')^2}{(1+2C')^2}$  \\

      $\frac{1}{\sqrt{2}}(B_1'\bar{B}+B\bar{B}_1')$     & $\frac{(1-2C')^2}{(1+2C')^2}$ & $\frac{(1-2C')^2}{(1+2C')^2}$  \\

      $\frac{1}{\sqrt{2}}(B_1\bar{B}+B\bar{B}_1)$     & $\frac{(2+C')^2}{(2-C')^2}$ & $\frac{(2+C')^2}{(2-C')^2}$  \\

      $\frac{1}{\sqrt{2}}(B_1'\bar{B}^\ast-B^\ast\bar{B}_1')$     & $\frac{1}{1}$  & $\frac{1}{1}$ \\

      $\frac{1}{\sqrt{2}}(B_1\bar{B}^\ast-B^\ast\bar{B}_1)$     & $\frac{(2+3C')^2}{(2-3C')^2}$  & $\frac{(2+3C')^2}{(2-3C')^2}$ \\

      $\frac{1}{\sqrt{2}}(B_2\bar{B}^\ast+B^\ast\bar{B}_2)$     & $\frac{(2-C')^2}{(2+C')^2}$ & $\frac{(2-C')^2}{(2+C')^2}$  \\\midrule[1pt]

      {Initial state\,\,$1^+(1^{--})$}  & {$0^-(1^{+-})+\pi$}  & {$1^+(1^{+-})+\eta$}  \\\midrule[1pt]

      & $\frac{1}{\sqrt{2}}(B\bar{B}^\ast-B^\ast\bar{B})\pi:B^\ast\bar{B}^\ast\pi$ & $\frac{1}{\sqrt{2}}(B\bar{B}^\ast-B^\ast\bar{B})\eta:B^\ast\bar{B}^\ast\eta$\\

      $\frac{1}{\sqrt{2}}(B_0\bar{B}^\ast-B^\ast\bar{B}_0)$     & $1:1$  & $1:1$ \\
      $\frac{1}{\sqrt{2}}(B_1'\bar{B}-B\bar{B}_1')$     & $1:1$  & $1:1$ \\
      $\frac{1}{\sqrt{2}}(B_1\bar{B}-B\bar{B}_1)$    & $1:1$  & $1:1$   \\
      $\frac{1}{\sqrt{2}}(B_1'\bar{B}^\ast+B^\ast\bar{B}_1')$     & $1:1$ & $1:1$ \\
      $\frac{1}{\sqrt{2}}(B_1\bar{B}^\ast+B^\ast\bar{B}_1)$     & $1:1$   & $1:1$  \\
      $\frac{1}{\sqrt{2}}(B_2\bar{B}^\ast-B^\ast\bar{B}_2)$     & $1:1$ & $1:1$ \\\midrule[1pt]

      {Initial state\,\,$0^-(1^{--})$}  & {$1^+(1^{+-})+\pi$} & {$0^-(1^{+-})+\eta$} \\\midrule[1pt]

      & $\frac{1}{\sqrt{2}}(B\bar{B}^\ast-B^\ast\bar{B})\pi:B^\ast\bar{B}^\ast\pi$ & $\frac{1}{\sqrt{2}}(B\bar{B}^\ast-B^\ast\bar{B})\eta:B^\ast\bar{B}^\ast\eta$\\

      $\frac{1}{\sqrt{2}}(B_0\bar{B}^\ast-B^\ast\bar{B}_0)$     & $1:1$ & $1:1$  \\
      $\frac{1}{\sqrt{2}}(B_1'\bar{B}-B\bar{B}_1')$     & $1:1$  & $1:1$\\
      $\frac{1}{\sqrt{2}}(B_1\bar{B}-B\bar{B}_1)$    & $1:1$   & $1:1$ \\
      $\frac{1}{\sqrt{2}}(B_1'\bar{B}^\ast+B^\ast\bar{B}_1')$     & $1:1$ & $1:1$ \\
      $\frac{1}{\sqrt{2}}(B_1\bar{B}^\ast+B^\ast\bar{B}_1)$     & $1:1$  & $1:1$   \\
      $\frac{1}{\sqrt{2}}(B_2\bar{B}^\ast-B^\ast\bar{B}_2)$     & $1:1$  & $1:1$\\\midrule[1pt]

    \end{tabular}
\end{center}
\end{table*}

With the help of the heavy quark symmetry, we further discuss the
strong decays between two hidden beauty molecular/resonant states.
Here, we only investigate the decay behaviors of those vector and
axial vector states with $J^{PC}=1^{--}$, $1^{-+}$, $1^{+-}$ and
$1^{++}$. The typical ratios of the
$B_{(1,2)}\bar{B}^{(\ast)}(B\bar{B}^\ast\,\,\mathrm{or}\,\,B^\ast\bar{B}^\ast)\rightarrow
B_{(1,2)}\bar{B}^{(\ast)}(B\bar{B}^\ast\,\,\mathrm{or}\,\,B^\ast\bar{B}^\ast)+light\,\,meson$
decay widths depend on the following parameters
\begin{eqnarray}
 A'=\frac{H_{10}(m)}{H_{11}(m)}, \,\,C'=\frac{H_{01}(m)}{H_{11}(m)},
\end{eqnarray}
where $H_{10}(m)=\langle Q,1\|H_{eff}(m)\|0\rangle$,
$H_{11}(m)=\langle Q,1\|H_{eff}(m)\|1\rangle$, and
$H_{01}(m)=\langle Q,0\|H_{eff}(m)\|1\rangle$. $m$ can be
$\pi,\eta,\rho,\omega,\sigma$, while $Q$ is the spin of the light
meson.

If the reduced matrix elements satisfiy the following relation
\begin{eqnarray*}
 H_{01}(m)&=&-\frac{1}{\sqrt{3}}H_{10}(m),
\end{eqnarray*}
the typical ratios satisfy the crossing symmetry, which is an
important test of our calculation. We list the obtained typical
ratios of the
$B_{(1,2)}\bar{B}^{(\ast)}(B\bar{B}^\ast\,\,\mathrm{or}\,\,B^\ast\bar{B}^\ast)\rightarrow
B_{(1,2)}\bar{B}^{(\ast)}(B\bar{B}^\ast\,\,\mathrm{or}\,\,B^\ast\bar{B}^\ast)+light\,\,meson$
decay widths in Tables \ref{tab:9}-\ref{tab:10}.

\section{The strong decays of the hidden-charm systems}\label{sec4}

The above discussion of the strong decays or production behaviors of
hidden beauty systems can be extended to investigate the strong
decays of the hidden charm systems. In the following, we combine the
experimental information of these observed charmonium-like states
with our numerical results.

\subsection{$Y(4260)$ and $Y(4360)$}

The charmonium-like state $Y(4260)$ with $J^{PC}=1^{--}$ was
reported by the BaBar Collaboration in the $e^+e^-\to \pi^+\pi^-
J/\psi$ process \cite{Aubert:2005rm}. Assuming $Y(4260)$ to be the
isoscalar state of $\frac{1}{\sqrt{2}}(D_1\bar{D}-D\bar{D}_1)$
system \cite{zhu-review,Ding:2008gr}, we can write down the spin
isospin wave function of $Y(4260)$ in heavy quark limit, i.e.,
\begin{eqnarray*}
  |Y(4260)\rangle &=& 
  \frac{1}{\sqrt{2}}\Big[\frac{\sqrt{6}}{6}(0_H^{-+}\otimes 1_l^{++})|_{J=1}^{-+}-\frac{\sqrt{3}}{6}(0_H^{-+}\otimes 1_l^{+-})|_{J=1}^{--}\\
  &&+\frac{\sqrt{3}}{6}(1_H^{--}\otimes 1_l^{++})|_{J=1}^{--}-\frac{\sqrt{6}}{12}(1_H^{--}\otimes 1_l^{+-})|_{J=1}^{-+}\\
  &&+\frac{\sqrt{10}}{4}(1_H^{--}\otimes 1_l^{+-})|_{J=1}^{-+}\Big]\bigg(\frac{|(c\bar{d})(\bar{c}d)\rangle+|(c\bar{u})(\bar{c}u)\rangle}{\sqrt{2}}\bigg) \\
  &&+\frac{1}{\sqrt{2}}\Big[\frac{\sqrt{6}}{6}(0_H^{-+}\otimes 1_l^{++})|_{J=1}^{-+}+\frac{\sqrt{3}}{6}(0_H^{-+}\otimes 1_l^{+-})|_{J=1}^{--}\\
  &&-\frac{\sqrt{3}}{6}(1_H^{--}\otimes 1_l^{++})|_{J=1}^{--}-\frac{\sqrt{6}}{12}(1_H^{--}\otimes 1_l^{+-})|_{J=1}^{-+}\\
  &&+\frac{\sqrt{10}}{4}(1_H^{--}\otimes 1_l^{+-})|_{J=1}^{-+}\Big]\bigg(\frac{|(\bar{c}d)(c\bar{d})\rangle+|(\bar{c}u)(c\bar{u})\rangle}{\sqrt{2}}\bigg).
\end{eqnarray*}

The discovery mode of $Y(4260)$ is $J/\psi\pi^+\pi^-$. If assuming
the $\pi^+\pi^-$ pair in the final state is from the intermediate
$\sigma$ resonance, we write down the spin isospin wave function of
$J/\psi\sigma$
\begin{eqnarray*}
  |J/\psi\sigma\rangle &=& |(1_H^{--}\otimes 0_l^{++})_0^{--}\rangle|(c\bar{c})\rangle|\frac{1}{\sqrt{2}}(d\bar{d}+u\bar{u})  \rangle,
\end{eqnarray*}
In the heavy quark symmetry limit, we find this decay mode is
suppressed.

However, the decay mode $\chi_{cJ}\omega$ is allowed with the spin
structures of the final states
\begin{eqnarray*}
  |\chi_{c0}\omega\rangle &=& \Big[\frac{1}{3}(1_H^{--}\otimes 0_l^{++})|_{J=1}^{--}-\frac{\sqrt{3}}{3}(1_H^{--}\otimes 1_l^{++})|_{J=1}^{--}\\
  &&+\frac{\sqrt{5}}{3}(1_H^{--}\otimes 2_l^{++})|_{J=1}^{--}\Big]|(c\bar{c})\rangle|\frac{1}{\sqrt{2}}(d\bar{d}+u\bar{u})  \rangle, \\
  |\chi_{c1}\omega\rangle &=& \Big[-\frac{\sqrt{3}}{3}(1_H^{--}\otimes 0_l^{++})|_{J=1}^{--}+\frac{1}{2}(1_H^{--}\otimes 1_l^{++})|_{J=1}^{--}\\
  &&+\frac{\sqrt{15}}{6}(1_H^{--}\otimes 2_l^{++})|_{J=1}^{--}\Big]|(c\bar{c})\rangle|\frac{1}{\sqrt{2}}(d\bar{d}+u\bar{u})  \rangle,  \\
  |\chi_{c2}\omega\rangle &=&\Big[\frac{\sqrt{5}}{3}(1_H^{--}\otimes 0_l^{++})|_{J=1}^{--}-\frac{\sqrt{15}}{6}(1_H^{--}\otimes 1_l^{++})|_{J=1}^{--}\\
  &&+\frac{1}{6}(1_H^{--}\otimes 2_l^{++})|_{J=1}^{--}\Big]|(c\bar{c})\rangle|\frac{1}{\sqrt{2}}(d\bar{d}+u\bar{u})  \rangle.
\end{eqnarray*}
In the heavy quark symmetry, we obtain the ratio of the strong
decays $Y(4260)\rightarrow \chi_{cJ}\omega$ ($J=0,1,2$), i.e.,
\begin{eqnarray*}
  &&\Gamma(\chi_{c0}\omega):\Gamma(\chi_{c1}\omega):\Gamma(\chi_{c2}\omega)\\
  &&= 4:3:5,
\end{eqnarray*}
where the phase space factors are ignored. Since the $\omega$ meson
can decay into $\pi^+\pi^-\pi^0$, then we can get the ratio of the
strong decays $Y(4260)\rightarrow \chi_{cJ}\pi^+\pi^-\pi^0$
($J=0,1,2$),
\begin{eqnarray*}
  &&\Gamma(\chi_{c0}\pi^+\pi^-\pi^0):\Gamma(\chi_{c1}\pi^+\pi^-\pi^0):\Gamma(\chi_{c2}\pi^+\pi^-\pi^0)\\
  &&= 4:3:5\, (2.11:1:1.28),
\end{eqnarray*}
where the ratio in the bracket is the result considering the phase
space factors. This ratio can also be used to test whether $Y(4260)$
has the $\frac{1}{\sqrt{2}}(D_1\bar{D}-D\bar{D}_1)$ structure.

The Belle Collaboration reported that there exists a charmonium-like
state $Y(4360)$ in the $\psi(2S)\pi^+\pi^-$ invariant mass spectrum
of the $e^+e^-\to \psi(2S)\pi^+\pi^-$ process \cite{Aubert:2007zz}.
$Y(4360)$ was suggested as an isoscalar state
$\frac{1}{\sqrt{2}}(D_1\bar{D}^\ast+D^\ast\bar{D}_1)$ state
\cite{zhu-review}. Then, its spin isospin wave function is
\begin{eqnarray*}
&&  |Y(4360)\rangle\\ &&= 
   \frac{1}{\sqrt{2}}\Big[\frac{\sqrt{3}}{6}(0_H^{-+}\otimes 1_l^{++})|_{J=1}^{-+}-\frac{\sqrt{6}}{12}(0_H^{-+}\otimes 1_l^{+-})|_{J=1}^{--}\\
  &&\quad+\frac{\sqrt{6}}{4}(1_H^{--}\otimes 1_l^{++})|_{J=1}^{--}-\frac{\sqrt{3}}{4}(1_H^{--}\otimes 1_l^{+-})|_{J=1}^{-+}\\
  &&\quad-\frac{\sqrt{5}}{4}(1_H^{--}\otimes 1_l^{+-})|_{J=1}^{-+}\Big]\bigg(\frac{|(c\bar{d})(\bar{c}d)\rangle+|(c\bar{u})(\bar{c}u)\rangle}{\sqrt{2}}\bigg) \\
  &&\quad+\frac{1}{\sqrt{2}}\Big[-\frac{\sqrt{3}}{6}(0_H^{-+}\otimes 1_l^{++})|_{J=1}^{-+}-\frac{\sqrt{6}}{12}(0_H^{-+}\otimes 1_l^{+-})|_{J=1}^{--}\\
  &&\quad+\frac{\sqrt{6}}{4}(1_H^{--}\otimes 1_l^{++})|_{J=1}^{--}+\frac{\sqrt{3}}{4}(1_H^{--}\otimes 1_l^{+-})|_{J=1}^{-+}\\
  &&\quad+\frac{\sqrt{5}}{4}(1_H^{--}\otimes 1_l^{+-})|_{J=1}^{-+}\Big]\bigg(\frac{|(\bar{c}d)(c\bar{d})\rangle+|(\bar{c}u)(c\bar{u})\rangle}{\sqrt{2}}\bigg).
\end{eqnarray*}
The $\chi_{cJ}\omega$ are the allowed decay modes of $Y(4360)$. We
have the following ratio
\begin{eqnarray*}
  &&\Gamma(\chi_{c0}\omega):\Gamma(\chi_{c1}\omega):\Gamma(\chi_{c2}\omega)\\
  &&=4:3:5.
  \end{eqnarray*}
We can get the ratio of these strong decays $Y(4360)\rightarrow
\chi_{cJ}\pi^+\pi^-\pi^0$ ($J=0,1,2$)
\begin{eqnarray*}
  &&\Gamma(\chi_{c0}\pi^+\pi^-\pi^0):\Gamma(\chi_{c1}\pi^+\pi^-\pi^0):\Gamma(\chi_{c2}\pi^+\pi^-\pi^0)\\
  &&= 4:3:5\,(1.94:1:1.36).
\end{eqnarray*}
where we assume the $3\pi$ in the final states comes from the
intermediate $\omega$ contribution. The results in the bracket
include the phase space factors. We also suggest that future
experiments carry out the measurement of this ratio, which can be
applied to test the
$\frac{1}{\sqrt{2}}(D_1\bar{D}^\ast+D^\ast\bar{D}_1)$ assignment of
$Y(4360)$.

In addition, $h_c\eta$ and $\chi_{cJ}\omega$ are the allowed decay
modes of both $Y(4260)$ and $Y(4360)$. The spin isospin wave
function of $h_c\eta$ is
\begin{eqnarray*}
  |h_c\eta \rangle &=&| (0_H^{-+}\otimes 1_l^{+-})_0^{--}\rangle|(c\bar{c})\rangle|\frac{1}{\sqrt{2}}(d\bar{d}+u\bar{u})  \rangle.
\end{eqnarray*}
In the  heavy quark limit, the $D$ and $D^\ast$ mesons belong to the
same heavy spin multiplet. Hence, $Y(4260)$ and $Y(4360)$ have the
same spatial wave functions and the same spatial matrix elements of
these discussed strong decays, which leads to quite simple ratios
between their decay widths, i.e.,
\begin{eqnarray*}
  \frac{\Gamma(Y(4260)\rightarrow h_c\eta)}{\Gamma(Y(4360)\rightarrow h_c\eta)} &=& 2:1\,(1.65:1),\\
  \frac{\Gamma(Y(4260)\rightarrow \chi_{c0}\pi^+\pi^-\pi^0)}{\Gamma(Y(4360)\rightarrow \chi_{c0}\pi^+\pi^-\pi^0)} &=& 2:9\,(1:6.39), \\
  \frac{\Gamma(Y(4260)\rightarrow \chi_{c1}\pi^+\pi^-\pi^0)}{\Gamma(Y(4360)\rightarrow \chi_{c1}\pi^+\pi^-\pi^0)} &=& 2:9\,(1:6.98), \\
  \frac{\Gamma(Y(4260)\rightarrow \chi_{c2}\pi^+\pi^-\pi^0)}{\Gamma(Y(4360)\rightarrow \chi_{c2}\pi^+\pi^-\pi^0)} &=& 2:9\,(1:7.39),
\end{eqnarray*}
where the results in the brackets are from the consideration of the
phase space factors.

\subsection{$X(3872)$}

There were extensive discussions of $X(3872)$ as an isoscalar
$D\bar{D}^\ast$ molecular state with $J^{pc}=1^{++}$
\cite{Close:2003sg,Voloshin:2003nt,Wong:2003xk,Swanson:2003tb,Tornqvist:2004qy,Suzuki:2005ha,Liu:2008fh,Thomas:2008ja,Lee:2009hy,Li:2012cs}.
In this picture, its spin isospin wave function reads as
\begin{eqnarray*}
 && |X(3872)\rangle\nonumber\\ &&= 
  \frac{1}{\sqrt{2}}\Big[\frac{1}{2}(0_H^{-+}\otimes 1_l^{--})|_{J=1}^{+-}-\frac{1}{2}(1_H^{--}\otimes 0_l^{-+})|_{J=1}^{+-}\\
  &&\quad+\frac{1}{\sqrt{2}}(1_H^{--}\otimes 1_l^{--})|_{J=1}^{++}\Big]\bigg(\frac{|(c\bar{d})(\bar{c}d)\rangle+|(c\bar{u})(\bar{c}u)\rangle}{\sqrt{2}}\bigg) \\
  &&\quad+\frac{1}{\sqrt{2}}\Big[-\frac{1}{2}(0_H^{-+}\otimes 1_l^{--})|_{J=1}^{+-}+\frac{1}{2}(1_H^{--}\otimes 0_l^{-+})|_{J=1}^{+-}\\
  &&\quad+\frac{1}{\sqrt{2}}(1_H^{--}\otimes 1_l^{--})|_{J=1}^{++}\Big]\bigg(\frac{|(\bar{c}d)(c\bar{d})\rangle+|(\bar{c}u)(c\bar{u})\rangle}{\sqrt{2}}\bigg).
\end{eqnarray*}
$\psi(1^3D_1)\omega$ and $\psi(1^3D_2)\omega$ are its kinematically
forbidden modes. The $\chi_{c1}\sigma$ and $J/\psi\omega$ modes are
allowed with the corresponding spin isospin wave functions
\begin{eqnarray*}
  |\chi_{c1}\sigma \rangle &=&| (1_H^{--}\otimes 1_l^{--})_1^{++}\rangle|(c\bar{c})\rangle|\frac{1}{\sqrt{2}}(d\bar{d}+u\bar{u})  \rangle,\\
  |J/\psi\omega\rangle &=& (1_H^{--}\otimes 1_l^{--})|_{J=1}^{++}|(c\bar{c})\rangle|\frac{1}{\sqrt{2}}(d\bar{d}+u\bar{u})  \rangle. \\
\end{eqnarray*}
In the heavy quark symmetry limit, we find that the decay modes
$\chi_{c1}\sigma$ and $J/\psi\omega$ are related to the spin
configurations $(1_H^{--}\otimes 1_l^{--})_1^{++}$ and
$(1_H^{--}\otimes 1_l^{--})|_{J=1}^{++}$.


\subsection{$Z_c(3900)$ and $Z_c(4020)$ }

$Z_c(3900)$ was first reported by BESIII in the $e^+e^-\rightarrow
J/\psi\pi^+\pi^-$ process at $\sqrt{s}=4.26\,\,\textmd{GeV}$
\cite{Ablikim:2013mio,Liu:2013dau}, which was
suggested as the charged isovector state of the $D\bar{D}^\ast$
system with $I^G(J^p)=1^+(1^+)$. $Z_c(4020)$ was observed in the
$h_c\pi^{\pm}$ invariant mass spectrum of $e^+e^-\rightarrow
h_c\pi^+\pi^-$ at $\sqrt{s}=4.26\,\,\textmd{GeV}$
\cite{Ablikim:2013wzq}. A similar state $Z_c(4025)$ was reported by
BESIII in $e^+e^-\rightarrow (D^\ast\bar{D}^\ast)^{\pm}\pi^{\mp}$ at
$\sqrt{s}=4.26\,\,\textmd{GeV}$ \cite{Ablikim:2013emm}. $Z_c(4020)$
(or $Z_c(4025)$) may be the charged isovector state of
$D^\ast\bar{D}^\ast$ system with $I^G(J^p)=1^+(1^+)$
\cite{He:2013nwa}. The spin isospin wave functions of their neutral
partners read as
\begin{eqnarray*}
  |Z_c(3900)\rangle &=& 
  \frac{1}{\sqrt{2}}\Big[\frac{1}{2}(0_H^{-+}\otimes 1_l^{--})|_{J=1}^{+-}-\frac{1}{2}(1_H^{--}\otimes 0_l^{-+})|_{J=1}^{+-}\\
  &&+\frac{1}{\sqrt{2}}(1_H^{--}\otimes 1_l^{--})|_{J=1}^{++}\Big]\bigg(\frac{|(c\bar{d})(\bar{c}d)\rangle-|(c\bar{u})(\bar{c}u)\rangle}{\sqrt{2}}\bigg)\\
  &&-\frac{1}{\sqrt{2}}\Big[-\frac{1}{2}(0_H^{-+}\otimes 1_l^{--})|_{J=1}^{+-}+\frac{1}{2}(1_H^{--}\otimes 0_l^{-+})|_{J=1}^{+-}\\
  &&+\frac{1}{\sqrt{2}}(1_H^{--}\otimes 1_l^{--})|_{J=1}^{++}\Big]\bigg(\frac{|(\bar{c}d)(c\bar{d})\rangle-|(\bar{c}u)(c\bar{u})\rangle}{\sqrt{2}}\bigg),\\
  |Z_c(4020)\rangle &=&
  \frac{1}{\sqrt{2}}\Big[\frac{1}{\sqrt{2}}(0_H^{-+}\otimes 1_l^{--})|_{J=1}^{+-}+\frac{1}{\sqrt{2}}(1_H^{--}\otimes 0_l^{-+})|_{J=1}^{+-}
  \Big]\\
  &&\times\Bigg(\bigg(\frac{|(c\bar{d})(\bar{c}d)\rangle-|(c\bar{u})(\bar{c}u)\rangle}{\sqrt{2}}\bigg)\\
  &&+\bigg(\frac{|(\bar{c}d)(c\bar{d})\rangle-|(\bar{c}u)(c\bar{u})\rangle}{\sqrt{2}}\bigg)\Bigg).
\end{eqnarray*}
If ignoring the heavy quark symmetry, their allowed decay modes are
$J/\psi\pi$, $\psi(1^3D_1)\pi$ and $\eta_c\rho$ with the spin
structures
\begin{eqnarray*}
  |J/\psi\pi^0 \rangle &=&| (1_H^{--}\otimes 0_l^{-+})_0^{+-}\rangle|(c\bar{c})\rangle|\frac{1}{\sqrt{2}}(d\bar{d}-u\bar{u})  \rangle,\\
  |\psi(1^3D_1)\pi^0 \rangle &=&| (1_H^{--}\otimes 2_l^{-+})_0^{+-}\rangle|(c\bar{c})\rangle|\frac{1}{\sqrt{2}}(d\bar{d}-u\bar{u})  \rangle,\\
  |\eta_c\rho^0\rangle &=& (0_H^{-+}\otimes 1_l^{+-})|_{J=1}^{--}|(c\bar{c})\rangle|\frac{1}{\sqrt{2}}(d\bar{d}-u\bar{u})  \rangle.
\end{eqnarray*}
Considering the heavy quark symmetry, we find that the decay mode
$\psi(1^3D_1)\pi$ is suppressed. While $J/\psi\pi$, $\eta_c\rho$ and
$\eta_{c2}\rho$ are still allowed, which are consistent with the
conclusion in Ref. \cite{He:2013nwa}. We obtain the ratios between
the decay widths of $Z_c(3900)$ and $Z_c(4020)$, i.e.,
\begin{eqnarray*}
  \frac{\Gamma(Z_c(3900)\rightarrow J/\psi\pi^0)}{\Gamma(Z_c(4020)\rightarrow J/\psi\pi^0)} &=& 1:1\,(1:1.07), \\
  \frac{\Gamma(Z_c(3900)\rightarrow \eta_c\rho^0)}{\Gamma(Z_c(4020)\rightarrow \eta_c\rho^0)} &=& 1:1\,(1:2.47),
\end{eqnarray*}
where the values in the brackets are the results after considering
the phase space factors.

\subsection{$Y(4274)$}

$Y(4274)$ was observed in the $J/\psi\phi$ invariant mass spectrum
by CDF \cite{Yi:2010aa}, which can be as an S-wave isoscalar state
$D_s\bar{D}_{s0}(2317)$ with $J^{PC}=0^{-+}$ \cite{Liu:2010hf},
whose spin structure reads as
\begin{eqnarray*}
  |Y(4274)\rangle &=& 
  \Big[-\frac{1}{2}(0_H^{-+}\otimes 0_l^{+-})|_{J=0}^{--}+\frac{1}{2}(1_H^{--}\otimes 1_l^{++})|_{J=0}^{--}\\
  &&+\frac{\sqrt{2}}{2}(1_H^{--}\otimes 1_l^{+-})|_{J=0}^{-+}\Big]|c\bar{s};\bar{c}s\rangle \\
  &&-\Big[\frac{1}{2}(0_H^{-+}\otimes 0_l^{+-})|_{J=0}^{--}-\frac{1}{2}(1_H^{--}\otimes 1_l^{++})|_{J=0}^{--}\\
  &&+\frac{\sqrt{2}}{2}(1_H^{--}\otimes 1_l^{+-})|_{J=0}^{-+}\Big]|\bar{c}s;c\bar{s}\rangle.\\
\end{eqnarray*}
In the heavy quark spin symmetry, $Y(4274)$ can decay into
$J/\psi\phi$ via the P-wave transition, where the spin configuration
$(1_H^{--}\otimes 1_l^{+-})|_{J=0}^{-+}$ is dominant with the decay
width proportional to the reduced matrix element
$|\langle1,0\|H_{eff}(\phi)\|1\rangle|$.

\subsection{$Y(3940)$ and $Y(4140)$}

$Y(3940)$ was observed by the BaBar Collaboration \cite{Abe:2004zs}
in $B\to K J/\psi\omega$, while the CDF Collaboration observed
$Y(4140)$ in $B\to K J/\psi\phi$ \cite{Aaltonen:2009tz}. $Y(3940)$
and $Y(4140)$ can be the candidates of the $D^\ast \bar{D}^\ast$ and
$D_s^\ast \bar{D}_s^\ast$ molecular systems, respectively
\cite{Liu:2009ei,Liu:2008tn}. Their quantum numbers may be
$J^{PC}=0^{++}$ or $J^{PC}=2^{++}$ \cite{Liu:2009ei,Liu:2008tn}. In
the following, we discuss their decay behaviors in these two cases.

\subsubsection{$J^{PC}=0^{++}$}

If $Y(3940)$ and $Y(4140)$ are the $0^{++}$ molecular states, their
spin structures read as
\begin{eqnarray*}
 && |Y(3940)\rangle\nonumber\\&&=
  \Big[\frac{\sqrt{3}}{2}(0_H^{-+}\otimes 0_l^{-+})|_{J=0}^{++}-\frac{1}{2}(1_H^{--}\otimes 1_l^{--})|_{J=0}^{++}\Big]\\
  &&\quad\times\Bigg(\frac{|(c\bar{d})(\bar{c}d)\rangle-|(c\bar{u})(\bar{c}u)\rangle}{\sqrt{2}}+\frac{|(\bar{c}d)(c\bar{d})\rangle-|(\bar{c}u)(c\bar{u})\rangle}{\sqrt{2}}\Bigg),\\
 && |Y(4140)\rangle \nonumber\\&&=
  \Big[\frac{\sqrt{3}}{2}(0_H^{-+}\otimes 0_l^{-+})|_{J=0}^{++}-\frac{1}{2}(1_H^{--}\otimes 1_l^{--})|_{J=0}^{++}\Big]|(c\bar{s})(\bar{c}s)\rangle.
\end{eqnarray*}
The kinematically allowed decay modes of $Y(3940)$ are
$\chi_{c0}\sigma$, $\eta_c\eta$ and $J/\psi\omega$, whose decay
widths are proportional to the reduced matrix elements
$|\langle0,1\|H_{eff}(\sigma)\|1\rangle|$,
$|\langle0,0\|H_{eff}(\eta)\|0\rangle|$ and
$|\langle1,2\|H_{eff}(\omega)\|1\rangle|$, respectively. $Y(4140)$
can decay into $J/\psi\phi$ through the spin configuration
$(1_H^{--}\otimes 1_l^{--})|_{J=0}^{++}$, whose decay width is
proportional to the reduced matrix element
$|\langle1,0\|H_{eff}(\phi)\|1\rangle|$.

\subsubsection{$J^{PC}=2^{++}$}

If both $Y(3940)$ and $Y(4140)$ are the $2^{++}$ tensor states, we
can write down their spin structures in heavy quark limit, i.e.,
\begin{eqnarray*}
 && |Y(3940)\rangle \nonumber\\&&= (1_H^{--}\otimes1_l^{--})|_{J=2}^{++}\\
  &&\quad \times\Bigg(\frac{|(c\bar{d})(\bar{c}d)\rangle-|(c\bar{u})(\bar{c}u)\rangle}{\sqrt{2}}+\frac{|(\bar{c}d)(c\bar{d})\rangle-|(\bar{c}u)(c\bar{u})\rangle}{\sqrt{2}}\Bigg),\\
 && |Y(4140)\rangle =
  (1_H^{--}\otimes1_l^{--})|_{J=2}^{++}|c\bar{s};\bar{c}s\rangle.
\end{eqnarray*}
The kinematically allowed decay modes of $Y(3940)$ are
$\chi_{c2}\sigma$, $J/\psi\omega$ and $\eta_{c2}\eta$. In the heavy
quark symmetry limit, the decay modes $\chi_{c2}\sigma$ and
$J/\psi\omega$ are allowed with their widths proportional to the
reduced matrix elements $|\langle0,0\|H_{eff}(\sigma)\|1\rangle|$
and $|\langle1,0\|H_{eff}(\omega)\|1\rangle|$ respectively. The
decay mode $\eta_{c2}\eta$ is suppressed due to the conservations of
heavy and light spin. In this case, $Y(4140)$ can also decay into
$J/\psi\phi$ through the spin configuration $(1_H^{--}\otimes
1_l^{--})|_{J=2}^{++}$, whose decay width is proportional to the
reduced matrix element $|\langle1,0\|H_{eff}(\phi)\|1\rangle|$.

\section{Summary}\label{sec5}

More and more charmonium-like and bottomium-like states were
reported in the past twelve years. Many of them are very close to
the open-charm or open-bottom threshold. Some are even charged. Many
the so-called XYZ states do not fit into the traditional quark model
spectrum easily. Many theoretical speculations were proposed to
understand their inner structures. Among them, the molecule picture
is quite popular. Historically, the deuteron has been identified to
be a very loosely bound molecular state composed of a proton and
neutron. It is very natural to investigate whether the loosely bound
di-meson molecular states exist or not. Dynamical calculation based
on the one boson exchange model may explore the possible existence
of the di-meson molecular states. On the other hand, the decay
pattern and production mechanism of these di-meson systems may also
shed light on their inner structures.

Under the heavy quark symmetry, the QED and QCD interactions don't
flip the heavy quark spin. The conservation of the heavy spin
together with the isospin, total angular momentum and other quantum
numbers such as parity, C parity and G parity provide an effective
scheme to probe the inner structures of the XYZ states through their
decay and production behaviors. We have extensively discussed the
three classes of strong decays
$B_{(1,2)}\bar{B}^{(\ast)}\rightarrow(b\bar{b})+light\,\,meson$,
$(b\bar{b})\rightarrow B_{(1,2)}\bar{B}^{(\ast)}+light\,\,meson$,
$B_{(1,2)}\bar{B}^{(\ast)}\rightarrow
B_{(1,2)}\bar{B}^{(\ast)}+light\,\,meson$, corresponding to the
strong decays of one molecular (resonant) state into a bottomonia,
one bottomonia into a molecular (resonant) state, and strong decays
of one molecular (resonant) state into another respectively. With
the same formalism, we also give detailed discussions on the
possible hidden-charm molecules (resonances).

If either the initial systems or final states belong to the same
heavy spin multiplet, the spatial matrix elements of these strong
decays are the same, which leads to quite simple ratios between
their decay widths. Different assumptions of the underlying
structures will give different decay ratios and different production
behaviors, which will help probe the inner structures of the XYZ
states after comparison with experiment measurements. For instance,
there are theoretical speculations that $Y(4260)$ may be an
isoscalar $\frac{1}{\sqrt{2}}(D_1\bar{D}-D\bar{D}_1)$ molecule. In
the heavy quark symmetry limit, $Y(4260)$ does not decay into
$J/\psi\pi^+\pi^-$. In other words, the discovery mode
$J/\psi\pi^+\pi^-$ of $Y(4260)$ disfavors the
$\frac{1}{\sqrt{2}}(D_1\bar{D}-D\bar{D}_1)$ molecule scheme.

In short summary, the strong decay behaviors of the $XYZ$ states
encode important information on their underlying structures.
Systematical experimental measurement of these decay behaviors will
be helpful to judge the various theoretical interpretations of the
$XYZ$ states. Hopefully the present extensive investigations will be
useful to illuminate the future strong decay data.

\subsection*{Acknowledgments}

This project is supported by the National Natural Science Foundation
of China under Grants No. 11222547, No. 11175073, No. 11035006, No.
11375240 and No. 11261130311, the Ministry of Education of China
(FANEDD under Grant No. 200924, SRFDP under Grant No. 2012021111000,
and NCET), the China Postdoctoral Science Foundation under Grant No.
2013M530461, the Fok Ying Tung Education Foundation (Grant No.
131006).

\end{document}